\documentclass[longauth]{aa}

\usepackage{graphicx}
\usepackage{txfonts}
\usepackage{gensymb,amssymb}
\usepackage{xcolor}
\usepackage[version=3]{mhchem}
\usepackage{caption,subcaption}
\usepackage{multirow}

\usepackage[colorlinks = true,
            linkcolor = blue,
            urlcolor  = blue,
            citecolor = blue,
            anchorcolor = blue]{hyperref}

\begin{document}

\title{MINDS. The DR~Tau disk II: probing the hot and cold \ce{H_2O} reservoirs in the JWST-MIRI spectrum}
\author{Milou Temmink\orcid{0000-0002-7935-7445}\inst{1} \and
        Ewine F. van Dishoeck\orcid{0000-0001-7591-1907}\inst{1,2} \and
        Danny Gasman\orcid{0000-0002-1257-7742}\inst{3} \and
        Sierra L. Grant\orcid{0000-0002-4022-4899}\inst{2} \and 
        Beno\^it Tabone\inst{4} \and
        Manuel G\"udel\orcid{0000-0001-9818-0588}\inst{5,6} \and
        Thomas Henning\orcid{0000-0002-1493-300X}\inst{7} \and
        David Barrado\orcid{0000-0002-5971-9242}\inst{8} \and 
        Alessio Caratti o Garatti\orcid{0000-0001-8876-6614}\inst{9,10} \and
        Adrian M. Glauser\orcid{0000-0001-9250-1547}\inst{6} \and
        Inga Kamp\orcid{0000-0001-7455-5349}\inst{11} \and
        Aditya M. Arabhavi\orcid{0000-0001-8407-4020}\inst{11} \and
        Hyerin Jang\orcid{0000-0002-6592-690X}\inst{12} \and
        Nicolas Kurtovic\orcid{?}\inst{2} \and
        Giulia Perotti\orcid{0000-0002-8545-6175}\inst{7} \and
        Kamber Schwarz\orcid{0000-0002-6429-9457}\inst{7} \and
        Marissa Vlasblom\orcid{0000-0002-3135-2477}\inst{1}}
\institute{Leiden Observatory, Leiden University, 2333 CC Leiden, the Netherlands \\
          \email{temmink@strw.leidenuniv.nl} \and
          Max-Planck-Institut f\"ur Extraterrestrische Physik, Giessenbachstraße 1, D-85748 Garching, Germany \and
          Institute of Astronomy, KU Leuven, Celestijnenlaan 200D, 3001 Leuven, Belgium \and
          Universit\'e Paris-Saclay, CNRS, Institut d’Astrophysique Spatiale, 91405, Orsay, France \and
          Dept. of Astrophysics, University of Vienna, T\"urkenschanzstr. 17, A-1180 Vienna, Austria \and
          ETH Z\"urich, Institute for Particle Physics and Astrophysics, Wolfgang-Pauli-Str. 27, 8093 Z\"urich, Switzerland \and 
          Max-Planck-Institut f\"{u}r Astronomie (MPIA), K\"{o}nigstuhl 17, 69117 Heidelberg, Germany \and 
          Centro de Astrobiolog\'ia (CAB), CSIC-INTA, ESAC Campus, Camino Bajo del Castillo s/n, 28692 Villanueva de la Ca\~nada, Madrid, Spain \and 
          INAF – Osservatorio Astronomico di Capodimonte, Salita Moiariello 16, 80131 Napoli, Italy \and 
          Dublin Institute for Advanced Studies, 31 Fitzwilliam Place, D02 XF86 Dublin, Ireland \and 
          Kapteyn Astronomical Institute, Rijksuniversiteit Groningen, Postbus 800, 9700AV Groningen, The Netherlands \and 
          Department of Astrophysics/IMAPP, Radboud University, PO Box 9010, 6500 GL Nijmegen, The Netherlands}
\date{Received 12/04/2024; accepted 05/07/2024}


\abstract
{The Medium Resolution Spectrometer (MRS) of the Mid-InfraRed Instrument (MIRI) on the James Webb Space Telescope (JWST) gives insights into the chemical richness and complexity of the inner regions of planet-forming disks. Several disks that are compact in the millimetre dust emission have been found by Spitzer to be particularly bright in \ce{H_2O}, which is thought to be caused by the inward drift of icy pebbles. Here, we analyse the \ce{H_2O}-rich spectrum of the compact disk DR~Tau using high-quality JWST-MIRI observations.}
{We infer the \ce{H_2O} column densities (in cm$^{-2}$) using methods presented in previous works, as well as introducing a new method to fully characterise the pure rotational spectrum. We aim to further characterise the abundances of \ce{H_2O} in the inner regions of this disk and its abundance relative to \ce{CO}. We also search for emission of other molecular species, such as \ce{CH_4}, \ce{NH_3}, \ce{CS}, \ce{H_2}, \ce{SO_2}, and larger hydrocarbons; commonly detected species, such as \ce{CO}, \ce{CO_2}, \ce{HCN}, and \ce{C_2H_2}, have been investigated in our previous paper.}
{We first use 0D local thermodynamic equilibrium (LTE) slab models to investigate the excitation properties observed in different wavelength regions across the entire spectrum, probing both the ro-vibrational and rotational transitions. To further analyse the pure rotational spectrum ($\geq$10 $\mathrm{\mu}$m), we use the spectrum of a large, structured disk (CI~Tau) as a template to search for differences with our compact disk. Finally, we fit multiple components to characterise the radial (and vertical) temperature gradient(s) present in the spectrum of DR~Tau. }
{The 0D slab models indicate a radial gradient in the disk, as the excitation temperature (emitting radius) decreases (increases) with increasing wavelength, which is confirmed by the analysis involving the large disk template. To explain the derived emitting radii, we need a larger inclination for the inner disk ($i\sim$10-23\degree), agreeing with our previous analysis on \ce{CO}. From our multi-component fit, we find that at least three temperature components ($T_1\sim$800 K, $T_2\sim$470 K, and $T_3\sim$180 K) are required to reproduce the observed rotational spectrum of \ce{H_2O} arising from the inner $R_\textnormal{em}\sim$0.3-8 au. By comparing line ratios, we derived an upper limit on the column densities (in cm$^{-2}$) for the first two components of $\log_{10}(N)\leq$18.4 within $\sim$1.2 au. We note that the models with a pure temperature gradient provide as robust results as the more complex models, which include spatial line shielding. No robust detection of the isotopologue \ce{H_2 ^{18}O} can be made and upper limits are provided for other molecular species.}
{Our analysis confirms the presence of a pure radial temperature gradient present in the inner disk of DR Tau, which can be described by at least three components. This gradient scales roughly as $\sim R_\textnormal{em}^{-0.5}$ in the emitting layers, in the inner 2 au. As the observed \ce{H_2O} is mainly optically thick, a lower limit on the abundance ratio of \ce{H_2O}/\ce{CO}$\sim$0.17 is derived, suggesting a potential depletion of \ce{H_2O}. Similarly to previous work, we detect a cold \ce{H_2O} component ($T\sim$180 K) originating from near the snowline, now with a multi-component analysis. Yet, we cannot conclude whether an enhancement of the \ce{H_2O} reservoir is observed following radial drift. A consistent analysis of a larger sample is necessary to study the importance of drift in enhancing the \ce{H_2O} abundances.}


\maketitle

\section{Introduction}

\begin{figure*}[ht!]
    \centering
    \includegraphics[width=\textwidth]{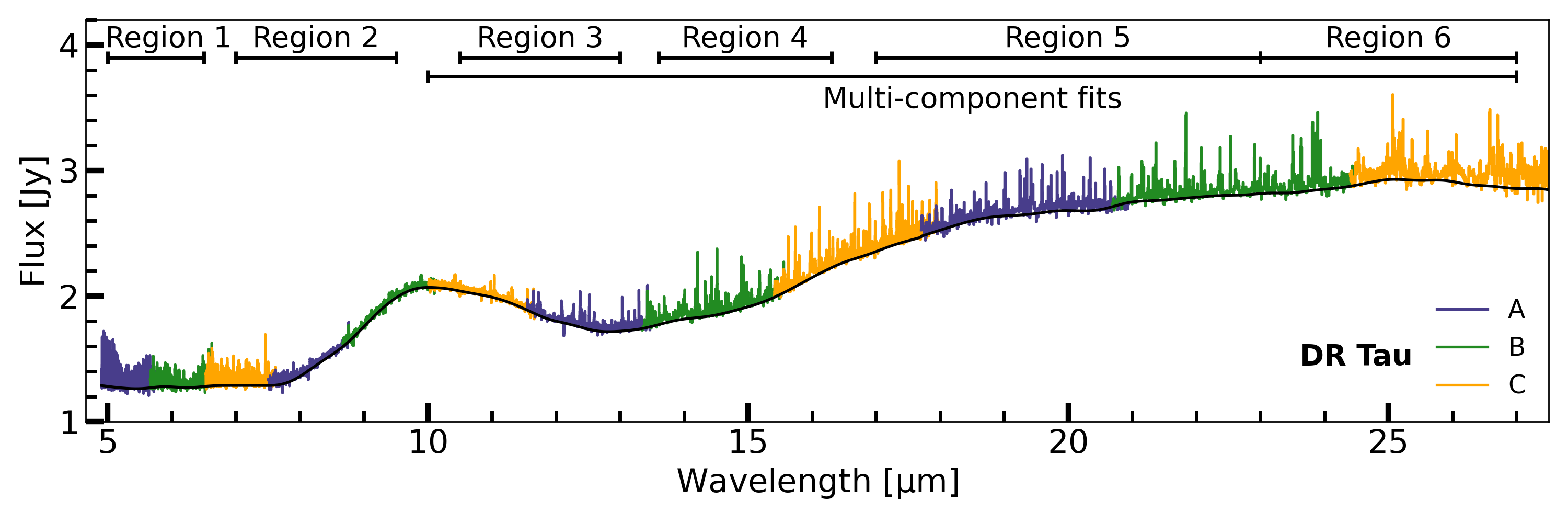}
    \caption{The full JWST-MIRI spectrum (4.9-27.5 $\mathrm{\mu}$m) of DR~Tau. The different wavelength ranges (subbands) of each MIRI/MRS channel are shown in blue ('A'), green ('B'), and orange ('C'), respectively. The wavelength regions over which \ce{H_2O} will be analysed are indicated by the horizontal bars. The continuum fit is indicated by the black line.}
    \label{fig:FullSpectrum}
\end{figure*}

\ce{H_2O} is a key ingredient in making habitable planets. As the bulk of exoplanets form in the inner-most regions ($<$10 au; \citealt{MorbidelliEA12,DJ18}) of planet-forming disks, their (atmospheric) composition is determined by the available elemental abundances. Therefore, it is of importance to characterise these regions in detail and, in particular, to analyse the \ce{H_2O} emission present. \\
\indent The recently launched \textit{James Webb Space Telescope} (JWST; \citealt{RigbyEA23}) and, in particular, the Medium Resolution Spectroscopy mode (MRS; \citealt{WellsEA15}) of the Mid-Infrared Instrument (MIRI; \citealt{RiekeEA15,WrightEA15}), provides the best opportunity to study the chemical composition of these inner-most regions of planet-forming disks. The wide wavelength range of JWST-MIRI (4.9-27.9 $\mathrm{\mu}$m)  covers many transitions of \ce{H_2O}. These include the ro-vibrational lines at the shorter wavelengths between $\sim$5.0 and 8.0 $\mathrm{\mu}$m and the rotational transitions longward of 10.0 $\mathrm{\mu}$m \citep{MeijerinkEA09}. This forest of \ce{H_2O} lines across the mid-infrared wavelength provides insights into the different parts of the inner disk: the lines at shorter wavelengths are thought to probe the innermost gas, whereas those at longer wavelengths probe colder gas located at larger radial distances (e.g. \citealt{BlevinsEA16,BanzattiEA17,BanzattiEA23}, and \citealt{GasmanEA23Subm}). Observations suggest that \ce{H_2O} vapour is prevalent in the spectra of T-Tauri disks, even in those with inner, dust-depleted cavities, albeit with lower line strengths \citep{PerottiEA23,SchwarzEA24}. \\ 
\indent Furthermore, \citet{BanzattiEA20} have found a correlation between the flux of strong \ce{H_2O} lines measured with the \textit{Spitzer Space Telescope} and the radial dust disk size ($R_\textnormal{dust}$) as measured with ALMA. The proposed explanation for the {correlation is driven by the \ce{H_2O} abundances expected for three types of disks: compact disks ($R_\textnormal{dust}\lesssim$60 au) with efficient radial drift; large disks (60$\lesssim R_\textnormal{dust}\lesssim$300 au) with substructures; and large disks with substructures and an inner cavity. Substructures are able to trap icy pebbles in the outer disk, preventing them from drifting inside the \ce{H_2O} snowline and enhancing the \ce{H_2O} column density in the inner disk. Subsequently, these large disks are thought to have low \ce{H_2O} column densities in the inner disk, which are expected to be further depleted in the presence of an inner cavity. For the compact dust disks, where radial drift is thought to be very efficient (e.g., \citealt{FacchiniEA19}) and substructures are found to be less common \citep{LongEA19}, the \ce{H_2O} column density is thought to be enhanced through the sublimation of \ce{H_2O}-ice. However, not all compact disks show strong \ce{H_2O} emission, such as DN~Tau, for example \citep{PontoppidanEA10,SalykEA11,BanzattiEA20}. \\
\indent In this work we analyse the \ce{H_2O} emission in the JWST-MIRI spectrum of DR~Tau, a compact T-Tauri (K6) disk ($<60$ au, \citealt{LongEA19}) located at a distance of $\sim$195 pc \citep{GC18} in the Taurus star-forming region. DR~Tau has a mass of $M$=0.93 M$_\odot$, an effective temperature of $T_\textnormal{eff}$=4205 K, and a luminosity of $L$=0.63 L$_\odot$ \citep{LongEA19}. Observations with \textit{Spitzer} have shown that DR~Tau has one of the highest line-to-continuum ratios and contains a rich \ce{H_2O} reservoir (e.g., \citealt{SalykEA11} and \citealt{BanzattiEA20}). Using ground-based observatories, various previous studies have analysed some of the bright \ce{H_2O} lines at much higher spectral resolution at both near- and mid-infrared wavelengths \citep{SalykEA08,NajitaEA18,SalykEA19,BanzattiEA23}. High-resolution, ground-based ro-vibrational \ce{CO} observations at $\sim$4.6-5.3 $\mathrm{\mu}$m are also available for this disk, which have been analysed in \citet{BastEA11,BrownEA13,BanzattiEA22} and \citet{TemminkEA24}. \\
\indent \citet{TemminkEA24} used rotational diagrams to investigate the excitation properties of the various \ce{CO} isotopologues: \ce{^{12}CO}, \ce{^{13}CO}, and \ce{C^{18}O}. For the optically thin isotopologue \ce{C^{18}O} an excitation temperature of $T\sim$975 K, a column density of $N\sim$2.0$\times$10$^{16}$ cm$^{-2}$, and an emitting radius of $R_\textnormal{\ce{CO}}\sim$0.23 au were derived. Using these parameters and accounting for the involved isotopologue ratio, a total number of molecules of $\mathcal{N}_\textnormal{\ce{CO}}\sim$4.1$\times$10$^{44}$ was found for \ce{CO}. The physical parameters derived from the high spectral resolution \ce{CO} observations were shown to also provide an explanation for the \ce{CO} emission observed with JWST-MIRI. In addition, \citet{TemminkEA24} investigated the emission of \ce{CO_2}, \ce{HCN}, and \ce{C_2H_2}. Left unanalysed, however, was the plethora of \ce{H_2O} lines present in the spectrum. Due to the strength of the \ce{H_2O} lines, the high line-to-continuum ratio, and the disk having the most ground-based, ro-vibrational and rotational \ce{H_2O} transitions observed to date (see, for example, \citealt{NajitaEA18,SalykEA19,BanzattiEA23}), DR~Tau is one of the best candidates for an extensive analysis of the \ce{H_2O} emission and to make a detailed comparison with the physical structure and abundance of \ce{CO}, under the assumption that these molecules are at least partially co-spatial. In addition, due to its compactness, DR~Tau is also a good candidate to study the effects of radial drift on the observable \ce{H_2O} abundance. \\
\indent In this paper, we provide a deep analysis of the \ce{H_2O}-rich spectrum using three different methods: we analyse various wavelength regions in the spectrum using 0D local thermal equilibrium (LTE) slab models. Additionally, we investigate the difference between the spectrum of a compact disk with that of a large, structured disk. Finally, we introduce a multi-component, radial and vertical gradient slab model fitting technique, effectively providing a 1D LTE slab model, to derive the temperature and column density gradients of the emitting layers in the inner disk. \\
\indent This paper is structured as follows: in Section \ref{sec:Obs} we describe the JWST-MIRI observations of DR~Tau. Section \ref{sec:AR} provides the used analysis methods and the accompanying results, which are further discussed in Section \ref{sec:Disc}. Finally, our conclusions are summarised in Section \ref{sec:CS}. 

\section{Observations} \label{sec:Obs}
DR~Tau has been observed with JWST-MIRI as part of the JWST Guaranteed Time Observations Program MIRI mid-INfrared Disk Survey (MINDS, PID: 1282, PI: T. Henning; \citealt{HenningEA24}). The details of the observations are described in \citet{TemminkEA24}. For the analysis in this work, we have re-reduced the observations using a standard pipeline reduction (Version 1.12.3, \citealt{BushouseEA23}) and the photom updates released in November 2023, which significantly improve the reduction of Channel 4. This reduction uses the residual fringe correction implemented in the pipeline, uses an annulus background subtraction, and the spectrum is extracted using an aperture with a size of 2$\times$ the full width at half maximum (FWHM). We note that the photom updates do not have a significant impact on the other channels, indicating that the results listed in \citet{TemminkEA24} still hold. \\
\indent We have used the continuum subtraction method as described in \citet{TemminkEA24}, which includes Channel 4 now as well. This continuum subtraction method first filters out downwards spikes before estimating the baseline with the `Iterative Reweighted Spline Quantile Regression' (IRSQR) method included in the python-package \textsc{pybaselines} \citep{PyBaselines}. The estimated continuum is also visible in Figure \ref{fig:FullSpectrum}.

\subsection{Wavelength calibration}
Our results, the multi-component slab model fits (see Section \ref{sec:MCFits}) and the potential (non-)detection of \ce{H_2 ^{18}O} (see Section \ref{sec:H2-18-O}) may be partially influenced by the wavelength calibration of JWST-MIRI/MRS. The wavelength accuracy, as shown by \citet{ArgyriouEA23} and \citet{PontoppidanEA23subm}, can be off by 40-90 km s$^{-1}$ and the offset is the largest in Channel 4 as clearly shown in Figure 5 of \citet{PontoppidanEA23subm}. As Channel 4 covers a large range of important \ce{H_2O} transitions, as well as the strongest \ce{H_2 ^{18}O} transitions observable with JWST-MIRI, the velocity shifts can significantly impact the slab models and/or ones ability of detecting \ce{H_2 ^{18}O}. \\
\indent To account for the wavelength correction, we have used, similar to \citet{PontoppidanEA23subm}, a plethora of \ce{CO} and \ce{H_2O} lines to estimate the offset between the spectrum and an initial model through a cross-correlation technique. In our estimates, we have accounted for the heliocentric velocity of DR~Tau ($v_\textnormal{hel}\sim$27.6 km s$^{-1}$; \citealt{ArdilaEA02}). In total, we have used 483 (blended) transitions, yielding corrections between $\sim$-100 km s$^{-1}$ and $\sim$120 km s$^{-1}$ with the largest offsets found in Channel 4. As the silicate feature is dominated by noise and no strong \ce{H_2O} transitions are found between $\sim$8 and $\sim$10 $\mathrm{\mu}$m, we have not determined the offsets for any possible transitions present in this region. In addition, as the offsets for the shortest wavelengths are mostly concentrated around 0, we have also not applied corrections for these regions. The corrections are thus only applied for the longer wavelengths ($>$10 $\mathrm{\mu}$m), where the offsets are found to be the largest. Figure \ref{fig:WC} displays our offsets found for the different \ce{CO} and \ce{H_2O} transitions.

\begin{figure}[ht!]
    \centering
    \includegraphics[width=\columnwidth]{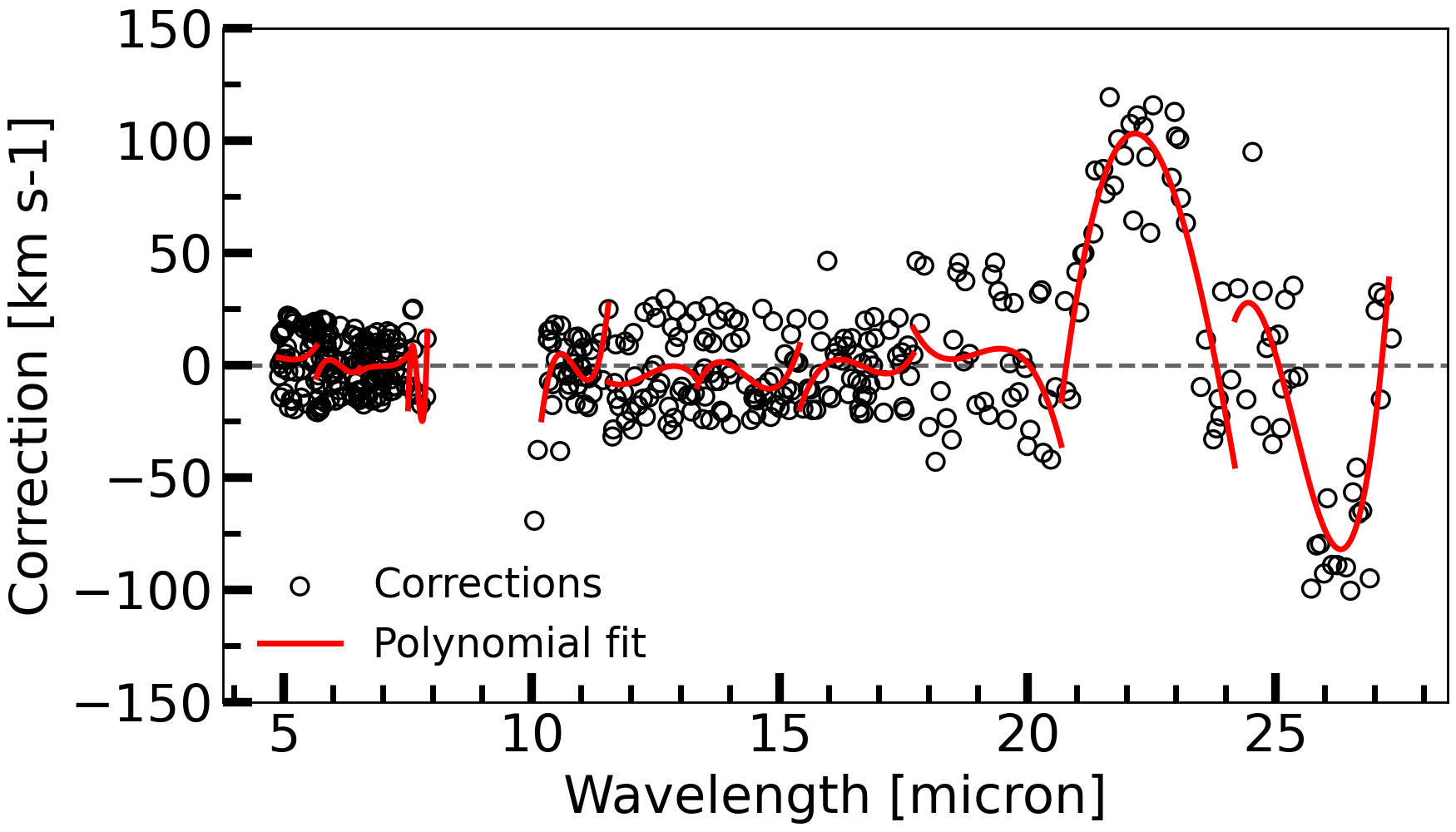}
    \caption{The estimated wavelength corrections (black circles) for the selected lines. The red lines are the polynomials of degree 3 fitted for each JWST-MIRI subband and are used in correcting the wavelength calibration.}
    \label{fig:WC}
\end{figure}

\indent To properly determine the wavelength corrections across the spectrum of DR~Tau, we have fitted polynomials of degree 3 through the found offsets for each JWST-MIRI subband. These polynomials, indicated by the red lines in Figure \ref{fig:WC}, have been used to correct the spectrum. We note that the corrections of \citet{PontoppidanEA23subm} have been implemented for the longest wavelengths in a newer version of the JWST reduction pipeline than used in this work (priv. comm.).

\section{Analysis \& results} \label{sec:AR}
\begin{figure*}[ht!]
    \centering
    \includegraphics[width=\textwidth]{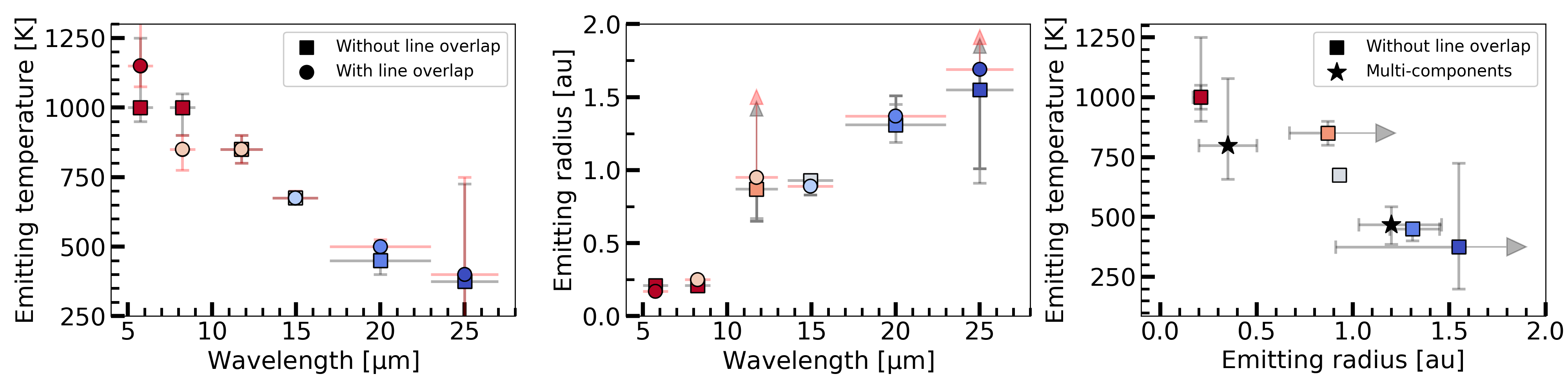}
    \caption{The inferred gas temperature (left) and emitting radii (middle) for the different wavelength regions, and the emitting temperature as a function of the emitting radius (right). We show the results for the slab model fits without (squares) and with (circles) line overlap. The colours indicate the varying gas temperatures, whereas the horizontal bars denote the wavelength ranges of the regions. Errorbars are shown for the derived quantities, where those for the models without line overlap are shown in black and those for the models with line overlap are shown in red. In addition, we show for comparison the results (blacks stars) of the three component fit (approach 3, see Section \ref{sec:RVG}) in the right panel.}
    \label{fig:TempRadius}
\end{figure*}

We analyse the \ce{H_2O} emission in DR~Tau using four different methods: first, in Section \ref{sec:H2O-Regions}, similar to \citet{GasmanEA23Subm}, we will investigate the \ce{H_2O} emission across separate regions in the spectrum using 0D local-thermal equilibrium (LTE) slab models (see Section \ref{sec:H2O-Regions}). The slab model fitting procedure is described in detail in \citet{TemminkEA24} (see also \citealt{GrantEA23} and \citealt{TaboneEA23}). As the new reduction yields slightly different fluxes, especially in Channel 4, we report in Table \ref{tab:Subband-Sigma} new values for the median flux, the signal-to-noise (S/N) ratio as given by the JWST \textit{Exposure Time Calculator} (ETC)\footnote{ETC: \url{https://jwst.etc.stsci.edu/}}, which accounts for the degrading transmission\footnote{For more information on the degrading transmission, see: \url{https://www.stsci.edu/contents/news/jwst/2023/miri-mrs-reduced-count-rate-update}} of JWST-MIRI/MRS, and for the estimated noise on the continuum used in our fitting procedure. For the \ce{H_2O} slab models, we use grids where the column density ($N$ in cm$^{-2}$) was allowed to vary between 12 and 22 in $\log_{10}$-spacing using a of $\Delta N$=0.1. We note that these column densities probe the upper layers of the disk, above the height where the dust becomes optically thick, or, in the case of optically thick lines, above the height where the emission lines become optically thick \citep{BrudererEA15,WoitkeEA18}. The temperature ($T$) was allowed to vary between 150 and 2500 K in steps of $\Delta T$=25 K, whereas the emitting radius ($R_\textnormal{em}$) was allowed to take on values between 0.01 and 10 au, using step sizes of $\Delta R$=0.02 au. The emitting radius is used as a parametrisation of the emitting area, assuming $A=\pi R_\textnormal{em}^2$. We note that the emitting area does not have to be a circle, it can have any shape (i.e. an annulus) at any radial distance from the star. In addition to regions listed in \citet{GasmanEA23Subm} (5.0-6.5, 13.6-16.3, 17.0-23.0, and 23.0-27.0 $\mathrm{\mu}$m), we also investigate the \ce{H_2O} emission in the regions between 7.0-9.5 and 10.5-13.0 $\mathrm{\mu}$m. The different wavelength regions are also indicated in Figure \ref{fig:FullSpectrum}. \\
\indent Second (see Section \ref{sec:Excess}), we follow the method described in \citet{BanzattiEA23Subm}, where a continuum-subtracted spectrum of a large disk (CI~Tau) is used to investigate a second, colder \ce{H_2O} component in a compact disk (GK~Tau), after scaling for the differences in distances and luminosity of the \ce{H_2O} lines with highest upper energy levels (6000-8000 K). They expect this second component to trace the enrichment of the \ce{H_2O} gas near the snowline, following the sublimation of icy mantles of drifting grains. As DR~Tau is a compact disk, we expect pebble drift to be efficient and, subsequently, we expect such a second \ce{H_2O} component should also be visible. Here, we will also use the spectrum of CI~Tau as the template spectrum to investigate the presence of an excess, cold second component in DR~Tau (see Section \ref{sec:LDT-CITau}). For consistency, the CI~Tau observations have been reduced in the same way as done for DR~Tau (see Section \ref{sec:Obs}). \\
\indent As a third method (Section \ref{sec:MCFits}), we investigate how well 0D LTE slab models can reproduce the observed spectrum when implementing radial and vertical temperature gradients. \\
\indent Finally, we obtain limits on the \ce{H_2O} column density (in cm$^{-2}$) by comparing the ratios of the Einstein-A coefficients and the line fluxes of \ce{H_2O} line pairs with the same upper energy level (Section \ref{sec:FEACR}, see also \citealt{GasmanEA23Subm}). We also search for emission of the rare isotopologue \ce{H_2 ^{18}O} in Section \ref{sec:H2-18-O}, using the isolated lines listed in \citet{CalahanEA22}. \\
\indent Additionally, in Section \ref{sec:OtherMols}, we fit an LTE slab model to the \ce{OH} emission over the wavelength region between 10.0 and 27.5 $\mathrm{\mu}$m and investigate the presence of other molecular species, such as \ce{CH_4}, \ce{NH_3}, and larger hydrocarbons. We compare their emission properties with those of \ce{H_2O}, \ce{CO_2}, \ce{HCN}, and \ce{C_2H_2}, of which the final three are presented in \citet{TemminkEA24}.

\subsection{Single \ce{H_2O} LTE slab models} \label{sec:H2O-Regions}
The JWST-MIRI observations of DR~Tau show a plethora of both ro-vibrational ($<$10 $\mathrm{\mu}$m) and rotational ($>$10 $\mathrm{\mu}$m) transitions of \ce{H_2O}. Following \citet{GasmanEA23Subm}, we investigate the excitation properties of \ce{H_2O} across different wavelength regions, covering both ro-vibrational and rotational transitions, using 0D-LTE slab models.  We have used slab models without and with line overlap, i.e. mutual shielding of \ce{H_2O} lines (i.e. no shielding by other molecules, see \citealt{TaboneEA23} for a full description). Table \ref{tab:Regions} lists the best fitting slab model parameters for both the models without line overlap (top half) and with line overlap (bottom half). The best fitting slab models are also displayed in Figures \ref{fig:RegionSpectra} and \ref{fig:RegionSpectra-Overlap} for those without and with line overlap, respectively. \\
\begin{table}[ht!]
    \centering
    \caption{The best fit parameters of the \ce{H_2O} slab models without (top) and with line overlap (bottom).}
    \begin{tabular}{c c c c c}
        \hline
        \hline
        Region & $\log_{10}\left(N\right)^\alpha$ & $T$ & $R$ & $\mathcal{N}$ \\
        
        [$\mathrm{\mu}$m] &  & [K] & [au] &  \\
        \hline
        \multicolumn{5}{c}{Without line overlap} \\
        5.0-6.5 & 18.4$^{+0.2}_{-0.1}$ & 1000$^{+250}_{-50}$ & 0.21$^{+0.02}_{-0.04}$ & 7.79$\times$10$^{43}$ \\
        7.0-9.5 & 18.3$^{+0.1}_{-0.1}$ & 1000$^{+50}_{-100}$ & 0.21$^{+0.02}_{-0.00}$ & 6.19$\times$10$^{43}$ \\
        10.5-13.0 & 18.1$^{+0.3}_{-1.8}$ & 850$^{+50}_{-50}$ & 0.87$^{+5.18}_{-0.20}$ & 6.70$\times$10$^{44}$ \\
        13.6-16.3 & 18.4$^{+0.0}_{-0.0}$ & 675$^{+0}_{-0}$ & 0.93$^{+0.00}_{-0.00}$ & 1.53$\times$10$^{45}$ \\
        17.0-23.0 & 18.9$^{+0.6}_{-0.3}$ & 450$^{+25}_{-50}$ & 1.31$^{+0.14}_{-0.12}$ & 9.58$\times$10$^{45}$ \\
        23.0-27.0 & 20.1$^{+2.0}_{-2.1}$ & 375$^{+350}_{-175}$ & 1.55$^{+1.92}_{-0.64}$ & 2.13$\times$10$^{47}$ \\
        \hline
        \multicolumn{5}{c}{With line overlap} \\
        5.0-6.5 & 18.5$^{+0.1}_{-0.0}$ & 1150$^{+100}_{-75}$ & 0.17$^{+0.00}_{-0.00}$ & 6.43$\times$10$^{43}$ \\
        7.0-9.5 & 18.5$^{+0.2}_{-0.1}$ & 850$^{+75}_{-75}$ & 0.25$^{+0.02}_{-0.02}$ & 1.39$\times$10$^{44}$ \\
        10.5-13.0 & 18.0$^{+0.4}_{-1.3}$ & 850$^{+50}_{-50}$ & 0.95$^{+2.86}_{-0.30}$ & 6.35$\times$10$^{44}$ \\
        13.6-16.3 & 18.5$^{+0.0}_{-0.0}$ & 675$^{+25}_{-0}$ & 0.89$^{+0.00}_{-0.06}$ & 1.76$\times$10$^{45}$ \\
        17.0-23.0 & 18.5$^{+0.3}_{-0.2}$ & 500$^{+25}_{-50}$ & 1.35$^{+0.14}_{-0.08}$ & 4.05$\times$10$^{45}$ \\
        23.0-27.0 & 19.5$^{+2.6}_{-1.8}$ & 400$^{+350}_{-200}$ & 1.69$^{+2.36}_{-0.68}$ & 6.35$\times$10$^{46}$ \\
        \hline
    \end{tabular}
    \label{tab:Regions}
    \tablefoot{$^\alpha$: $N$ is given in units of in cm$^{-2}$.}
\end{table}
\indent The best fitting models were determined using a reduced $\chi^2$-minimisation method, following, for example, \citet{GrantEA23}, \citet{TaboneEA23}, and \citet{TemminkEA24}. The reduced $\chi^2$ is determined using the following equation: 
\begin{align} \label{eq:Chi2}
    \chi^2_\textnormal{red} = \frac{1}{N_\textnormal{obs}}\sum^{N_\textnormal{obs}}_{i=1}\frac{\left(F_{\textnormal{obs},i}-F_{\textnormal{mod},i}\right)^2}{\sigma^2}.
\end{align}
Here, $N_\textnormal{obs}$ is the number of spectral resolution elements, covering isolated \ce{H_2O} transitions, used in the fitting, $F_{\textnormal{obs}}$ and $F_{\textnormal{mod}}$ are the corresponding observed and model flux, respectively, and $\sigma$ denotes the uncertainty on the flux (see Table \ref{tab:Subband-Sigma}). The uncertainties on the best-fit parameters are taken from the confidence intervals, which are determined for, respectively, 1$\sigma$, 2$\sigma$, and 3$\sigma$ as $\chi^2_\textnormal{red,min}$+2.3, $\chi^2_\textnormal{red,min}$+6.2 and $\chi^2_\textnormal{red,min}$+11.8 \citep{Avni76,PressEA92}. \\
\indent In accordance with previous works (e.g. \citealt{BlevinsEA16,BanzattiEA23Subm,GasmanEA23Subm}), the slab models probe the inner, hotter regions at the shortest wavelengths, whereas the outer, colder regions are traced by the longer wavelengths. The trends hold for both the models with and without line overlap. Figure \ref{fig:TempRadius} visualises the decreasing (increasing) trend of temperature (emitting radius) with wavelength for the models without overlap (squares) and those with line overlap (circles). We find that the temperatures decrease from 1000 K to 400 K over a span of 0.2 to 2.0 au, if $R_\textnormal{em}$ corresponds to the actual radius of the emitting area. Additionally, we show the temperature as a function of radius in the right panel of Figure \ref{fig:TempRadius}. This panel also includes the results from our three component fit (using method 3, see Section \ref{sec:MCFits}) and the 1$\sigma$ uncertainties. \\
\indent All fits yield generally well constrained values for the column densities, excitation temperatures and emitting areas, parameterised by the emitting radius $R_\textnormal{em}$. The uncertainties, which are mostly within $\pm$0.5 for the logarithm of the column density (in cm$^{-2}$), $\pm$100 K for the excitation temperature, and $\pm$0.20 au for the emitting radius, are the largest for the fits in region 6 (23.0-27.0 $\mathrm{\mu}$m), likely the effect of this wavelength region having the lowest sensitivity. In addition, the upper uncertainty on the emitting radius for the third region is found to be higher compared to the other uncertainties. This is likely the result of this region having fewer bright, optically thick lines. Using the results for the inner two regions, those traced by the ro-vibrational \ce{H_2O} transitions, we can make a comparison with those for \ce{CO} as analysed in \citet{TemminkEA24}, since those arise from a similar region of $\sim$0.2 au. This comparison is made in Section \ref{sec:CO-H2O}. \\
\indent We find that our slab models to the JWST-MIRI spectrum of DR~Tau yield very similar results for the excitation temperatures and emitting radii for both the vibrational and rotational transitions of \ce{H_2O} compared to the previously mentioned ground-based, high spectral resolution observations \citep{NajitaEA18,SalykEA19,BanzattiEA23}. Furthermore, \citet{BanzattiEA23} provides linewidths for the vibrational \ce{H_2O} lines at 5.0 $\mathrm{\mu}$m ($FWHM\sim$22 km~s$^{-1}$ and the rotational transitions at 12.4 $\mathrm{\mu}$m ($FWHM\sim$16 km~s$^{-1}$), so we can estimate the emitting radii from these line profiles using Kepler's law: $R=GM_*\left(\sin(i)/\Delta v\right)^2$, where $i$ is the inclination of the disk ($i\sim$5.4, as derived from ALMA observations; \citealt{LongEA19}) and $\Delta v$ is the linewidth, here taken equal to half the width at half maximum (HWHM). We derive emitting radii of $R_\textnormal{em}\sim$0.06 au and $\sim$0.11 au for the 5.0 and 12.4 $\mu$m high-resolution transitions, respectively. Both of these are significantly smaller than the derived emitting radii from our JWST observations and the slab models fitted to the ground-based, high-resolution observations. A similar discrepancy was found in our previous work, analysing the high spectral resolution CO observations \citep{TemminkEA24}. There we found that the emitting radii derived from the slab models and those using the linewidths for \ce{CO} agree if the (inner) disk inclinations are increased to $i_\textnormal{inner}\sim$10-23\degree. Notably, observations of the VLTI-GRAVITY instrument suggest at similar inclinations for the inner disk ($i_\textnormal{inner}\sim$18$^{+10}_{-18}$\degree; \citealt{GravityEA23}). Using these inclinations, we derive emitting radii in the ranges of $R_\textnormal{em}\sim$0.21-0.96 au for vibrational transitions at 5 $\mathrm{\mu}$m and $R_\textnormal{em}\sim$0.39-1.81 au, which do agree with the emitting radii derived from the slab models. As mentioned in \citet{TemminkEA24}, these derivations hold for the assumption that the emitting radii correspond to a circular region enclosing the host star and may suggest a misalignment between the inner and outer disk of DR~Tau.

\subsection{Cold and warm \ce{H_2O} reservoirs: large disk template} \label{sec:Excess}
\subsubsection{Identifying unblended lines}
Evidence for the presence of multiple rotational \ce{H_2O} components in the JWST-MIRI spectra was introduced by \citet{BanzattiEA23Subm} and \citet{GasmanEA23Subm}. In a similar fashion to \citet{BanzattiEA23Subm}, we first identify the unblended, high-energy (6000 $\leq E_\textnormal{up} \leq$ 8000 K) \ce{H_2O} lines, which all happen to have Einstein-A coefficients $A_\textnormal{ul}>$ 10 s$^{-1}$. Figure \ref{fig:H2O-Lines} shows $A_\textnormal{ul}$ as a function of $E_\textnormal{up}$ (left panel) and wavelength (right panel), where only the lines with $A_\textnormal{ul}\geq$10$^{-2}$ s$^{-}$ and wavelengths above 10 $\mathrm{\mu}$m are shown. Figure \ref{fig:H2O-Lines-Full}, on the other hand, shows all lines available for our analysis without a lower limit imposed on $A_\textnormal{ul}$. As our analysis includes many more \ce{H_2O} lines compared to \citet{BanzattiEA23Subm}, we have identified fewer isolated lines. This is mainly due to the inclusion of transitions with low Einstein-A coefficients ($<$10$^{-2}$ s$^{-1}$). These low Einstein-A coefficient transitions are expected to have a negligible contribution to the spectrum, unless the transitions with high Einstein-A values have high optical depths of $\tau>$100. \\ 
\indent To identify unblended, high-energy lines or high-energy lines that are blended together, we cross-listed these lines with all the transitions available (including Einstein-A coefficients $<$10$^{-2}$ s$^{-1}$) for the molecules detected in DR~Tau. As mentioned above, the transitions with low Einstein-A coefficients may have a significant contribution to the spectra if the transitions with higher Einstein-A values have high optical depths. Consequently, we are also considering these transitions when identifying the unblended lines. In this process, we only included these lines in the wavelength regions in which molecules have been detected. For \ce{C_2H2}, \ce{CO_2}, and \ce{HCN} this means that we cross-listed with all the available transitions in the region between 13.6 and 16.3 $\mathrm{\mu}$m, whereas for \ce{OH} we cross-listed with all transitions at wavelengths $>$13.6 $\mathrm{\mu}$m. In addition, we cross-listed with all the low-energy ($<$6000 K) \ce{H_2O} transitions. If no transitions of any of the molecules are located within 5 resolution elements enclosing a high-energy \ce{H_2O} transition, we considered the transition to be useful for this analysis. Our cross-listing yielded a total of 24 potentially useful transitions, of which 11 are, upon visual inspection, strongly detected in both DR~Tau and CI~Tau. Of those 11 lines, 6 are blended together in 3 pairs and are, subsequently, treated as one. The lines are highlighted in Figure \ref{fig:H2O-Lines} by the black circles and will be used for the scaling of CI~Tau. 

\begin{figure*}[ht!]
    \centering
    \includegraphics[width=\textwidth]{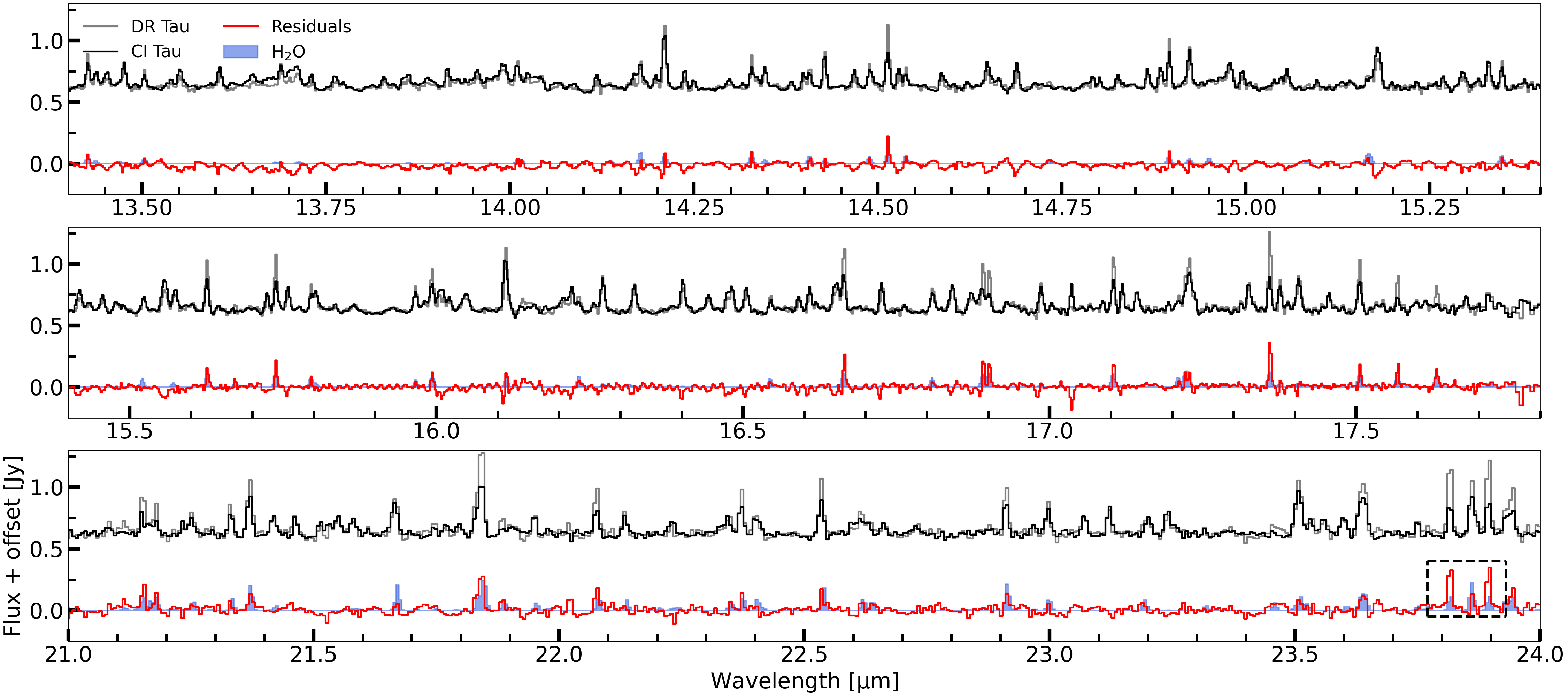}
    \caption{The spectra (across 13.4-24.0 $\mathrm{\mu}$m) of DR~Tau (grey) and CI~Tau (black, scaled; see Section \ref{sec:LDT-CITau}) shown together with the residual spectrum (in red) of DR~Tau after subtraction of the scaled spectrum of CI~Tau. The best fitting \ce{H_2O} slab model ($T$=375 K) to the residuals is shown in blue. The black dashed box just shortward of $\sim$24.0 $\mathrm{\mu}$m indicates the pair of lines identified by \citet{BanzattiEA23Subm}, hinting at a third component ($\sim$170 K) needed to fully explain the observed \ce{H_2O} reservoir.}
    \label{fig:LDTemplate}
\end{figure*}

\subsubsection{Large disk template: CI~Tau} \label{sec:LDT-CITau}
We scale the spectrum of CI~Tau to match the flux level of the hot \ce{H_2O} transitions in DR~Tau by accounting for the different distances and scaling by the averaged \ce{H_2O} line luminosity of the unblended lines of $L_\textnormal{\ce{H_2O},DR~Tau}/L_\textnormal{\ce{H_2O},CI~Tau}\simeq$4.88. Figure \ref{fig:LDTemplate} displays the (scaled) spectra of DR~Tau (grey) and CI~Tau (black) placed at a small flux offset with respect to the residuals in red over a large portion of the 10.0-27.5 $\mathrm{\mu}$m wavelength region. Similar to \citet{BanzattiEA23Subm}, various cold \ce{H_2O} lines have excess flux present in the residuals, suggestive of a second, colder \ce{H_2O} emission component, likely the result of efficient radial drift. By fitting a simple LTE slab model to the residuals, which is also displayed in Figure \ref{fig:LDTemplate}, we find that the second component has an excitation temperature of $T$=375 K, traces a column density of $N$=2.0$\times$10$^{19}$ cm$^{-2}$, and has an emitting radius of $R_\textnormal{em}$=0.91 au. \\
\indent Two lines just short of $\sim$24 $\mathrm{\mu}$m, indicated by the black, dashed rectangle in Figure \ref{fig:LDTemplate}, are not well fitted by the residual slab model. These lines are also visible in the work of \citet{BanzattiEA23Subm}, who found that these are best fitted by an even colder component at $T\sim$170 K, which most prominently emits at even larger wavelengths, $\geq$30 $\mathrm{\mu}$m \citep{ZhangEA13,BlevinsEA16,BanzattiEA23Subm}.

\subsection{Cold and warm \ce{H_2O} reservoirs: multi-component slab models} \label{sec:MCFits}
\begin{figure*}[h!]
    \centering
    \includegraphics[width=0.9\textwidth]{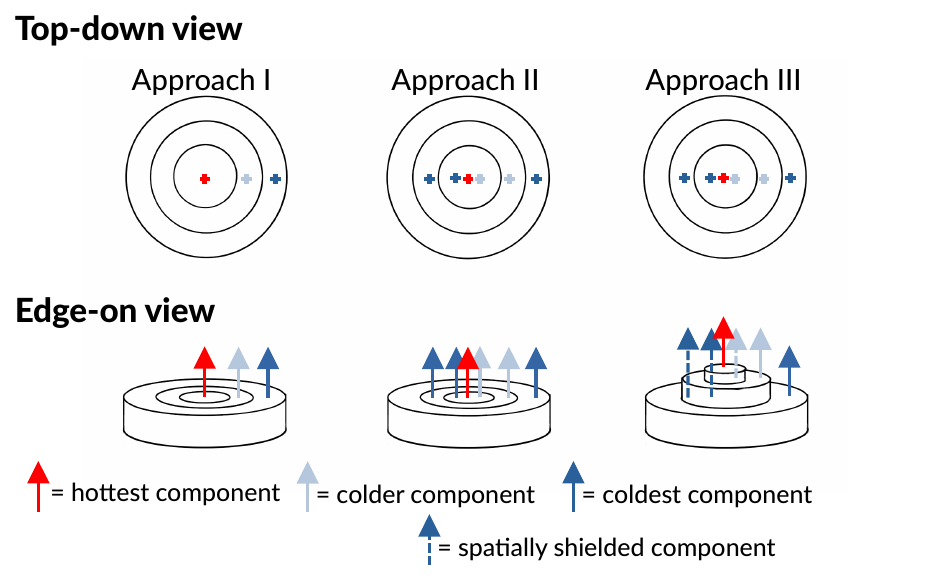}
    \caption{Cartoon visualising the different radial and vertical temperature gradients tested in this work. A top down version is shown in the top half of the figure, whereas as an edge-on view is shown in the bottom half. The coloured arrows indicate the emission from the different components, with red showing the emission from the hottest component, light blue for the component with the intermediate temperature, and dark blue for the coldest component. The dashed arrows indicate emission originating from spatially shielded regions.}
    \label{fig:MethodsCartoon}
\end{figure*}
As shown in Section \ref{sec:LDT-CITau}, the pure rotational \ce{H_2O} spectrum ($\geq$10 $\mathrm{\mu}$m) of DR~Tau is best described by multiple, at least 2, temperature components. In addition, following Section \ref{sec:H2O-Regions}, the hotter (colder) components are thought to trace emission at shorter (longer) wavelengths with smaller (larger) emitting areas. A next step in the analysis of \ce{H_2O} would be to combine these notions: i.e., we fit multiple slab models to the spectrum, consisting of components with decreasing temperatures and increasing emitting areas. The fitting is carried out using the Monte-Carlo Markov Chain implementation \textsc{emcee} \citep{EMCEE}, using 250 walkers and 150,000 iterations. We test three different approaches for fitting multiple components, see also Figure \ref{fig:MethodsCartoon}. \\
\indent In the first approach (I) we assume that the temperature only varies radially, accounting for the decreasing temperature gradient found in planet-forming disks. We consider that each component has a separate emitting region, with hotter components emitting from the smallest regions, closest to the host star. A simple radial temperature gradient is introduced by considering a weighted sum of the components, where the weights correspond to the emitting area ($A$), $F_\textnormal{total}$=$\sum_iF_iA_i$. The flux of each component, $F_i$, is determined by a temperature $T_i$ (in K) and a column density $N_i$ (in cm$^{-2}$), whereas the emitting area is parameterised by an emitting radius $R_i$ (in au). For our three components, this yields the following total flux:
\begin{align} \label{eq:RadGr}
    F_\textnormal{total} = & F_1\pi\left(\frac{R_1}{1\textnormal{ au}}\right)^2 + F_2\pi\left[\left(\frac{R_2}{1\textnormal{ au}}\right)^2-\left(\frac{R_1}{1\textnormal{ au}}\right)^2\right] + \\
    & F_3\pi\left[\left(\frac{R_3}{1\textnormal{ au}}\right)^2-\left(\frac{R_2}{1\textnormal{ au}}\right)^2\right]. \nonumber
\end{align} \\
\indent As planet-forming disks also have a vertical temperature gradient, we allow the colder components in the second approach (II) to also emit from the overlapping regions with the hotter components. If we are not taking shielding into account and assume that the various components have overlapping emitting areas, the total flux simply becomes:
\begin{align} \label{eg:RadGr_NS}
    F_\textnormal{total} = F_1\pi \left(\frac{R_1}{1\textnormal{ au}}\right)^2 + F_2\pi \left(\frac{R_2}{1\textnormal{ au}}\right)^2 + F_3\pi \left(\frac{R_3}{1\textnormal{ au}}\right)^2.
\end{align} \\
\indent In the final approach (III), we account for shielding of the colder, deeper components by the optical depth of the hotter components. In the case of shielding, the flux is attenuated by a factor of $\exp\left(-\sum_i\tau_i\right)$, where the sum is taken over the optical depths of all the components with a higher temperature. The optical depth is taken to be wavelength-dependent determined for all transitions combined using the corresponding excitation temperature and column density. As shown in Sections \ref{sec:H2O-Regions} and \ref{sec:disc-LO}, and in Table \ref{tab:Regions}, the mutual line shielding by neighbouring \ce{H_2O} transitions does not have a significant impact on the slab models and, hence, we only account for the spatial shielding of colder components by the hotter ones (see Figure \ref{fig:MethodsCartoon}). The total flux of the this approach, in which we account for spatial line shielding of the colder components by the hotter one, becomes:
\begin{align} \label{eq:RadGr_Sh}
    F_\textnormal{total} = & F_1\pi \left(\frac{R_1}{1\textnormal{ au}}\right)^2 + F_2\pi \left(\frac{R_1}{1\textnormal{ au}}\right)^2\exp(-\tau_1) + \\
    & F_2\pi\left[\left(\frac{R_2}{1\textnormal{ au}}\right)^2-\left(\frac{R_1}{1\textnormal{ au}}\right)^2\right] + F_3\pi \left(\frac{R_1}{1\textnormal{ au}}\right)^2\exp\left(-(\tau_1+\tau_2)\right) + \nonumber \\
    & F_3\pi \left[\left(\frac{R_2}{1\textnormal{ au}}\right)^2-\left(\frac{R_1}{1\textnormal{ au}}\right)^2\right]\exp(-\tau_2) + \nonumber \\
    & F_3\pi\left[\left(\frac{R_3}{1\textnormal{ au}}\right)^2-\left(\frac{R_2}{1\textnormal{ au}}\right)^2\right]. \nonumber
\end{align}
For the two component version, all terms including $F_3$ can be neglected. \\
\indent For all three approaches, we consider three components and impose the following constraints: the temperature must decrease with each component, whereas the emitting radius must increase. We keep the column density as a completely free parameter. In addition, we also show a model with two components for the third method, which allows for a thorough comparison between the models. The results for the different approaches are presented in the following subsections.

\begin{table}[ht!]
    \centering
    \caption{The best-fit parameters for the multiple component slab model fits, including radial and vertical temperature gradients.}
    \begin{tabular}{c c c c c}
        \hline\hline
        Component & $T$ [K] & $\log_{10}\left(N\right)^\alpha$ & $R_\textnormal{em}$ [au] & $\mathcal{N}$ \\
        \hline
        \multicolumn{5}{c}{Approach I} \\
        1 & 806$^{+289}_{-154}$ & 19.2$^{+1.5}_{-0.5}$ & 0.35$\pm0.15$ & 1.40$\times$10$^{45}$ \\
        2 & 468$^{+79}_{-85}$ & 18.5$^{+0.9}_{-0.4}$ & 1.18$^{+0.27}_{-0.18}$ & 3.03$\times$10$^{45}$ \\
        3 & 181$^{+36}_{-43}$ & 17.9$^{+2.2}_{-0.7}$ & 6.45$^{+2.26}_{-2.64}$ & 2.30$\times$10$^{46}$ \\
        \hline
        \multicolumn{5}{c}{Approach II} \\
        1 & 796$^{+268}_{-141}$ & 19.2$^{+1.4}_{-0.5}$ & 0.36$^{+0.14}_{-0.15}$ & 1.50$\times$10$^{45}$ \\
        2 & 466$^{+77}_{-84}$ & 18.5$^{+0.9}_{-0.4}$ & 1.14$^{+0.24}_{-0.17}$ & 2.78$\times$10$^{45}$ \\
        3 & 184$^{+29}_{-42}$ & 17.6$^{+2.2}_{-0.7}$ & 7.98$^{+4.49}_{-3.52}$ & 1.68$\times$10$^{46}$ \\
        \hline
        \multicolumn{5}{c}{Approach III} \\
        1 & 798$^{+281}_{-141}$ & 19.2$^{+1.5}_{-0.5}$ & 0.35$\pm0.15$ & 1.50$\times$10$^{45}$ \\
        2 & 467$^{+76}_{-81}$ & 18.5$^{+0.9}_{-0.4}$ & 1.20$^{+0.26}_{-0.17}$ & 3.03$\times$10$^{45}$ \\
        3 & 184$^{+30}_{-43}$ & 17.6$^{+2.2}_{-0.7}$ & 8.08$^{+4.43}_{-3.42}$ & 1.70$\times$10$^{46}$ \\
        \hline
        \multicolumn{5}{c}{Approach III - two components} \\
        1 & 712$^{+130}_{-100}$ & 19.1$^{+0.8}_{-0.3}$ & 0.50$^{+0.14}_{-0.12}$ & 2.02$\times$10$^{45}$ \\
        2 & 341$^{+74}_{-78}$ & 18.5$^{+1.0}_{-0.4}$ & 1.83$^{+0.54}_{-0.32}$ & 7.97$\times$10$^{45}$ \\
        \hline
    \end{tabular}
    \label{tab:Grad-Results}
    \tablefoot{$^\alpha$: $N$ is given in units of cm$^{-2}$.}
\end{table}

\subsubsection{Radial temperature gradient}
The top part of Table \ref{tab:Grad-Results} contains median values yielded by the MCMC for the first approach. The uncertainties are given, respectively, by the 16$^\textnormal{th}$ and 84$^\textnormal{th}$ percentiles. For this first approach, we retrieve temperatures of 806$^{+289}_{-154}$ K, 468$^{+79}_{85}$ K, and 181$^{+36}_{-43}$ K for the different components. In addition, we obtain combinations for the column density ($\log_{10}\left(N\right)$, with $N$ in units of cm$^{-2}$) and emitting radius of, respectively, 19.2$^{+1.5}_{-0.5}$ and 0.35$\pm0.15$ au, 18.5$^{+0.9}_{-0.4}$ and 1.18$^{+0.27}_{-0.18}$ au, and 17.9$^{+2.2}_{-0.7}$ and 6.45$^{+2.26}_{-2.64}$ au. The best fitting model is displayed in the top panel of Figure \ref{fig:Gradients}, while the corner plots are shown in Figure \ref{fig:CP1}.

\begin{figure*}[ht!]
    \centering
    \includegraphics[width=\textwidth]{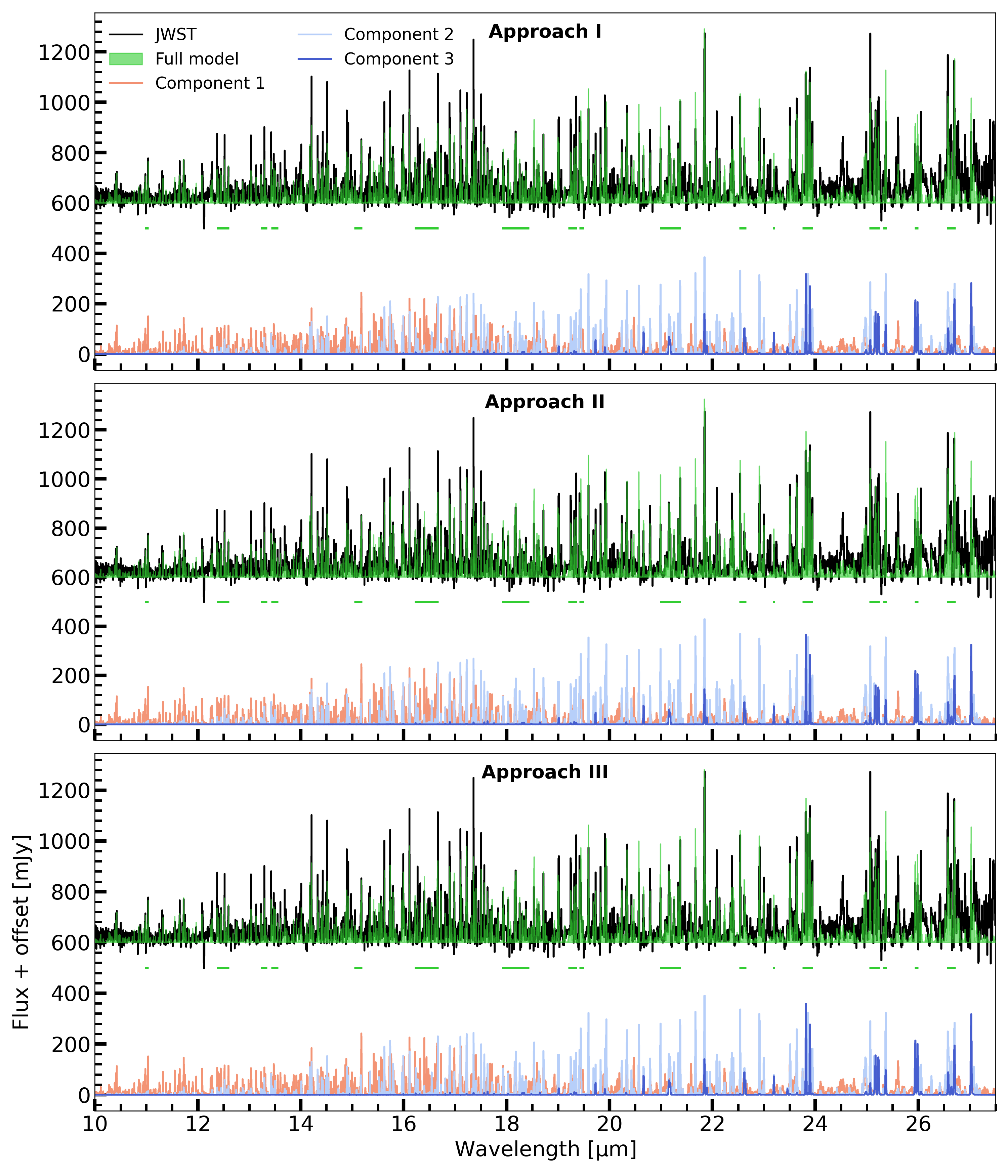}
    \caption{Multi-component slab model fits for the different methods of radial gradients: method 1 (radial temperature gradient, without overlapping regions) is shown in the top panel, whereas the methods 2 (radial gradients with overlapping regions, but no line shielding) and 3 (radial and vertical temperature gradients) are displayed in the middle and bottom panels, respectively. The full model is in each panel shown in green, whereas the components are coloured depending on the corresponding temperatures. Red indicates the hottest component, followed by the light blue and dark blue ones. The green horizontal bars indicate the regions used in the $\chi^2_\textnormal{red}$-fits.}
    \label{fig:Gradients}
\end{figure*}

\subsubsection{Radial and vertical temperature gradient: no shielding}
The resulting values for the second approach are shown in the second part of Table \ref{tab:Grad-Results}. Compared with the method 1, assuming only a radial gradient with no spatial overlap, our best fit parameters do not significantly change. In particular, the main difference can be found in the emitting radius of the third, coldest component. However, the values still agree within the given uncertainties and show a clear gradient. The final model is shown in the middle panel of Figure \ref{fig:Gradients} and their corresponding corner plots are shown in Figure \ref{fig:CP2}.

\subsubsection{Radial and vertical temperature gradient: accounting for spatial shielding} \label{sec:RVG}
\begin{figure*}[ht!]
    \centering
    \includegraphics[width=0.9\textwidth]{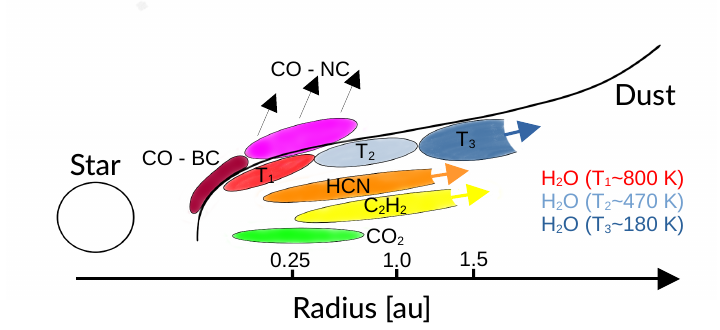}
    \caption{Cartoon visualising the expected emission regions, based on the derived excitation temperatures, emitting radii, and optical depths, of the multiple components of \ce{H_2O}. Shown are also the expected emission locations of \ce{CO}, \ce{CO_2}, \ce{HCN}, and \ce{C_2H_2} adopted from \citet{TemminkEA24}. The \ce{CO} is decomposed in two components: one broad component (BC) tracing the Keplerian rotation of the disk, and a narrow component (NC) tracing a disk wind.}
    \label{fig:Cartoon}    
\end{figure*}
The results for the three component and two component fits, using the third approach, are shown in respectively the third and bottom parts of Table \ref{tab:Grad-Results}. For the three component fit (also visible in the bottom panel of Figure \ref{fig:Gradients}), we find that the best fit parameters are similar to those of the second approach. That means that the main difference can once more be found in the emitting radius of the third component. It must be noted that the values for all three approaches agree with one another within the given uncertainties. In accordance with \citet{PontoppidanEA23subm}, we find that the results for this approach can be well described by the power-law $T(R_\textnormal{em})\sim$500$\left(\frac{R_\textnormal{em}}{1\textnormal{ [au]}}\right)^{-0.5}$ K for the emitting layer. For the two component fit, we see that the temperature of the first (hottest) component and the second component are lower by $\gtrsim$100 K. It is clear that the coldest component, $T\sim$180 K, is not considered by the two component fit. The lack of this component is also visible in Figure \ref{fig:3Cvs2C}, where the lines just short of 24 $\mathrm{\mu}$m are not well fitted by the model. The column density and the emitting radius of the first component have, respectively, slightly decreased and increased with respect to the three component fit. On the other hand, only the emitting radius of the second component increased from $\sim$1.20 au to $\sim$1.83 au. The corner plots for the three component and two component fits are shown in Figures \ref{fig:CP3} and \ref{fig:CP3-2C}, respectively. \\
\indent We find that all models underproduce the flux for a range of \ce{H_2O} transitions between $\sim$14 $\mathrm{\mu}$m and $\sim$18 $\mathrm{\mu}$m. This region is dominated by both the first (hottest) and second (warm) component, suggesting that either one or both components do not properly describe this region and slightly different values for the parameters would yield a better fit. Nonetheless, the fit provides an overall good description of the observed rotational spectrum of \ce{H_2O} in DR~Tau. \\
\indent Finally, Figure \ref{fig:Cartoon} displays an cartoon of the disk, indicating the expected emission locations from our multi-component fit. We have also included the expected locations from \ce{CO}, \ce{CO_2}, \ce{HCN}, \ce{C_2H_2} as found in \citet{TemminkEA24} (see also their Figure 12), based on their excitation temperatures and emitting areas.

\subsection{Line pair ratios} \label{sec:FEACR}
The inferred column density of the slab models can be tested and further constrained by comparing the peak fluxes of \ce{H_2O} line pairs which have the same value for $E_\textnormal{up}$, but different values for $A_\textnormal{ul}$ \citep{GasmanEA23Subm}. As the strength of these lines and, therefore, their ratio depends primarily on the line opacity, they can be used to approximate as independent constraints on the column density and provide a sanity check for the column densities retrieved with the slab models. If both lines are optically thin, the flux ratio is expected to converge to the ratio of the respective values for $A_\textnormal{ul}$, whereas the flux ratio will deviate from the $A_\textnormal{ul}$ ratio if one of the lines becomes optically thick. We use the same, pure rotational line pairs as \citet{GasmanEA23Subm} (see their Table 2) to investigate the \ce{H_2O} column density. Some of the lines are part of line clusters and are, subsequently, blended, which means that it cannot be concluded from the flux ratio alone which of the lines are optically thick, if the flux ratio deviates from that of $A_\textnormal{ul}$.\\
\indent Figure \ref{fig:FR-EAR} shows the model line ratios for different temperatures (300, 500, and 900 K) and column densities (14$\leq\log_{10}(N)\leq$22, with $N$ in units of cm$^{-2}$). The ratios become constant for column densities of $\lesssim$10$^{17}$ cm$^{-2}$. For these column densities we can expect both lines to be optically thin and, subsequently, their ratios to be equal to the values for $A_\textnormal{ul}$. At larger column densities (at least) one of the lines becomes optically thick and the ratio deviates from that for $A_\textnormal{ul}$. For even larger column densities ($\geq$10$^{20}$ cm$^{-2}$) the ratio becomes constant again, the effect of both lines being fully saturated. \\
\begin{figure}[ht!]
    \centering
    \includegraphics[width=\columnwidth]{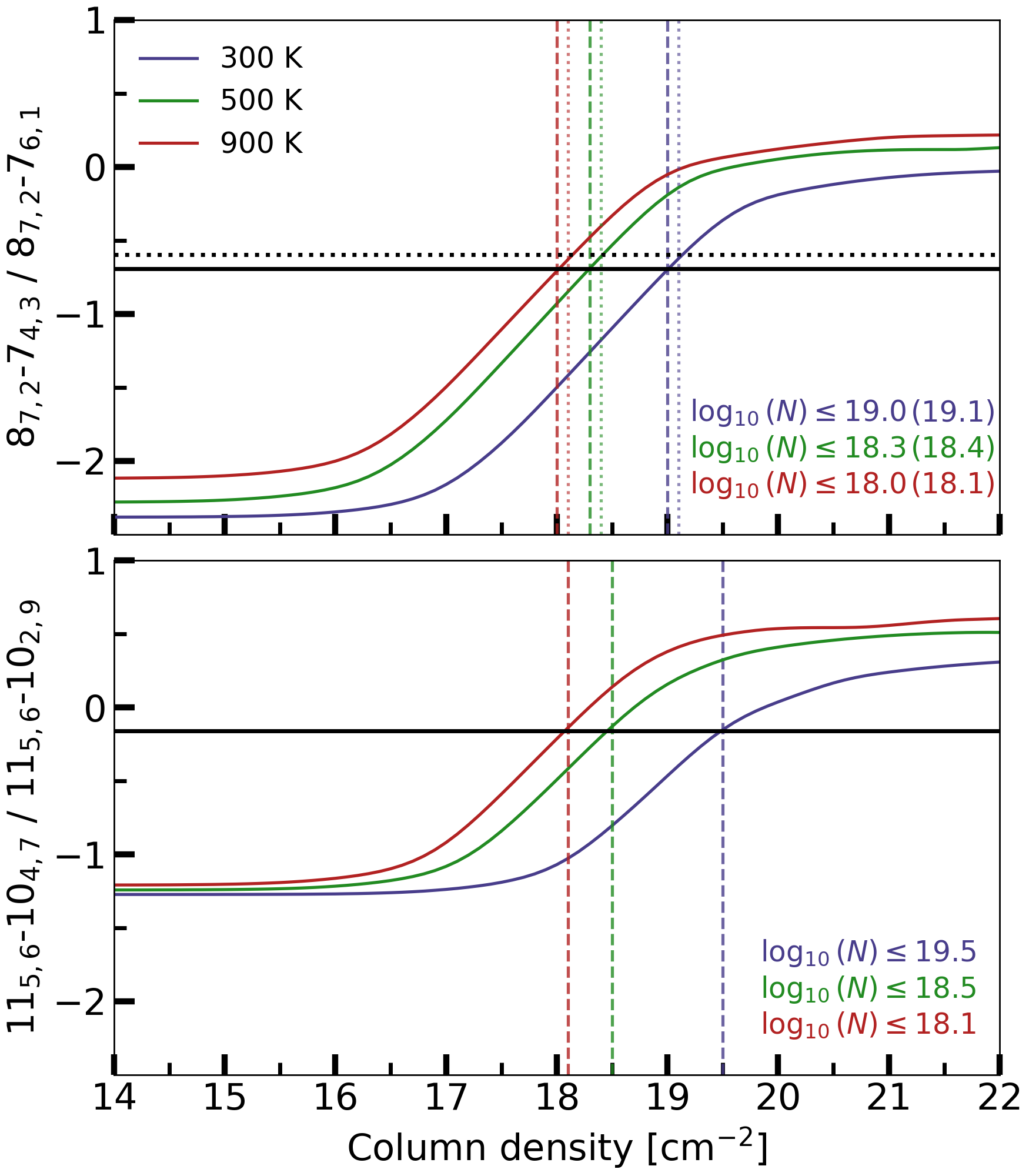}
    \caption{The logarithm of the line flux ratios as a function of the logarithm of the column densities (in cm$^{-2}$) for the line pairs with similar upper level energies. The ratios for the slab models with $T=$300 K are shown in blue, whereas those for $T$=500 K and $T$=900 are shown in green and red, respectively. The horizontal black line indicates the line ratio taken from the continuum-subtracted spectrum of DR~Tau. The dashed, vertical lines indicate where the different model ratios equal the observed line ratio. In addition, the dotted lines and the column densities listed in parenthesis indicate the results for the top line pair after subtracting off the contribution of the component with the lowest temperature (see Section \ref{sec:MCFits}).}
    \label{fig:FR-EAR}
\end{figure}
\indent The black horizontal line shown in Figure \ref{fig:FR-EAR} is the line ratio obtained from the JWST-MIRI spectrum. The coloured, dotted vertical lines are the intercepts between the black horizontal line and the model ratio curves. Based on the intercept, we can expect the actual \ce{H_2O} column density to take on values of $\leq$10$^{19.4}$ cm$^{-2}$, $\leq$10$^{18.4}$ cm$^{-2}$, and/or $\leq$10$^{18.0}$ cm$^{-2}$ for the different temperatures ($T$=300, 500, and 900 K), respectively. Using the results from the multi-component fits, we can infer the contributions from each component to the lines used for both line pairs. For the bottom pair (11$_{5,6}$-10$_{4,7}$/11$_{5,6}$-10$_{2,9}$), both transitions consist of contributions from the hotter two components, so the $T\sim$300 K models can be ignored. For both transitions, the lower temperature ($T\sim$475 K) has a slightly larger contribution to the flux. Based on this line pair and the temperature from both contributing components ($T\sim$830 K and $T\sim$475 K), we infer that the column density of the combination of these two \ce{H_2O} components must be $\log_{10}\left(N\right)\leq$18.5 within $\sim$1.2 au. For the top pair (8$_{7,2}$-7$_{4,3}$/8$_{7,2}$-7$_{6,1}$), however, the lowest temperature component ($T\sim$185 K) also has a significant contribution to the line flux of one of the lines. As only one of these lines has a contribution from the lowest temperature component, this pair does not provide strong constraints. After subtracting off the model corresponding to this temperature, we do infer a slightly more stringent constraint on the \ce{H_2O} column density (in cm$^{-2}$) of $\log_{10}\left(N\right)\leq$18.4 for the first two components combined, compared to value derived for the other line pair. Even though \ce{H_2O} is optically thick, these line pair ratios provide good constraints on the column densities for the first two components. We note that the derived limit agrees well with those obtained for the second, warm component, whereas it falls just outside the uncertainties for the hottest, most optically thick component.

\subsection{Searching for \ce{H_2 ^{18}O}} \label{sec:H2-18-O}
\begin{figure*}[ht!]
    \centering
    \includegraphics[width=\textwidth]{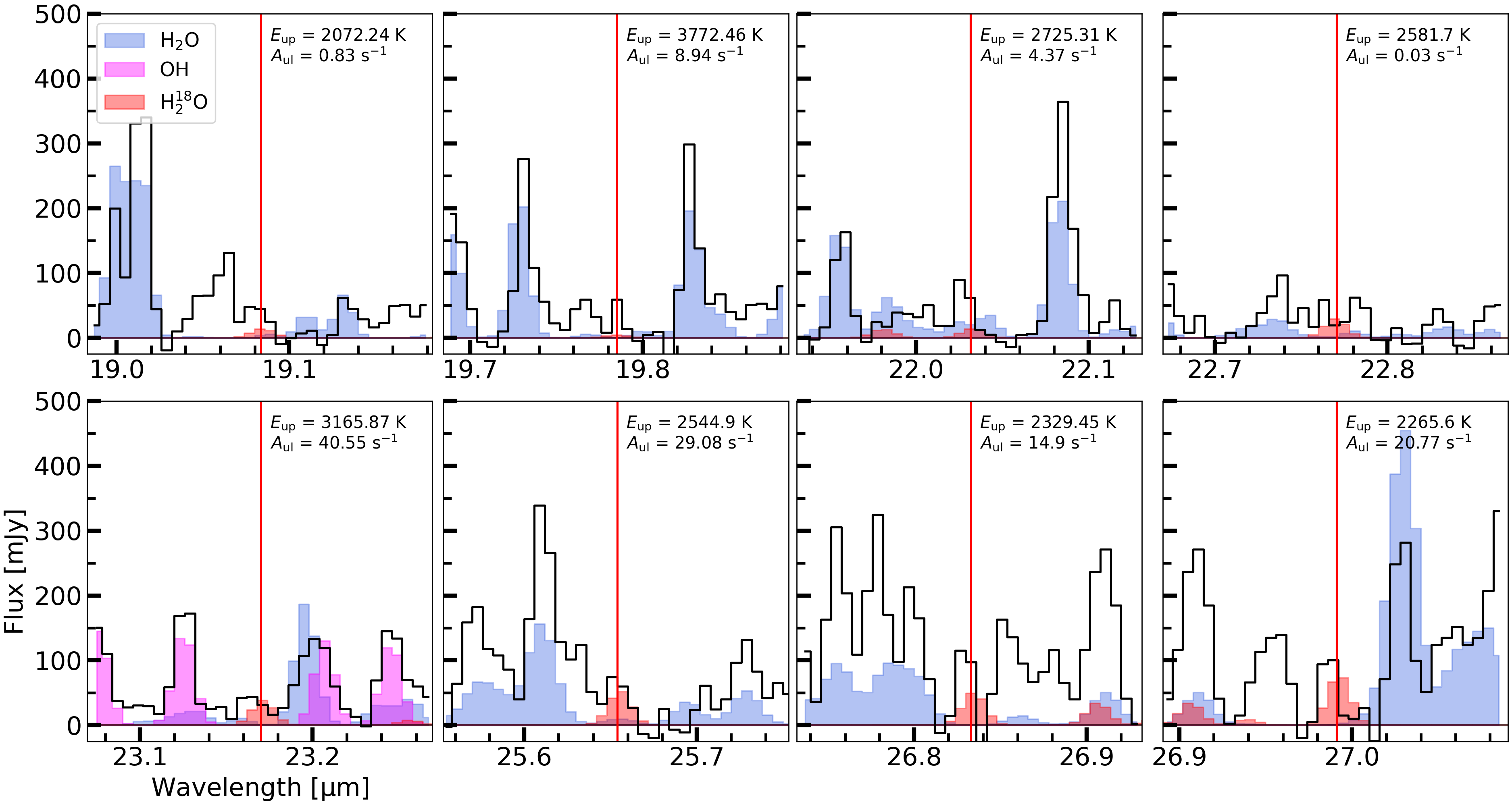}
    \caption{Zoom-ins on the isolated \ce{H_2 ^{18}O} lines, highlighted by the red vertical line. A simple \ce{H_2 ^{18}O} slab model ($T$=350 K, $\log_{10}\left(N\right)$=17.66, and $R_\textnormal{em}$=1.0 au) is shown in red, whereas the best models of \ce{H_2O} (three component fit accounting for both the radial and vertical temperature gradient) and \ce{OH} are shown in blue and magenta, respectively. For each transition, we have listed the upper level energies and Einstein-A coefficients.}
    \label{fig:H2-18-O}
\end{figure*}

\indent We are inconclusive about the detection of \ce{H_2 ^{18}O} in DR~Tau. Some of the transitions, in particular those at 25.65, 26.83, and 26.99 $\mathrm{\mu}$m, look promising (see Figure \ref{fig:H2-18-O}). Accounting for the large error in subband 4C (see Table \ref{tab:Subband-Sigma}), the peak fluxes suggest maximum detection levels of $\lesssim$4$\sigma$ and should be statistically significant for the confirmation of a detection. However, the lack of flux seen for the transitions at $\sim$22.03 $\mathrm{\mu}$m and $\sim$23.17 $\mathrm{\mu}$m provide the strongest argument against a detection of \ce{H_2 ^{18}O} in DR~Tau. Re-observing DR~Tau with JWST-MIRI with the aim of achieving higher signal-to-noise ratios across Channel 4 may improve our chances of detecting \ce{H_2 ^{18}O}. \\
\indent Using our three component models, we find that the majority of the \ce{H_2O} emission features are optically thick. While transitions with low Einstein-A values may be optically thin and can be used to derive better constraints on the total \ce{H_2O} vapour reservoir present in the inner disk, we note that these lines are very scarce in the spectrum of DR~Tau. In total, we identify four clean emission features (between 12.496-12.506, 14.377-14.392, 16.362-16.375, and 19.370-19.392 $\mathrm{\mu}$m) that are optically thin ($\tau<$1) according to our models and not blended with other (unidentified) emission features. All these transitions trace mainly the hottest component. Subsequently, a fit to these transitions only may provide a better constraint on the column density for the hottest component and observations of the optically thin isotopologues, such as \ce{H_2 ^{18}O}, could help to infer a better estimate of the column density and the total number of \ce{H_2O} molecules present in the inner disk. \\
\indent \citet{CalahanEA22} list 8 lines in the 19.0-27.0 $\mathrm{\mu}$m wavelength range that may be used for the detection of \ce{H_2 ^{18}O} in JWST-MIRI spectra (see their Table 1). We, however, note that the line at 20.01 $\mathrm{\mu}$m is blended with \ce{OH} and can likely not be used for the detection of \ce{H_2 ^{18}O} in many T-Tauri disks. On the other hand, through slab model exploration we discovered a blended pair of \ce{H_2 ^{18}O} lines that can be used instead. The line pair (8$_{8,0}$-7$_{7,1}$ and 8$_{8,1}$-7$_{7,0}$) is located at 25.65 $\mathrm{\mu}$m and have the same upper level energies ($E_\textnormal{up}\sim$2544.9 K) and Einstein-A coefficients ($A_\textnormal{ul}\sim$29.08 s$^{-1}$), suggesting that both lines contribute half of the observed line flux. The individual lines are displayed in Figure \ref{fig:H2-18-O}. The \ce{H_2 ^{18}O} emission is highlighted by a simple \ce{H_2^{18}O} slab model ($T$=350 K, $\log_{10}\left(N\right)$=17.36, and $R_\textnormal{em}$=1.0 au), where we have accounted for the isotopologue ratio in the local interstellar medium (ISM) of \ce{^{16}O}/\ce{^{18}O}$\sim$550 \citep{Wilson99} in the column density, i.e. we have divided the found column density of \ce{H_2O} in the wavelength region of 23.0-27.0 $\mathrm{\mu}$m ($\log\left(N\right)\sim$20.1, see Table \ref{tab:Regions}) by 550. We show our best fitting \ce{H_2 ^{16}O} and \ce{OH} slab models (see Table \ref{tab:Regions} and Section \ref{sec:OH}) for further identification. We show the full 17.0-27.5 $\mathrm{\mu}$m wavelength region, including the aforementioned \ce{H_2 ^{16}O}, \ce{H_2 ^{18}O}, and \ce{OH} slab models in Figure \ref{fig:H218O-Spectrum}. \\
\indent Even though we are inconclusive about the detection of \ce{H_2 ^{18}O}, the used slab model yields underproduced fluxes for the most promising peaks. This may suggest that, for the used temperature, the input column density is not high enough. As the column density was set to account for the isotopologue ratio (\ce{^{16}O}/\ce{^{18}O}$\sim$550), this indicates optically thick \ce{H_2O} emission, which is in agreement with our models. We note that an underestimate of the column density for \ce{H_2 ^{18}O} may also point to a deviation of the isotopologue from the ISM value. As discussed in \citet{CalahanEA22}, the gas may be enriched in \ce{H_2 ^{18}O} in the disk layers where \ce{CO} becomes optically thick and self-shielding against photodissociation becomes important. As the main isotopologue, \ce{C^{16}O}, is more abundant than \ce{C^{18}O}, there will be a layer where \ce{C^{16}O} will be self-shielded, whereas \ce{C^{18}O} is still being dissociated. The gas in this layer will in that case be enriched in \ce{^{18}O} atoms, which may end up in \ce{H_2O}-molecules, enhancing the \ce{H_2 ^{18}O} column. Additionally, UV-shielding by the main isotopologue, \ce{H_2 ^{16}O}, can also enhance the \ce{H_2 ^{18}O} abundance.

\subsection{Other emission features} \label{sec:OtherMols}
Besides emission from \ce{H_2O}, the spectrum of DR~Tau is rather molecule rich. \ce{CO}, \ce{CO_2}, \ce{HCN}, and \ce{C_2H_2} have been thoroughly discussed been in \citet{TemminkEA24}, so the upcoming subsection discusses the other molecular features potentially present in the spectrum, including \ce{OH} and the larger hydrocarbons that have been observed in the disks around very low-mass stars (e.g. \citealt{TaboneEA23}).

\subsubsection{OH} \label{sec:OH}
Since we do not detect any \ce{OH} lines at wavelengths $<$13 $\mathrm{\mu}$m, similar to \citet{GasmanEA23Subm}, we do not expect prompt emission, following \ce{H_2O} photodissociation \citep{NajitaEA10}, to play a role in the observed \ce{OH} emission in DR~Tau. We fit the \ce{OH} emission, using a reduced $\chi^2$-minimisation (see also Section \ref{sec:H2O-Regions}), over the entire wavelength range of 10 $\mathrm{\mu}$m to 27.5 $\mathrm{\mu}$m. We have used the same grid for $T$, $\log_{10}\left(N\right)$, and $R_\textnormal{em}$ as in Section \ref{sec:AR}, except we extended the grid for the excitation temperature up to 4000 K. The emission was fit to the residual spectrum after subtracting the resulting \ce{H_2O} model spectrum taking both a radial and vertical temperature gradient into account (see Section \ref{sec:RVG}). For \ce{OH}, we retrieve a temperature of $T$=1875$^{+775}_{-725}$ K, a column density (in cm$^{-2}$) of $\log\left(N\right)$=13.5$^{+3.3}_{-0.3}$, and emitting radius of $R_\textnormal{em}$=9.49$^{+0.52}_{-9.06}$ au. As the emission is optically thin ($\tau\leq$0.005) for this combination of excitation temperature and column density, the emitting radius is not well constrained, as indicated by the large lower uncertainty. We note that the upper uncertainty of the emitting radius and, therefore, the lower constraints on the column density are set by the limits of our grids. For these parameters, we derive that the number of \ce{OH} molecules present in the inner disk of DR~Tau is $\mathcal{N}\sim$2.0$\times$10$^{42}$. Parts of the best fitting spectrum, including the wavelength regions used in the fitting, are displayed in the bottom part of Figure \ref{fig:RegionSpectra}. Similar to \citet{GasmanEA23Subm}, the \ce{OH} lines probe larger gas temperatures compared to \ce{H_2O}, but lower column densities. As the excitation of the \ce{OH} lines includes prompt emission \citep{TaboneEA21,TaboneEA24} and collisional excitation or chemical pumping \citep{ZanneseEA24} through the \ce{O}+\ce{H_2}$\rightarrow$\ce{OH}+\ce{H} reaction, it is unlikely that the observed transitions follow a single excitation temperature. Therefore, the inferred temperature from the slab model is not related to a kinetic temperature. While they cannot infer a kinetic temperature, we note that slab models are a powerful tool for the identification of \ce{OH} transitions in the JWST-MIRI/MRS spectra of disks. Detailed models, including the different excitation pathways, can be used to further explore the excitation properties and to infer other key disk parameters (see, for example, \citealt{TaboneEA24}).

\subsubsection{Atomic and molecular hydrogen}
We list the detection of three atomic hydrogen (\ce{H}) transitions (10-6, 6-5, and 8-6) and the potential detection of four molecular hydrogen (\ce{H_2}) transitions: 0,0 S(1), S(2), S(3), and S(4). As the \ce{H_2} lines are either blended with \ce{H_2O} or potentially blended with unidentified features, their detections are not fully secure. Higher transitions (e.g. S(5), S(6), S(7), and S(8)) may also be present, however, their detections are even less certain due to stronger blending with \ce{H_2O} and \ce{CO} emission features. Figures \ref{fig:AtomicH} and \ref{fig:MolecularH} display the (potentially) observed transitions of \ce{H} and \ce{H_2} (S(1), S(2), S(3), and S(4)), respectively. We also see no evidence of extended \ce{H_2} emission coming from the DR~Tau disk in the form of a disk wind or an outflow, however, we do observe extended \ce{H_2} emission (most prominent in the S(1) transition) from the cloud surrounding DR~Tau, as was previously observed by \citet{ThiEA01} in millimetre single dish \ce{CO} $J$=3-2 transitions offset by $\sim$2 km~s$^{-1}$. The emission coming from the background is visualised in Figure \ref{fig:MolecularH-Spatial}. Visual inspection of the spectral cube shows that the flux of the cloud emission ($F_\textnormal{S(1)}\sim$2$\times$10$^{2}$ MJy sr$^{-1}$) is a factor $\sim$100 lower than the flux at the position of DR~Tau ($F_\textnormal{S(1)}\sim$2$\times$10$^{4}$ MJy sr$^{-1}$). As we approximate our background using an annulus, the \ce{H_2} emission from the cloud will be captured in that annulus. However, as the flux is significantly lower than the source flux, the background subtraction does not impact our detections. \\ 
\indent Using the residual flux, after subtracting the corresponding \ce{H_2O} slab model fit, we determined an upper limit on the column density of \ce{H_2} based on the 0,0 S(4) transition ($E_\textnormal{up}$=3474.5 K). Assuming that \ce{H_2} comes from the inner most region, we adopt a temperature of $T$=1000 K and an emitting radius of $R_\textnormal{em}$=0.20~au, as this also allows us to make direct comparisons with the total number of molecules of both \ce{CO} and \ce{H_2O} (as obtained in regions 1 and 2 for the ro-vibrational transitions, see Section \ref{sec:H2O-Regions}) derived in these innermost regions. To calculate the column density under the assumption of optically thin emission, we follow the description given in \citet{GL99}:
\begin{align}
    \frac{4\pi F}{hc\nu\Omega g_\textnormal{up}A_\textnormal{ul}} = \frac{N_\textnormal{tot}}{Q(T)}\exp\left(-\frac{E_\textnormal{up}}{k_BT}\right),
\end{align}
where $F$ is the integrated flux (in erg s$^{-1}$ cm$^{-2}$), $\nu$ the frequency (in cm$^{-1}$), $g_\textnormal{up}$ the upper level degeneracy, $Q(T)$ the partition function at a temperature $T$, and $E_\textnormal{up}$ the upper level energy (in $K$). Using the resolution element at the location of the S(4) transition and the two resolution elements directly next to it, we derive $F$=3.9$\times$10$^{-15}$ erg s$^{-1}$ cm$^{-2}$. As mentioned above, the \ce{H_2} flux might be slightly affected by the annulus background subtraction due to cloud emission. However, as the background emission is significantly lower than the on-source emission, we expect the derived flux to be minimally impacted. This yields a column density of $N_\textnormal{tot}\sim$7.6$\times$10$^{25}$ cm$^{-2}$ and a total number of molecules, assuming an emitting radius of 0.20 au, of $\mathcal{N}_\textnormal{\ce{H_2}}\sim$2.1$\times$10$^{51}$. We note once more that these derived column densities and total number of molecules hold for the upper layers of the disk, above the region where the dust emission becomes optically thick, or, in the case of optically thick lines, above the height where the emission lines become optically thick (see for example \citealt{BosmanEA22}).

\subsubsection{Non-detections}
In an attempt to identify as many molecular features as possible, we have also looked for emission signatures of various other molecules and isotopologues. These species are comprised of \ce{^{13}CO_2}, \ce{CH_4}, \ce{NH_3}, \ce{CS}, \ce{H_2S}, \ce{SO_2}, and various larger hydrocarbons, including \ce{^{13}CCH_2}. We have obtained the required spectroscopic data for these species through the HITRAN database \citep{HITRAN}. Table \ref{tab:NonDs} summarises the parameters (assuming an emitting radius of $R_\textnormal{em}$=1.0 au) of simple LTE slab models with fixed temperature and column density that have been used to search for these species. In addition, the table lists the best fit parameters for \ce{CO_2}, \ce{HCN}, and \ce{C_2H_2}, as found in \citet{TemminkEA24}. The slab models are also displayed in Figures \ref{fig:NDSlabs-1} and \ref{fig:NDSlabs-3}. \\
\begin{table}[ht!]
    \centering
    \caption{Slab models parameters of the various detected (\ce{CO_2}, \ce{HCN}, \ce{C_2H_2}; \citealt{TemminkEA24}) and non-detected species, using $R_\textnormal{em}$=1 au.}
    \begin{tabular}{c c c c c}
        \hline
        \hline
        Mol. & Region & $T$ & $\log_{10}\left(N\right)^\alpha$ & $\mathcal{N}$ \\
        
        & [$\mathrm{\mu}$m] & [K] & &  \\
        \hline
        \ce{^{12}CO_2}$^\beta$ & 13.60-16.30 & 325$^{+50}_{-100}$ & 17.4$^{+0.7}_{-0.2}$ & 4.96$\times$10$^{43}$ \\
        \ce{HCN}$^\beta$ & 13.60-16.30 & 900$^{+50}_{-50}$ & 14.7$^{+2.1}_{-0.6}$ & 8.22$\times$10$^{42}$ \\
        \ce{C_2H_2}$^\beta$ & 13.60-16.30 & 775$^{+150}_{-625}$ & 15.0$^{+5.6}_{-1.7}$ & 1.48$\times$10$^{42}$ \\
        \hline
        \ce{^{13}CO_2} & 15.30-15.50 & 300 & 15.57 & $\leq$2.6$\times$10$^{42}$ \\
        \ce{CH_4} & 7.60-7.75 & 700 & 15.75 & $\leq$4.0$\times$10$^{42}$ \\
        \ce{NH_3} & 8.50-9.50 & 500 & 15.50 & $\leq$2.2$\times$10$^{42}$ \\
        \ce{CS} & 7.50-8.30 & 500 & 15.50 & $\leq$2.2$\times$10$^{42}$ \\
        \ce{H_2S} & 7.00-8.00 & 500 & 18.00 & $\leq$7.0$\times$10$^{44}$ \\
        \ce{SO_2} & 7.10-7.62 & 500 & 15.50 & $\leq$2.2$\times$10$^{42}$ \\
        \hline
        \ce{^{13}CCH_2} & 13.60-13.85 & 750 & 15.00 & $\leq$7.0$\times$10$^{41}$ \\
        \ce{C_2H_4} & 10.50-10.60 & 700 & 16.00 & $\leq$7.0$\times$10$^{42}$ \\
        \ce{C_2H_6} & 12.00-12.34 & 700 & 17.00 & $\leq$7.0$\times$10$^{43}$ \\
        \ce{C_4H_2} & 15.80-16.00 & 700 & 15.50 & $\leq$2.2$\times$10$^{42}$ \\
        \ce{C_6H_6} & 14.70-15.00 & 700 & 11.25 & $\leq$1.3$\times$10$^{38}$ \\
        \hline
    \end{tabular}
    \label{tab:NonDs}
    \tablefoot{$^\alpha$: $N$ is given in units of cm$^{-2}$. \\
    $^\beta$: The listed values for \ce{CO_2}, \ce{HCN}, and \ce{C_2H_2} are taken from \citet{TemminkEA24} and are shown here for completion.}
\end{table} 
\indent We are not able to confidently detect any of the listed molecular species. The non-detection of some the molecules, for example, \ce{^{13}CO_2}, \ce{CH_4}, \ce{SO_2}, and the larger hydrocarbons, becomes apparent from the slab models, mostly due to the lack of a molecular continuum. For the other molecules, \ce{NH_3}, \ce{CS}, \ce{H_2S}, and \ce{^{13}CCH_2}, a (non-)detection cannot be confirmed, largely due to leftover broad features (especially visible at the location of the silicate feature at $\sim$10 $\mathrm{\mu}$m), which are potentially of molecular nature, in the spectrum. We also do not detect the larger hydrocarbons, mostly due to the lack of molecular continua. While the slab model of \ce{C_6H_6} looks promising in the last panel of Figure \ref{fig:NDSlabs-4}, the presence of \ce{C_6H_6} cannot be confirmed in the spectrum of DR~Tau, as the lack of emission observed for the other hydrocarbons makes the presence of \ce{C_6H_6} less likely. \\
\indent For most of the non-detected molecules, the upper limit on the total number of molecules is on the order of $\mathcal{N}\sim$10$^{42}$-10$^{43}$. Only the upper limit for \ce{H_2S} is larger ($\mathcal{N}\sim$7.0$\times$10$^{44}$), whereas that for \ce{C_6H_6} is significantly lower ($\mathcal{N}\sim$1.3$\times$10$^{38}$). Using the number of molecules from our three component \ce{H_2O} fits ($\mathcal{N}\sim$10$^{45}$-10$^{46}$), most of the non-detected molecules are a factor $\geq$10$^{2}$-10$^{4}$ less abundant than \ce{H_2O}. \ce{H_2S} and \ce{C_6H_6} are factors of, respectively, $\geq$10-100 and $\geq$10$^{7}$-10$^{8}$ less abundant. On the other hand, the upper limits are of a similar level compared to the total number of molecules derived for \ce{CO_2}, \ce{HCN}, and \ce{C_2H_2}. Once again, the upper limit for \ce{H_2S} is slightly larger, whereas that of \ce{C_6H_6} is a few orders of magnitude lower. For comparison, the abundance of \ce{CO_2} is found to be a factor $\sim$60 lower than the second, warm component of \ce{H_2O}.

\section{Discussion} \label{sec:Disc}
\subsection{The need for \ce{H_2O} line overlap} \label{sec:disc-LO}
For slab model fits of molecules with a well defined $Q$-branch (see \citealt{TaboneEA23}), line overlap (i.e. mutual shielding of lines) is often included to properly fit the spectra. The spectrum of \ce{H_2O}, on the other hand, consists of well defined, separate/isolated transitions. In the disk of AS~209, \citet{MunozRomeroEA24subm} also tested the need for line overlap of multiple molecular species. They conclude that a proper treatment of the line overlap is necessary when analysing the observed molecular species together. \\
\indent In Section \ref{sec:H2O-Regions}, we test the need for mutual shielding by the \ce{H_2O} transitions alone. We test this by creating slab models without and with mutual line shielding. As can be seen in Table \ref{tab:Regions}, the differences between the slab models without (top part of the table) and those with (bottom part) mutual line overlap are negligible when fitting \ce{H_2O} alone. The largest differences are found for temperatures derived for the regions spanning the ro-vibrational transitions, which is likely due to the lines being more densely packed in this part of the spectrum compared to those in the pure rotational part. We note that the residuals for the models without and with mutual line shielding are quite similar. We find that one model will provide a better fit for some transitions, whereas other transitions are better fitted by the other model. Therefore, we conclude that the inclusion of mutual shielding of neighbouring \ce{H_2O} lines is not of particular importance for the pure rotational transitions of \ce{H_2O}, when fitting them alone, and that the slab models without mutual line shielding can be used to infer constraints on the excitation conditions. As our slab models yield different results for the rovibrational part of the spectrum, the inclusion of mutual shielding may be more important for these more densely packed transitions. Mutual line shielding is of particular importance for molecules with a well defined $Q$-branch, such as \ce{CO_2}, \ce{HCN}, and \ce{C_2H_2}.

\subsection{The need for a radial temperature gradient} \label{sec:disc-RG}
As expected, the various methods all indicate a radial temperature gradient visible in the spectrum. Fitting, however, the \ce{H_2O} spectrum across multiple, independent wavelength regions does not yield any information on the low temperature ($T\sim$170-200 K, see Table \ref{tab:Regions}) component, even though the fitted regions include the lines best suitable for finding this component. Subsequently, to properly investigate the various components, one could use the large disk template method implemented by \citet{BanzattiEA23Subm} or the multi-component slab model method introduced here. \\
\indent Rather surprisingly, the current approaches of the multi-component slab model fits yield the same results for all three methods: temperatures of $\sim$800 K, $\sim$470 K, and $\sim$180 K, column densities of $\sim$10$^{19.2}$ cm$^{-2}$, $\sim$10$^{18.5}$ cm$^{-2}$, and $\sim$10$^{17.4}$ cm$^{-2}$, and emitting radii of $\sim$0.3 au, $\sim$1.2 au, and $\sim$6.5-8.1 au. For comparison, the midplane \ce{H_2O} snowline in DR~Tau can be estimated from a simple power law \citep{CG97,DullemondEA01,vdMarelEA21} involving the total luminosity ($L_\textnormal{tot}$=$L_*$+$L_\textnormal{acc}\simeq$0.63 + 0.58 L$_\odot$; \citealt{LongEA19,BanzattiEA20}), providing a radius $R_\textnormal{\ce{H_2O}}\sim$0.75 au. The calculations of \citet{MuldersEA15} (see their Figure 1) suggest that the snowline is located at a radial distance of $\sim$1-2 au. Radiative transfer modelling of the dust disks may provide further constraints on the location of the \ce{H_2O} snowline. As the third component at $\sim$180 K is at larger distances than the midplane snow surface, this suggests a curved rather than vertical snow surface. We find that the decreasing temperature profile can be well described by the powerlaw $T(R_\textnormal{em})\sim$500$\left(\frac{R_\textnormal{em}}{1\textnormal{ [au]}}\right)^{-0.5}$ K. Only the emitting radius for the final component differs between method 1 and methods 2 and 3, but still agrees well within the given uncertainties. The resulting temperatures agree with those given by the large scale template method, in particular the third component ($T\sim$180 K) corresponds well with the one ($T\sim$170 K) that \citet{BanzattiEA23Subm} needed to explain the lines just shortward of 24 $\mathrm{\mu}$m. To extend upon this notion, Figure \ref{fig:3Cvs2C} shows a comparison between the multi-component fits including three and two components. The main difference is most clearly visible in the region of 23.50-24.25 $\mathrm{\mu}$m, which includes the aforementioned cold lines. The two component model is not able to properly fit this line cluster, whereas the three component model can. \\
\indent Compared to results from fitting \ce{H_2O} across the different wavelength regions (see Section \ref{sec:H2O-Regions}), we note that the derived column densities for these methods do not show the same behaviour. For the multi-component fit, the column densities decrease with lower temperatures, whereas they increase with temperature for the different regions. The difference in behaviour may have arisen from combining the contribution of every component in the multi-component fits, whereas contributions of other components and/or wavelength regions have not been accounted for in the fits to the different wavelength regions. \\
\indent A multi-component fit could, instead of a large disk model, also be used to investigate the enhanced reservoir of cold \ce{H_2O} through radial drift. We expect that a two component fit is sufficient to fit the spectra without enhancement of cold \ce{H_2O} through radial drift, whereas three component fits are required to properly fit those with the enhancement. Further work examining a large sample of disks, small and large, structured and smooth, is necessary to fully explore this thought. \\
\indent The reason why all three models yield the same results is due to line optical depths. These optical depths are too high (reaching values up to $\tau\sim$3000 for the innermost region and $\tau\sim$350 for the first annulus for the brightest lines) for any of the shielded regions to make a significant contribution to the total model. In other words, considering only the innermost region, the optical depth of the hottest components is so high that the contributions (i.e. those with an exponential in Equation \ref{eq:RadGr_Sh}) of the lower temperature components are negligible. As discussed in Section \ref{sec:FEACR}, the column density (in cm$^{-2}$) of the warmer two components (components 1 and 2) can be expected to be on the order of $\log\left(N\right)\sim$18.4. \\ 
\indent When using simple LTE slab models to describe the full rotational spectrum of \ce{H_2O}, it is clear that a vertical temperature gradient does not need to be included. Due to the negligible contributions from the overlapping region, a simple radial gradient (using two or three components) alone should be sufficient to represent the emitting layer in the upper region of the disk. For more sophisticated thermochemical modelling methods, more realistic column densities may be probed and the inclusion of vertical temperature gradients may become of significant importance.

\subsection{The importance of radial drift} \label{sec:disc-RD}
We find that our introduced multi-component models, with various levels of complexity, are able to yield similar results as those obtained with the method of \citet{BanzattiEA23Subm}. We note that at least three components are required to fully describe the pure rotational \ce{H_2O} of DR~Tau (see also Section \ref{sec:disc-RG}). In particular, these three components, including a cold component of $T\sim$180 K, are necessary to describe the quartet of \ce{H_2O} lines just shortward of $\sim$24 $\mathrm{\mu}$m. This cold component has been suggested by \citet{BanzattiEA23Subm} to trace an additional \ce{H_2O} reservoir near the snowline, following the inward drift of icy pebbles, the subsequent sublimation, and the diffusion of \ce{H_2O} vapour. Modelling works (e.g. \citealt{BosmanEA18,KalyaanEA21}) have shown that the inward drift of solids can increase the \ce{H_2O} abundance (or vapour mass) inside the snowline up to an order of magnitude during the first few million years. As our models require this cold component and the disk around DR~Tau is compact in the millimetre dust ($\lesssim$60 au), our multi-component analysis can be used to further study the importance of radial drift in setting the potentially enhanced observable \ce{H_2O} reservoirs in the inner regions of planet-forming disks. An upcoming paper will utilise this method on a larger sample of compact disks.

\subsection{\ce{CO} versus \ce{H_2O}} \label{sec:CO-H2O}
With both \ce{CO} (see \citealt{TemminkEA24}) and ro-vibrational \ce{H_2O} (Regions 1 and 2 in Table \ref{tab:Regions}) analysed, we can compare their respective total number of molecules, assuming the two molecules are co-spatial down to the layer where the $\sim$6 $\mathrm{\mu}$m dust continuum becomes optically thick. The assumption that the two molecules may be co-spatial is supported by the high spectral resolution study of \citet{BanzattiEA23}, where the similar line profiles suggest that the ro-vibrational \ce{H_2O} lines may originate from the same region as the broad component of \ce{CO}, which is thought to trace the Keplerian rotation of the disk. Using the optically thin \ce{C^{18}O} emission seen in the VLT-CRIRES observations of DR~Tau, \citet{TemminkEA24} derived a total number of molecules of $\mathcal{N}_\textnormal{\ce{CO}}$=4.1$\times$10$^{44}$. For the \ce{C^{18}O}, they retrieved an excitation temperature of $T\sim$975 K and an emitting radius of $R_\textnormal{em}$=0.23 au, which agree well with the best fit parameters found for the ro-vibrational \ce{H_2O} transitions at the shortest wavelengths (5.0-6.5 $\mathrm{\mu}$m, see Table \ref{tab:Regions}), suggesting that the probed emission may indeed come from the same region in the disk. For this wavelength range (5.0-6.5 $\mathrm{\mu}$m), we retrieve a total number of molecules of $\mathcal{N}_\textnormal{\ce{H_2O}}\sim$7$\times$10$^{43}$. These values yield a ratio of $\mathcal{N}_\textnormal{\ce{H_2O}}/\mathcal{N}_\textnormal{\ce{CO}}\sim$0.17. \\
\indent Taking typical ISM abundances of [O]=5.8$\times$10$^{-4}$ and [C]=3.0$\times$10$^{-4}$, and taking into account the amounts of oxygen and carbon that are locked up in solid form, the maximum \ce{H_2O}/\ce{CO} ratio is $\sim$1.4-2.0, if all volatile oxygen not locked up in \ce{CO} is driven into \ce{H_2O} \citep{vDishoeckEA21}. This ratio is significantly higher than we retrieve for DR~Tau, suggesting that the total number of \ce{H_2O} molecules may be low, instead of enhanced as expected for compact disks, in the inner disk. However, as discussed at previous points throughout this work (see Sections \ref{sec:H2-18-O} and \ref{sec:disc-RG}), many of the observed \ce{H_2O} transitions are optically thick. Subsequently, our derived ratio of $\mathcal{N}_\textnormal{\ce{H_2O}}/\mathcal{N}_\textnormal{\ce{CO}}\sim$0.16 acts as a lower limit. A factor $\sim$10 is required to meet the aforementioned expected ratio of \ce{H_2O}/\ce{CO}$\sim$1.4-2.0. Perhaps the \ce{H_2O} emission becomes optically thick well above the layer where the dust becomes optically thick, suggesting that we may probe less deep into disk compared to both \ce{CO} (probed with the optically thin \ce{C^{18}O} emission) and \ce{H_2}. Additionally, \ce{H_2O} may self-shield against the photodissociating UV-photons before \ce{C^{18}O} \citep{CalahanEA22}, leaving a thin emitting layer where both molecules are co-located. Higher signal-to-noise detections of one of the rarer \ce{H_2O} isotopologue (e.g. \ce{H_2 ^{18}O}) or detections of the optically thin, low Einstein-A \ce{H_2 ^{16}O} transitions (see Section \ref{sec:H2-18-O}), which probe deeper into the disk \citep{BosmanEA22}, are needed to further explore the abundance ratio of \ce{CO} and \ce{H_2O}. \\
\indent We note that the total number of molecules for both \ce{H_2O} and \ce{CO} appear to be low with respect to that of molecular hydrogen, \ce{H_2}. For \ce{CO} the ISM abundance is of the order of \ce{CO}/\ce{H_2}$\sim$10$^{-4}$ \citep{BW17,LacyEA17}, whereas that of \ce{H_2O} is \ce{H_2O}/\ce{H_2}$\sim$4$\times$10$^{-4}$ \citep{vDishoeckEA21}, if most volatile oxygen is used in the formation of \ce{H_2O}. Assuming a temperature of $T\sim$1000 K and an emitting radius of $R_\textnormal{em}$=0.20 au, we derive an upper limit on the total number of molecules of $\mathcal{N}_\textnormal{\ce{H_2}}\leq$2.1$\times$10$^{51}$. This number is a factor of, respectively, 5.1$\times$10$^6$ and 3.0$\times$10$^7$ higher than the total number of molecules derived for \ce{CO} and \ce{H_2O}, under the assumption that all molecules probe the same region of the disk. However, caution is warranted as \ce{H_2} is not strongly detected and we emphasise that derived number of molecules must be treated as an lower limit. 

\section{Conclusions \& summary} \label{sec:CS}
In this work, we thoroughly analysed the \ce{H_2O} excitation temperatures and column densities in the inner region of the planet-forming disk of DR~Tau, as seen with JWST-MIRI/MRS. We have used techniques presented in previous works, as well as tried, for the first time, a multi-component approach to fully describe the rotational \ce{H_2O} spectrum ($\geq$10 $\mathrm{\mu}$m). In addition, to the \ce{H_2O} analysis, we have also analysed the \ce{OH}, atomic and molecular hydrogen emission, and looked for other molecular species. We summarise our conclusions as follows:
\begin{itemize}
    \item By analysing the \ce{H_2O} over various wavelength regions, we find that the excitation temperature decreases with wavelength, whereas the emitting radius increases. This is a clear sign of a radial temperature gradient visible in the spectrum. The need for a temperature gradient, or multiple components, is further confirmed by using the spectrum of a large disk (CI~Tau) as comparison template for the spectrum of our compact disk. Various cold emission lines are not reproduced by this large disk template and are explained by slab models with low temperatures of $T\sim$400 K and $T\sim$200 K.
    \item To further test the need for multiple components, we used an MCMC approach to fit two and three components at the same time. Our methods include a simple radial gradient, a radial gradient with spatial overlap, and a radial and vertical gradient. The different methods yield no significant differences in the excitation properties (temperatures, column densities, and emitting radii) of the different components, likely due to the high optical depth of the hotter components in the overlapping regions. Consequently, we can state that simple models (i.e. models only considering a radial temperature gradient) are able to provide a proper description of the rotational \ce{H_2O} spectrum and are supported by the more complex models (i.e. those including a vertical gradient). Only for disks where the hotter components are not optically thick, the results will differ and a more complex model must be used.
    \item We find that the temperature for the third approach can be described roughly as $T(R_\textnormal{em})\sim$500$\left(\frac{R_\textnormal{em}}{1\textnormal{ [au]}}\right)^{-0.5}$ K. For DR~Tau, we emphasise the need for a minimum of three components ($T_1\sim$800 K, $T_2\sim$470 K, and $T_3\sim$180 K), as the two component model is unable to fit the transitions at the longest wavelengths, especially those around $\sim$24 $\mu$m. While the need for the third component around the \ce{H_2O} snowline is in agreement with previous works, it is not yet possible to conclude whether the \ce{H_2O} abundance is enhanced due to radial drift. A future study focusing on a consistent analysis of a larger sample of compact disks will provide more insights in the importance of radial drift.
    \item While \ce{OH} and various atomic and molecular hydrogen transitions are detected, we are inconclusive about the detection of the isotopologue \ce{H_2 ^{18}O}, mostly due to the (current) large uncertainties at the longest wavelengths. We also do not report the detections of various other species, including: \ce{CH_4}, \ce{NH_3}, \ce{CS}, \ce{H_2}, \ce{SO_2}, and larger hydrocarbons. The derived upper limits on the column densities of these non-detected species are almost all significantly lower, by factors $\sim$10$^2$-10$^4$, than those found for \ce{H_2O} in the different wavelength regions (see Section \ref{sec:H2O-Regions}). 
    \item Finally, we have compared the derived number of \ce{H_2O} molecules with that of \ce{CO}, derived from the optically thin \ce{C^{18}O} emission. The comparison yields a low ratio of $\sim$0.17, which, for the hot \ce{H_2O} component $T\sim$1000 K, may point to \ce{H_2O} having a relative low total number of molecules in the layer that we are able to trace. However, the strongest \ce{H_2O} lines are found to be optically thick, suggesting that we may be probing less deep into the disk compared to \ce{C^{18}O}, and so this ratio must be treated as a lower limit. A detection of one of the rarer isotopologues (e.g. \ce{H_2 ^{18}O}) is required to better constrain their relative number of molecules down to the layer where the dust continuum becomes optically thick. 
\end{itemize}
This work has shown the need of fitting multiple components to explain the rotational spectrum of \ce{H_2O}. Future works focusing on the multiple-component analysis of \ce{H_2O} in larger samples of disks can help us further understand the \ce{H_2O} reservoir present in the inner regions of disks. In addition, this method may shed more light on the importance of radial drift in setting the inner disk \ce{H_2O} reservoir by comparing small disks, where radial drift is expected to be very efficient, with large disks, in which drift is expected to be halted by substructures.


\begin{acknowledgements}
    The authors would like to thank Andrea Banzatti for many very useful discussions. We also thank the referee for many thoughtful, constructive comments that helped improve the manuscript. \\
    This work is based on observations made with the NASA/ESA/CSA James Webb Space Telescope. The data were obtained from the Mikulski Archive for Space Telescopes at the Space Telescope Science Institute, which is operated by the Association of Universities for Research in Astronomy, Inc., under NASA contract NAS 5-03127 for JWST. These observations are associated with program \#1282. The following National and International Funding Agencies funded and supported the MIRI development: NASA; ESA; Belgian Science Policy Office (BELSPO); Centre Nationale d’Etudes Spatiales (CNES); Danish National Space Centre; Deutsches Zentrum fur Luft- und Raumfahrt (DLR); Enterprise Ireland; Ministerio De Econom\'ia y Competividad; Netherlands Research School for Astronomy (NOVA); Netherlands Organisation for Scientific Research (NWO); Science and Technology Facilities Council; Swiss Space Office; Swedish National Space Agency; and UK Space Agency. \\
    \indent The authors acknowledge the GO \#1640 team for developing their observing program with a zero-exclusive-access period, which includes CI~Tau. \\
    \indent M.T., E.v.D., and M.V. acknowledge support from the ERC grant 101019751 MOLDISK. E.v.D. acknowledges support the Danish National Research Foundation through the Center of Excellence ``InterCat'' (DNRF150). E.v.D., I.K., and A.M.A., acknowledge support from grant TOP-1 614.001.751 from the Dutch Research Council (NWO). D.G. thank the Belgian Federal Science Policy Office (BELSPO) for the provision of financial support in the framework of the PRODEX Programme of the European Space Agency (ESA). B.T. is a Laureate of the Paris Region fellowship program, which is supported by the Ile-de-France Region and has received funding under the Horizon 2020 innovation framework program and Marie Sklodowska-Curie grant agreement No. 945298. T.H. and K.S. acknowledge support from the European Research Council under the Horizon 2020 Framework Program via the ERC Advanced Grant Origins 83 24 28. D.B. has been funded by Spanish MCIN/AEI/10.13039/501100011033 grants PID2019-107061GB-C61 and No. MDM-2017-0737. A.C.G. acknowledges from PRIN-MUR 2022 20228JPA3A “The path to star and planet formation in the JWST era (PATH)” funded by NextGeneration EU and by INAF-GoG 2022 “NIR-dark Accretion Outbursts in Massive Young stellar objects (NAOMY)” and Large Grant INAF 2022 “YSOs Outflows, Disks and Accretion: towards a global framework for the evolution of planet forming systems (YODA)”. I.K.  acknowledge funding from H2020-MSCA-ITN-2019, grant no. 860470 (CHAMELEON). G.P. gratefully acknowledges support from the Max Planck Society. \\
    \indent This work also has made use of the following software packags that have not been mentioned in the main text: NumPy, SciPy, Astropy, Matplotlib, pandas, IPython, Jupyter \citep{Numpy,Scipy,AstropyI,AstropyII,AstropyIII,Matplotlib,pandas,IPython,Jupyter}.
\end{acknowledgements}

\bibliographystyle{aa}
\bibliography{Bibliography}


\newpage
\onecolumn
\begin{appendix}
\section{Uncertainties for the JWST-MIRI subbands}
\begin{table}[ht!]
    \centering
    \caption{The median continuum flux, estimated $S/N$ from the ETC, and the obtained uncertainty $\sigma$ for each subband.}
    \begin{tabular}{c c c c}
    \hline\hline
    Subband ([$\mathrm{\mu}$m]) & Median flux [Jy] & ETC S/N & $\sigma$ [mJy] \\
    \hline
    1A (4.90-5.74) & 1.31 & 533.5 & 2.5 \\
    1B (5.66-6.63) & 1.31 & 613.5 & 2.1 \\
    1C (7.51-8.77) & 1.32 & 743.9 & 1.8 \\
    2A (7.51-8.77) & 1.43 & 745.2 & 1.9 \\
    2B (8.67-10.13) & 1.98 & 962.6 & 2.1 \\
    2C (10.02-11.70) & 2.03 & 1014.2 & 2.0 \\
    3A (11.55-13.47) & 1.79 & 956.5 & 2.0 \\
    3B (13.34-15.57) & 1.89 & 991.57 & 1.9 \\
    3C (15.41-17.98) & 2.35 & 1143.0 & 2.1 \\
    4A (17.70-20.95) & 2.69 & 497.3 & 5.4 \\
    4B (20.69-24.48) & 2.85 & 271.7 & 10.5 \\
    4C (24.19-27.90) & 2.97 & 75.5 & 39.4 \\
    \hline
    \end{tabular}
    \label{tab:Subband-Sigma}
\end{table}

\clearpage
\section{\ce{H_2O} slab models across the wavelength regions}
\subsection{Without line overlap}
\begin{figure*}[ht!]
    \centering
    \includegraphics[width=0.9\textwidth]{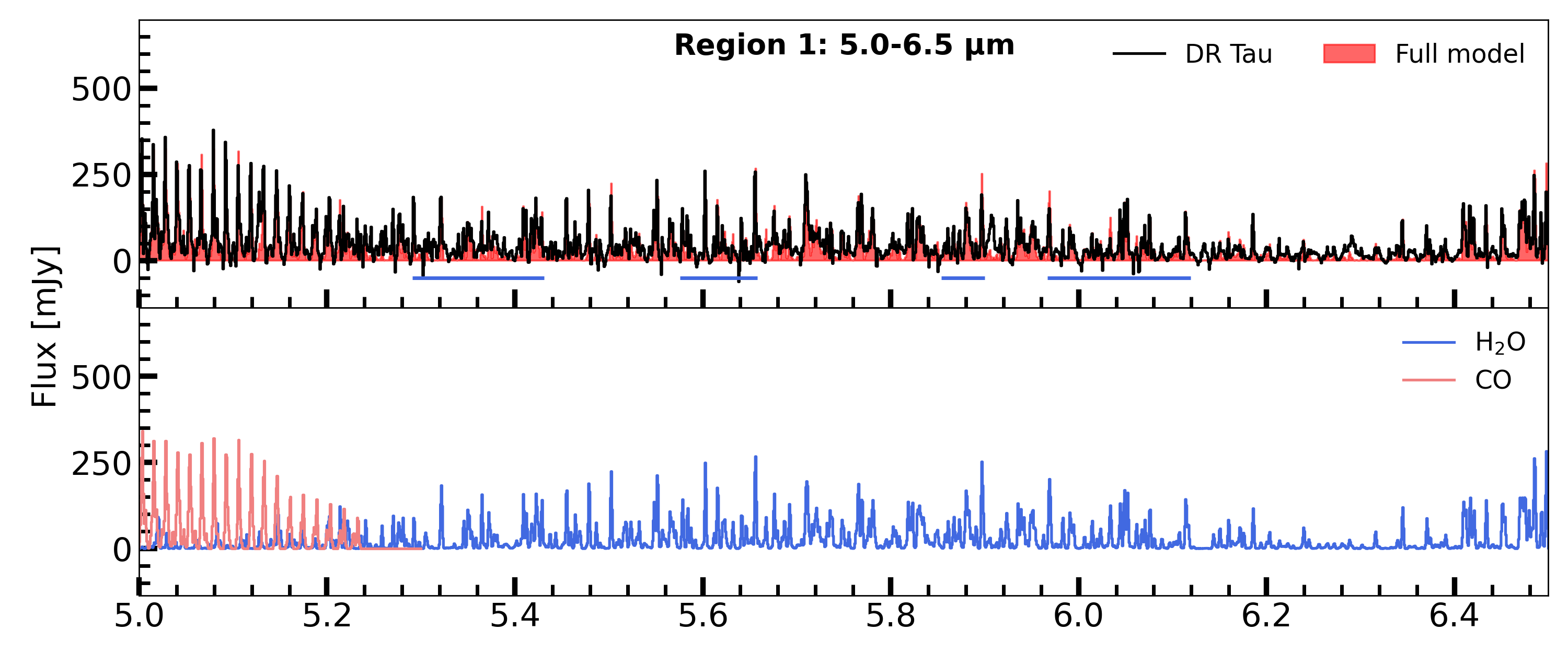}
    \includegraphics[width=0.9\textwidth]{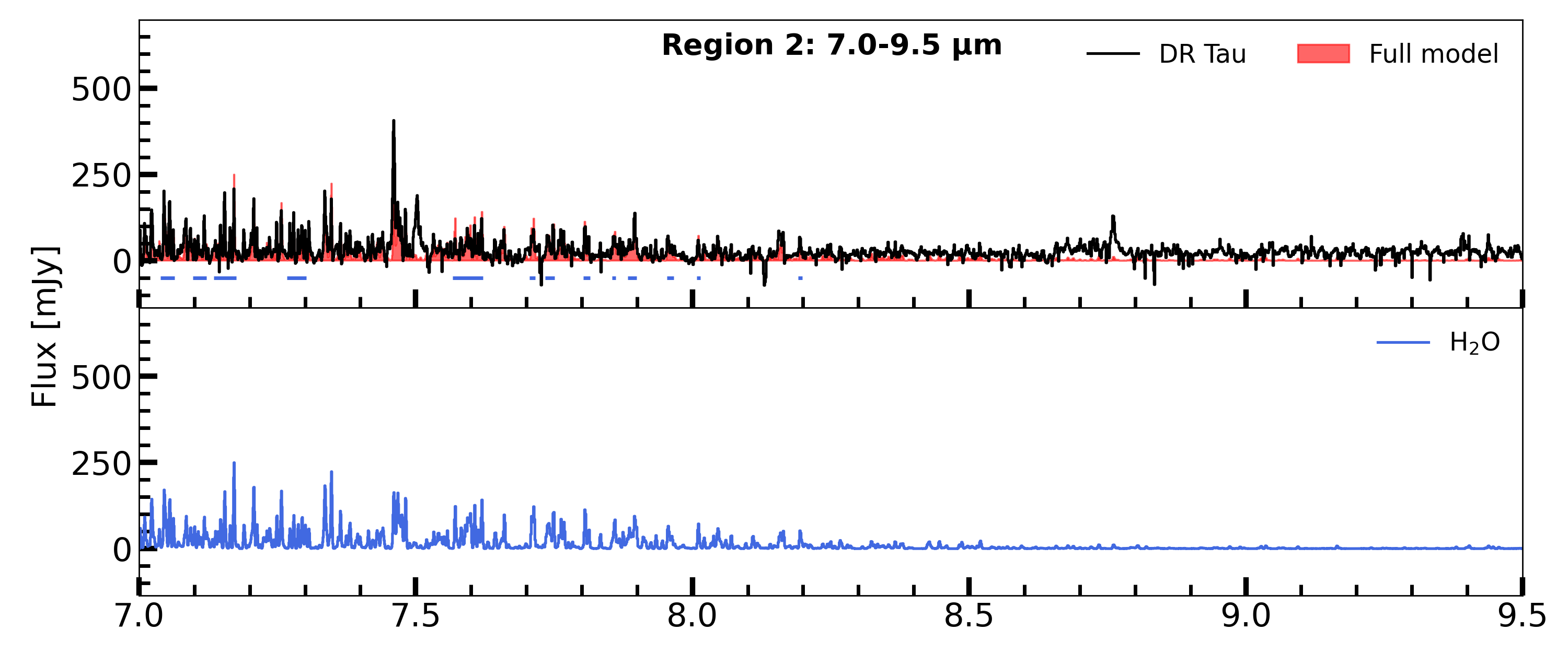}
    \includegraphics[width=0.9\textwidth]{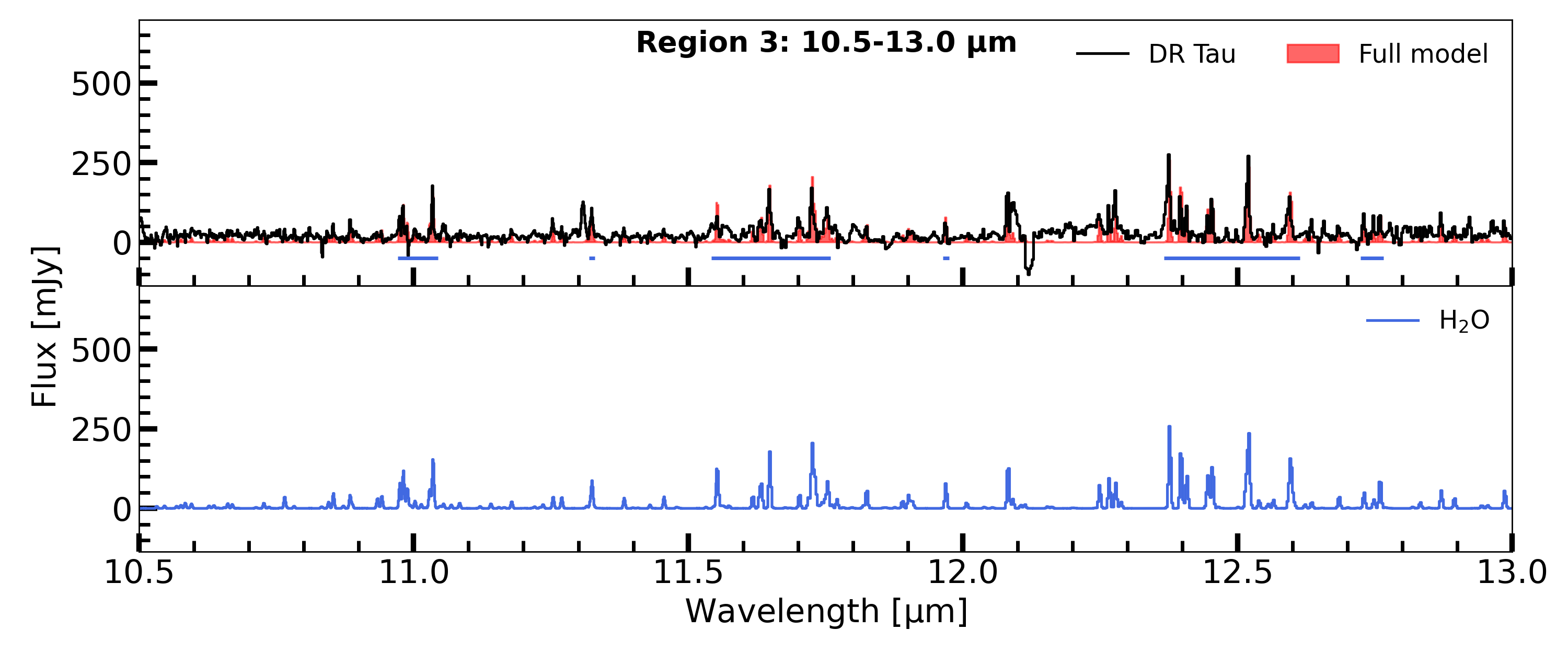}
    \caption{Best fitting slab models (without line overlap) for the different regions. In each subfigure, the top panel displays the continuum subtracted JWST spectrum in a specific region, while the full model spectrum is shown in red. The bottom panels show the models for the individually detected molecules. In addition, we show the \ce{CO} model in pink from \citet{TemminkEA24}. The horizontal bar in each top panel indicate the line regions used in the $\chi^2_\textnormal{red}$-fits.}
    \label{fig:RegionSpectra}
\end{figure*}

\begin{figure*}[ht!]
    \ContinuedFloat
    \centering
    \includegraphics[width=0.9\textwidth]{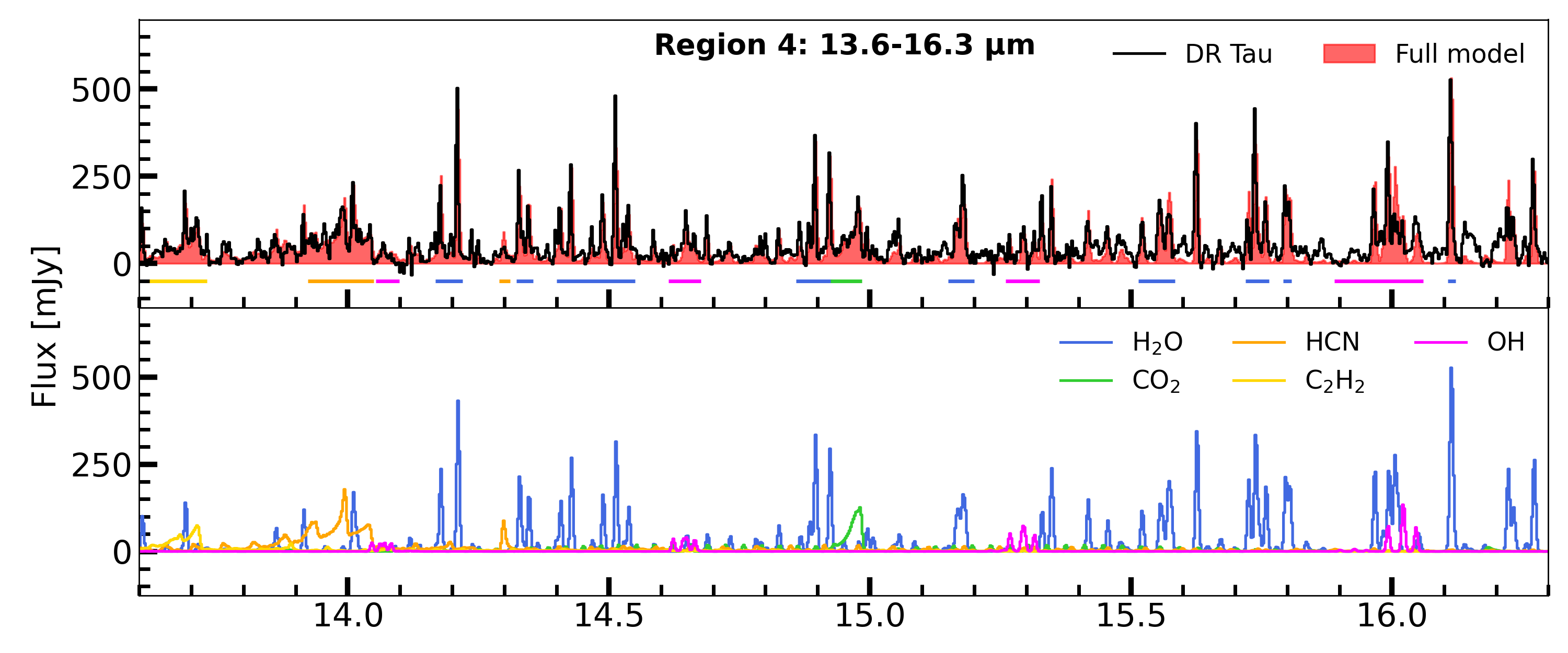}
    \includegraphics[width=0.9\textwidth]{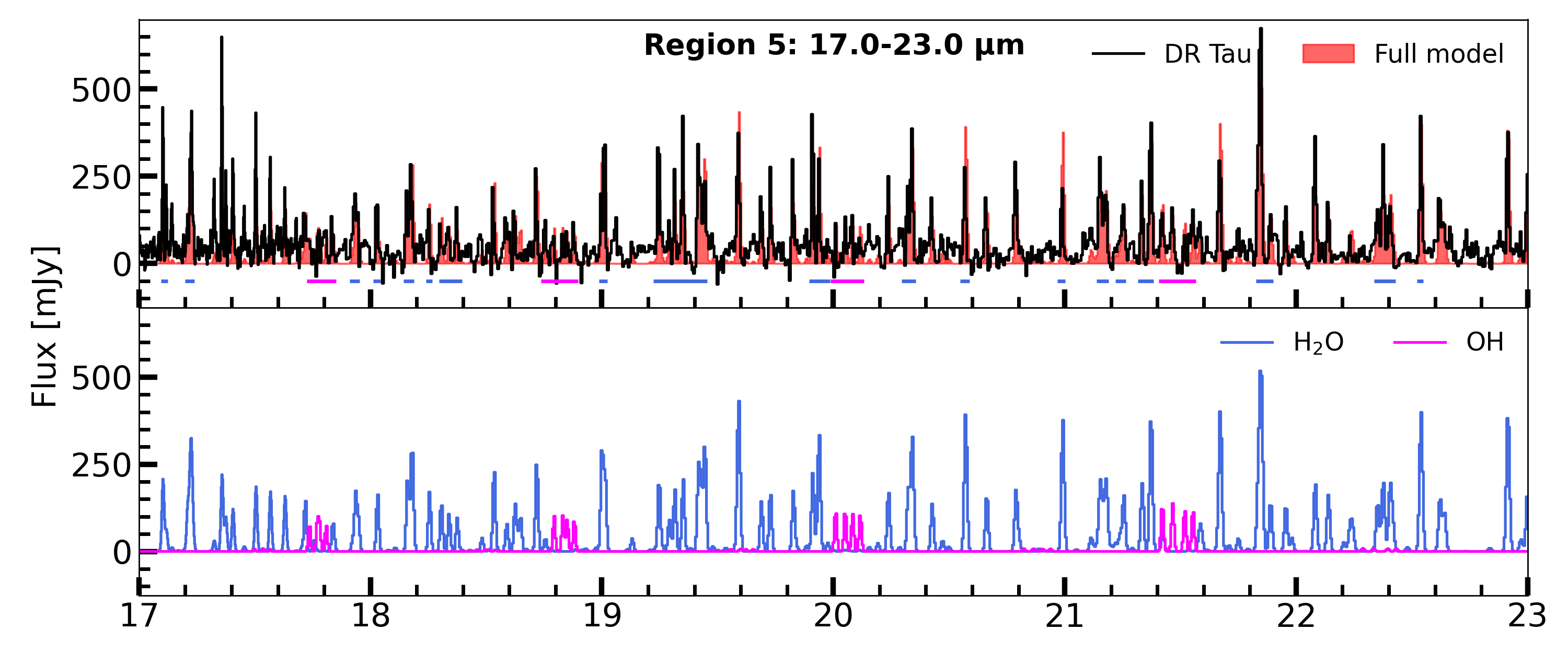}
    \includegraphics[width=0.9\textwidth]{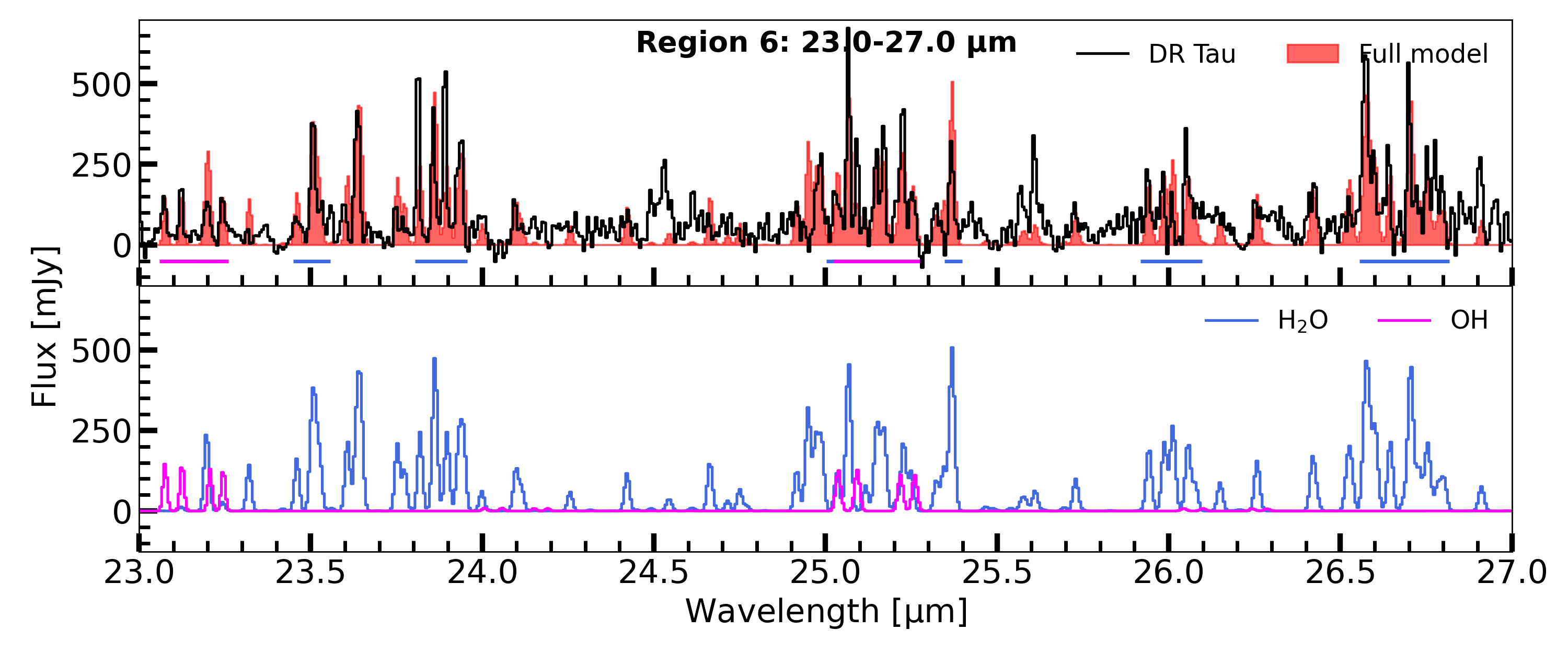}
    \caption{Continuation of Figure \ref{fig:RegionSpectra}. The best fit to the \ce{OH} emission is shown in magenta. In addition, we show the best fitting slab models, adopted from \citet{TemminkEA24}, for \ce{CO_2} (green), \ce{HCN} (orange), and \ce{C_2H_2} (yellow) in the wavelength region of 13.6-16.3 $\mathrm{\mu}$m. }
    \label{fig:RegionSpectra}
\end{figure*}

\clearpage
\subsection{With line overlap}
\begin{figure*}[ht!]
    \centering
    \includegraphics[width=0.9\textwidth]{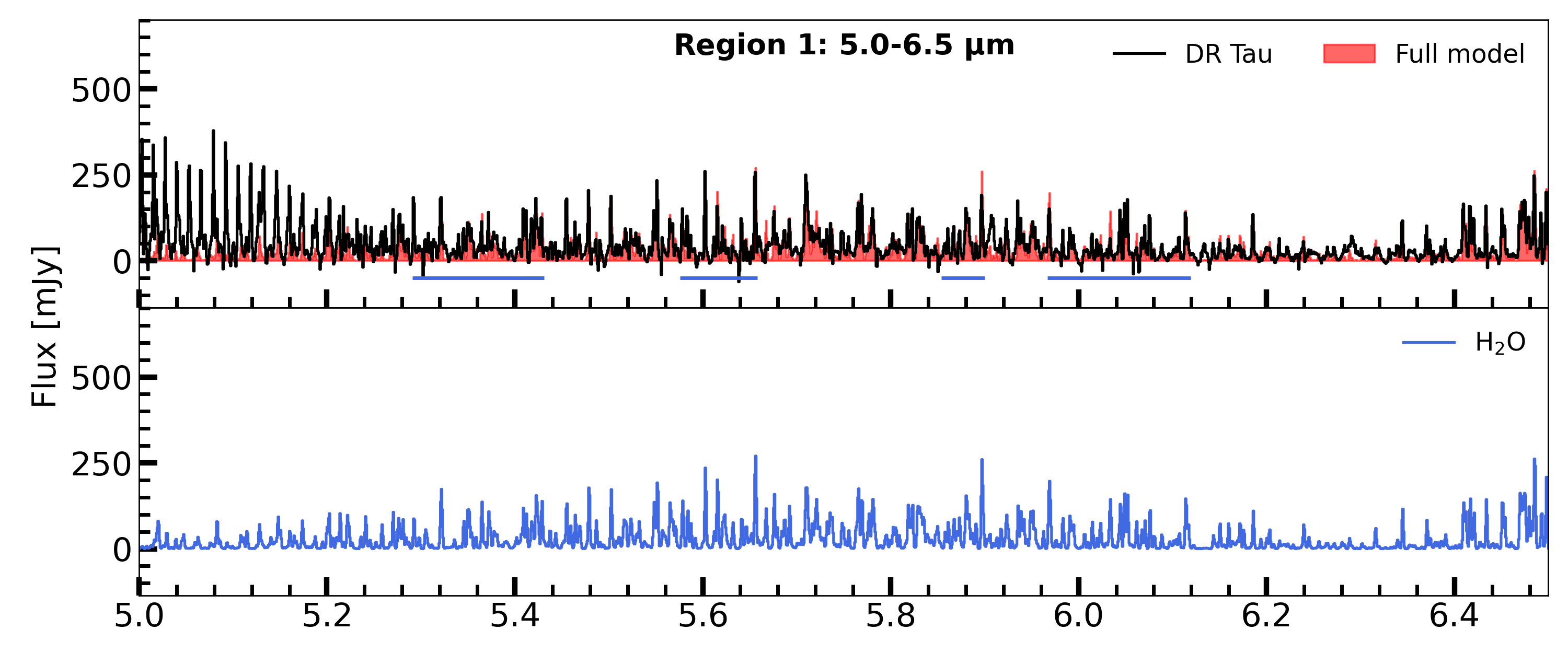}
    \includegraphics[width=0.9\textwidth]{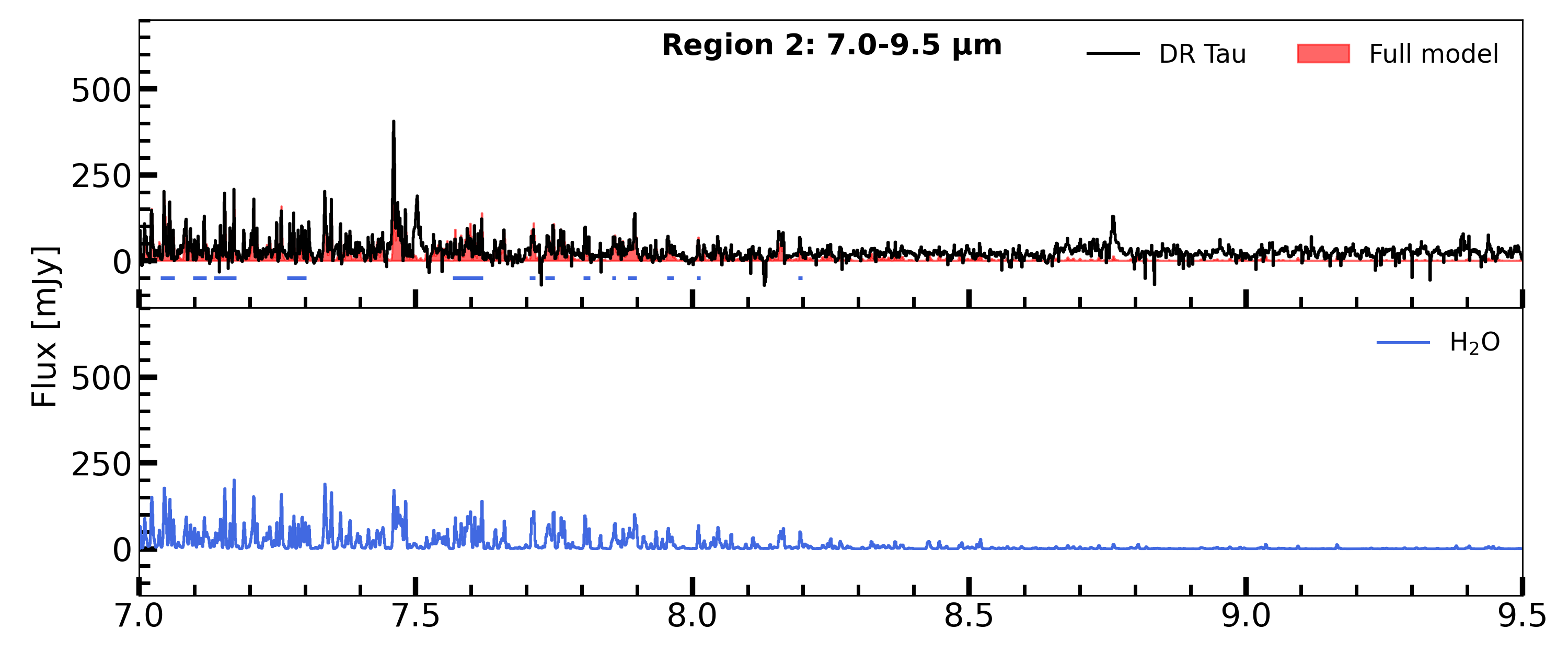}
    \includegraphics[width=0.9\textwidth]{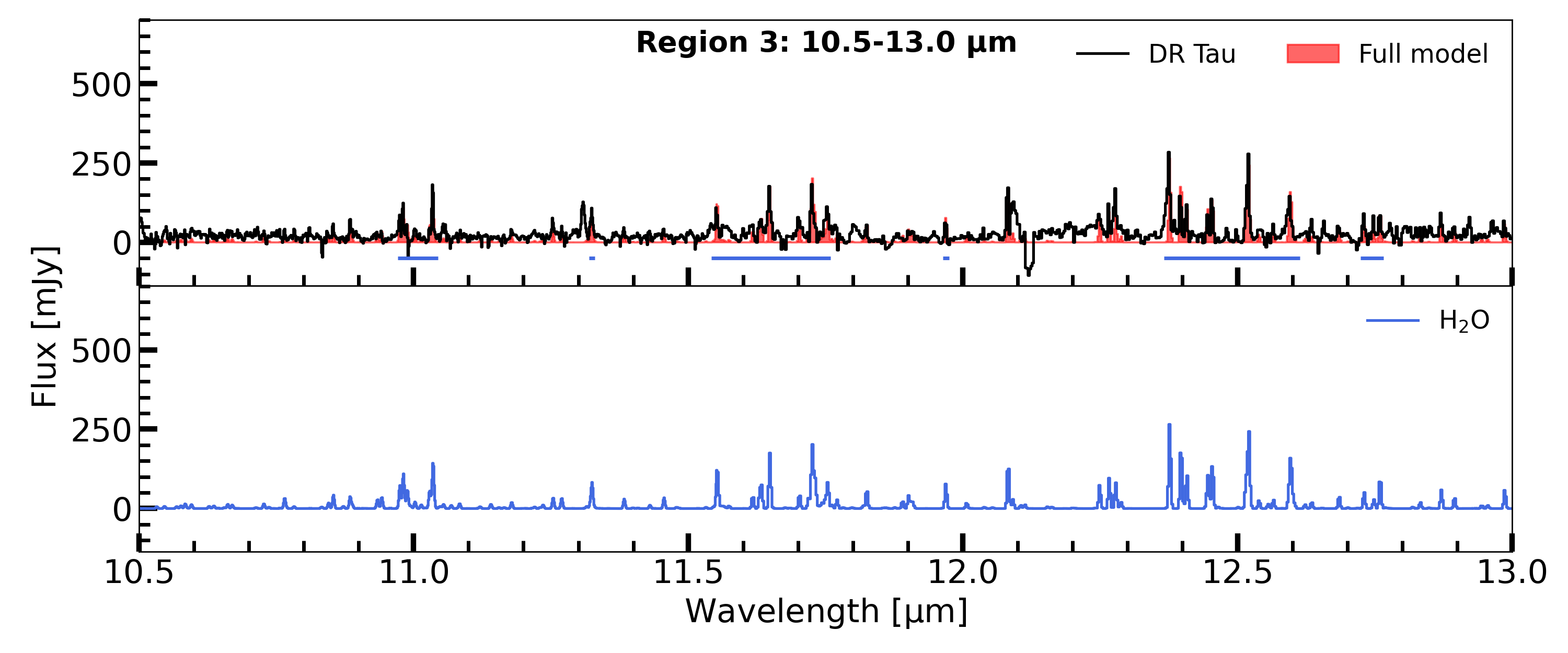}
    \caption{The same as Figure \ref{fig:RegionSpectra}, but only showing the fitted \ce{H_2O} slab models with mutual line overlap.}
    \label{fig:RegionSpectra-Overlap}
\end{figure*}

\begin{figure*}[ht!]
    \ContinuedFloat
    \centering
    \includegraphics[width=0.9\textwidth]{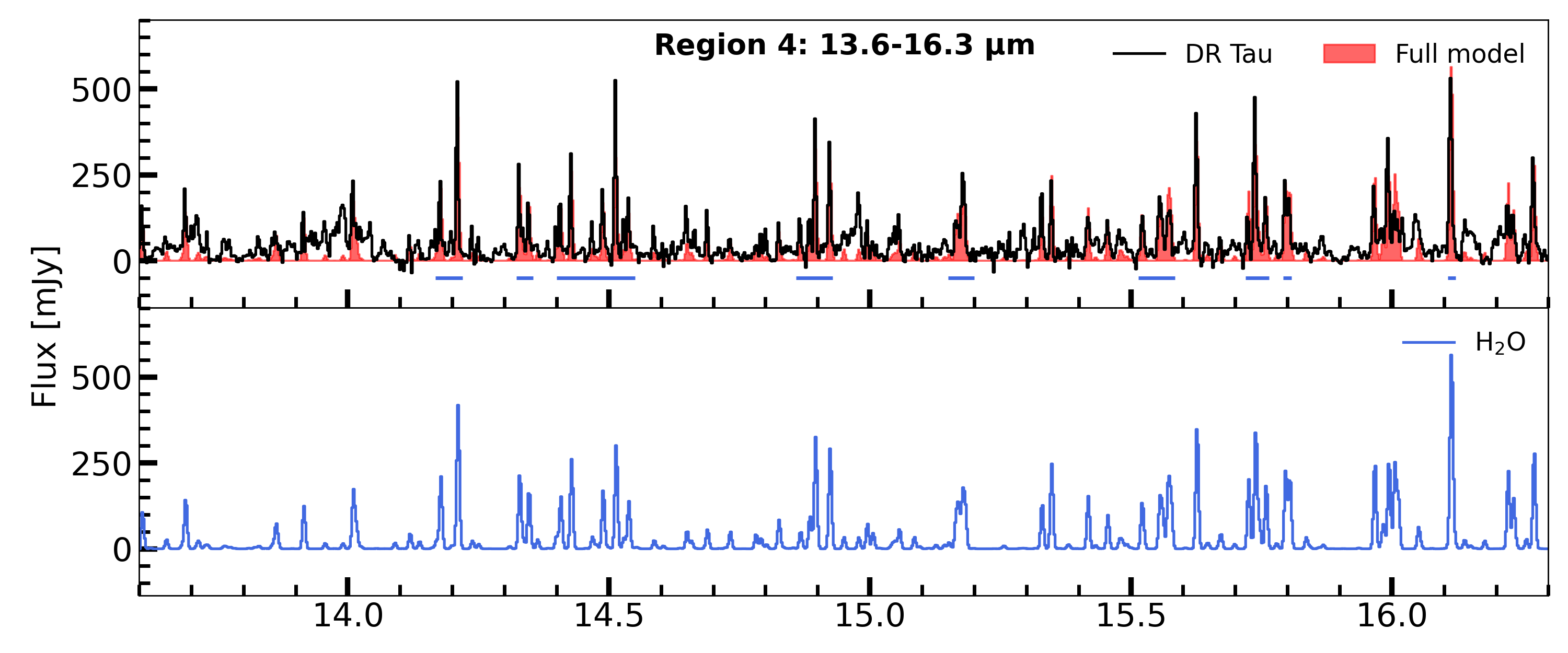}
    \includegraphics[width=0.9\textwidth]{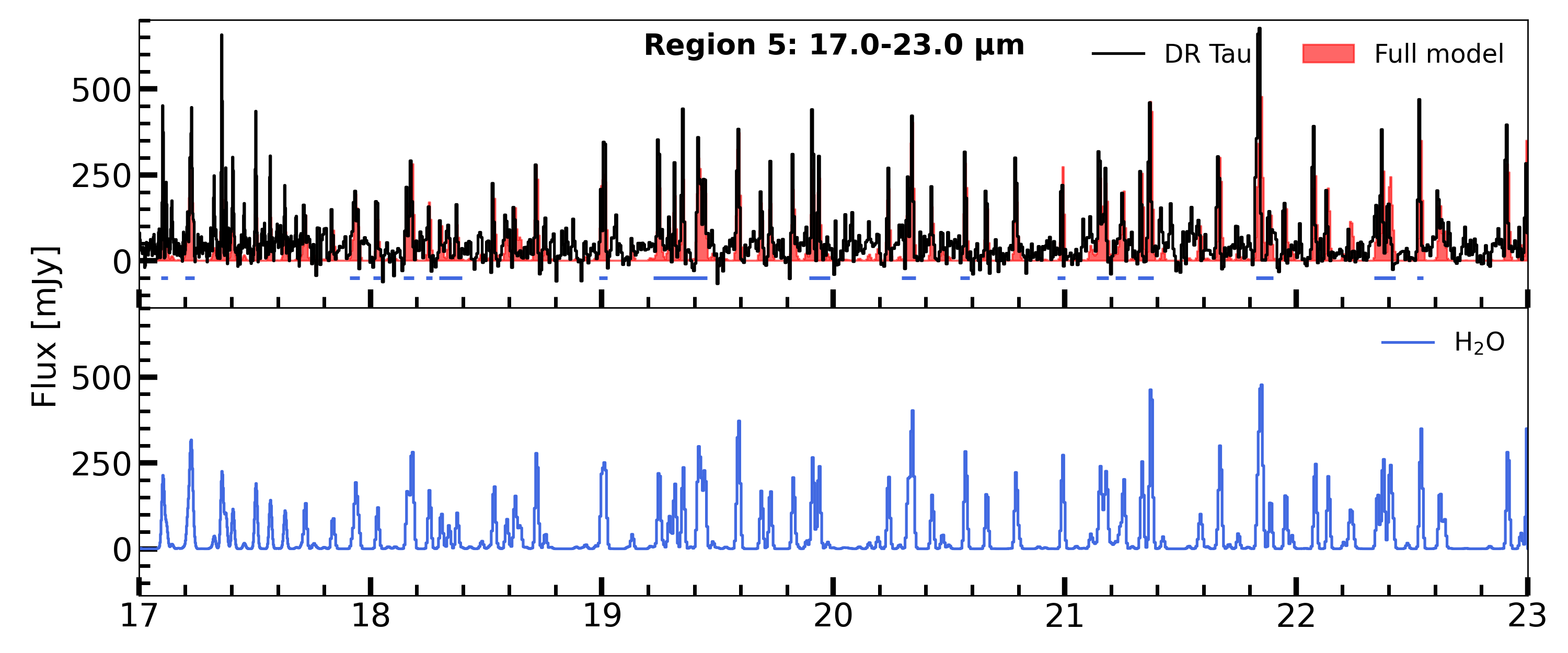}
    \includegraphics[width=0.9\textwidth]{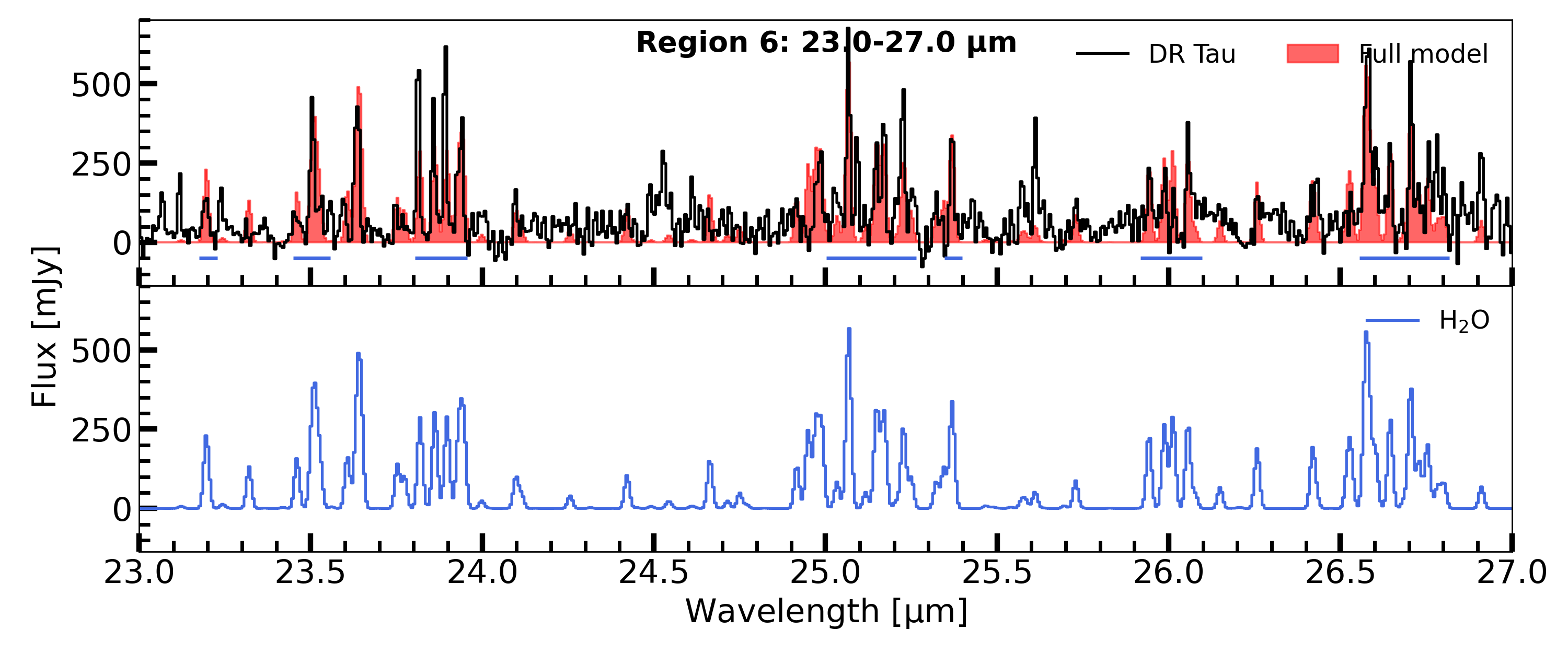}
    \caption{Continuation of Figure \ref{fig:RegionSpectra-Overlap}.}
    \label{fig:RegionSpectra-Overlap}
\end{figure*}

\clearpage
\section{Selected, isolated \ce{H_2O} transitions}
\begin{figure*}[ht!]
    \centering
    \includegraphics[width=\textwidth]{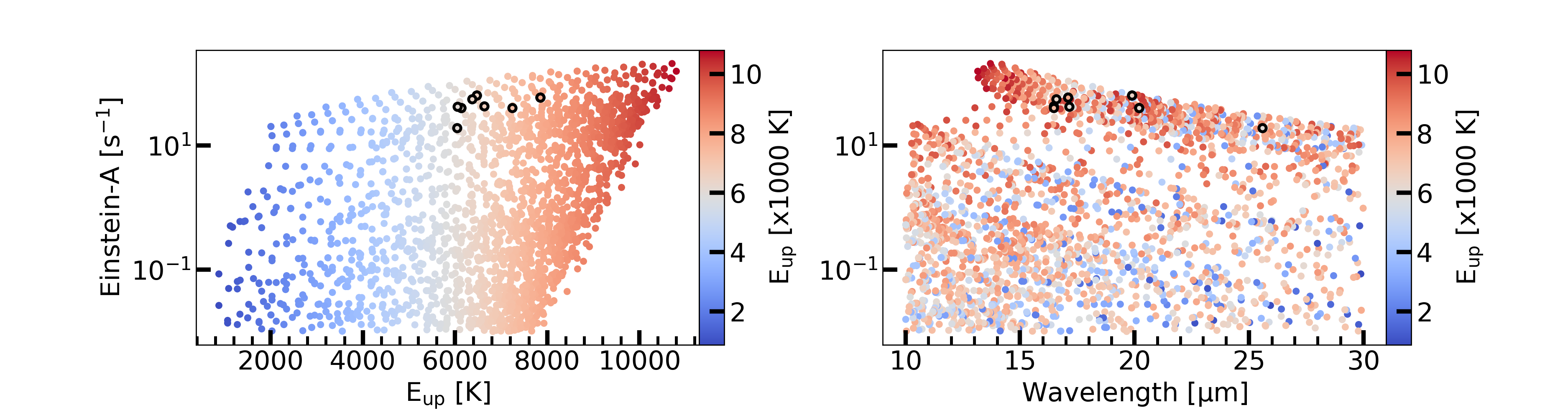}
    \caption{All the \ce{H_2O} transitions with $A_\textnormal{ul}\geq$10$^{-2}$ s$^{-1}$ used in our analysis. The 6 hot ($6000\leq E_\textnormal{up}\leq8000$ K), unblended transitions are indicated by the black circles. The colour-scale indicates the upper level energies.}
    \label{fig:H2O-Lines}
\end{figure*}

\begin{figure*}[ht!]
    \centering
    \includegraphics[width=\textwidth]{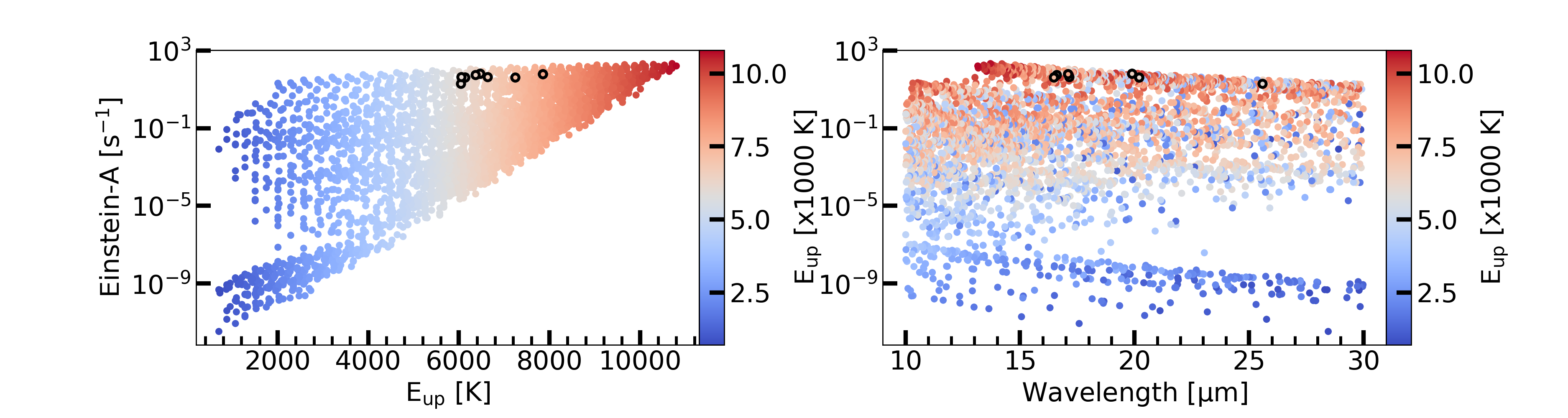}
    \caption{The same as Figure \ref{fig:H2O-Lines}, but without a lower limit imposed on the values for $A_\textnormal{ul}$.}
    \label{fig:H2O-Lines-Full}
\end{figure*}

\clearpage
\section{Two versus three component fits}
\begin{figure*}[ht!]
    \centering
    \includegraphics[width=\textwidth]{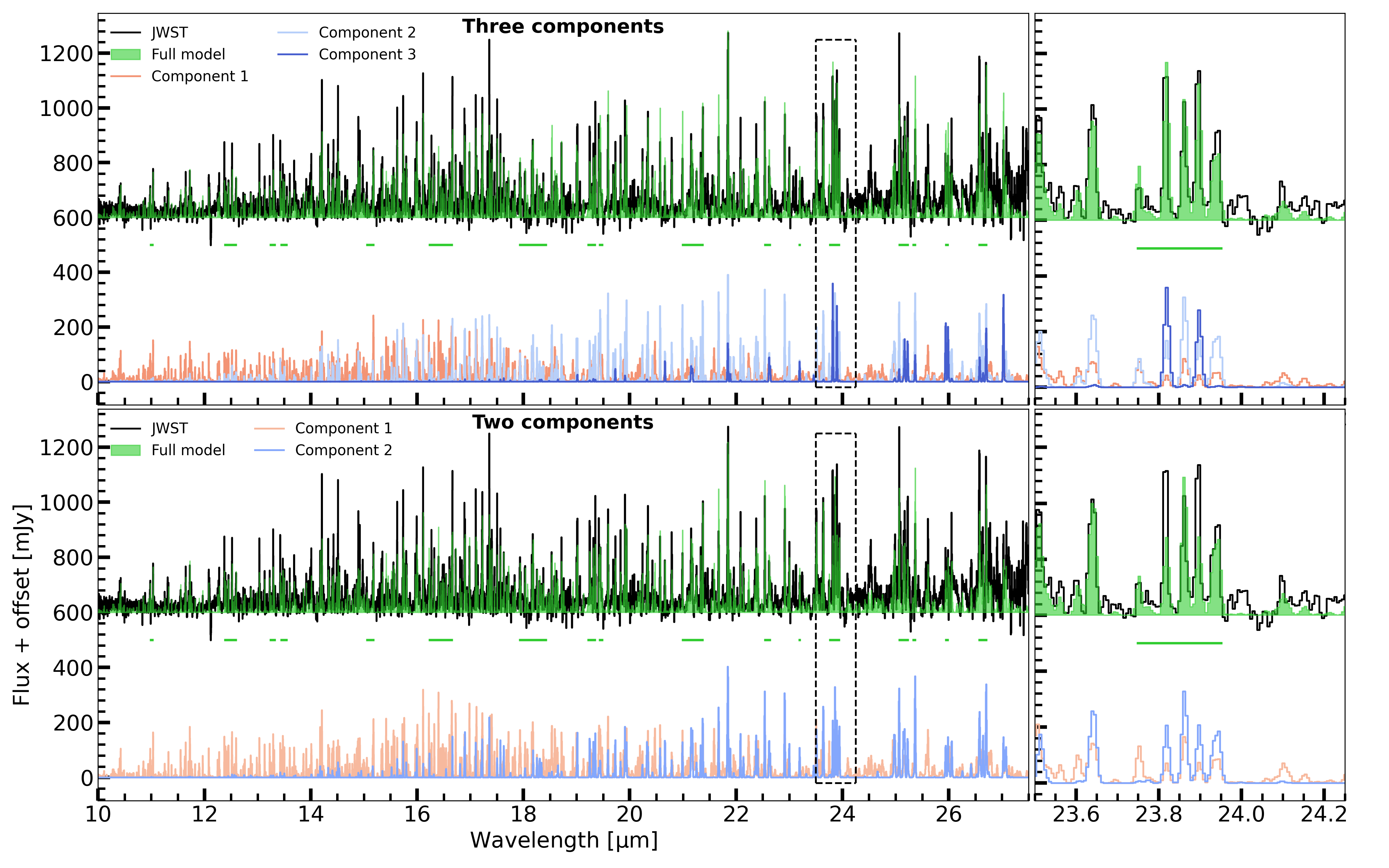}
    \caption{Fits of the three (top panels) and two (bottom panels) component slab models. The full models are shown in green, whereas the different components are shown in various shades of red and blue. The green horizontal bars indicate the regions used in the $\chi^2_\textnormal{red}$-fits. The panels on the right-side of the figure show a zoom-in on the 23.5-24.25 $\mathrm{\mu}$m region, which is also indicated by the dashed lines in the left panels. Table \ref{tab:Grad-Results} (see `Approach III') lists the parameters of each component; the coldest component (i.e. component 3) is not accounted for in the two-component fit.}
    \label{fig:3Cvs2C}
\end{figure*}

\clearpage
\section{Multi-component corner plots}
\begin{figure*}[ht!]
    \centering
    \includegraphics[width=\textwidth]{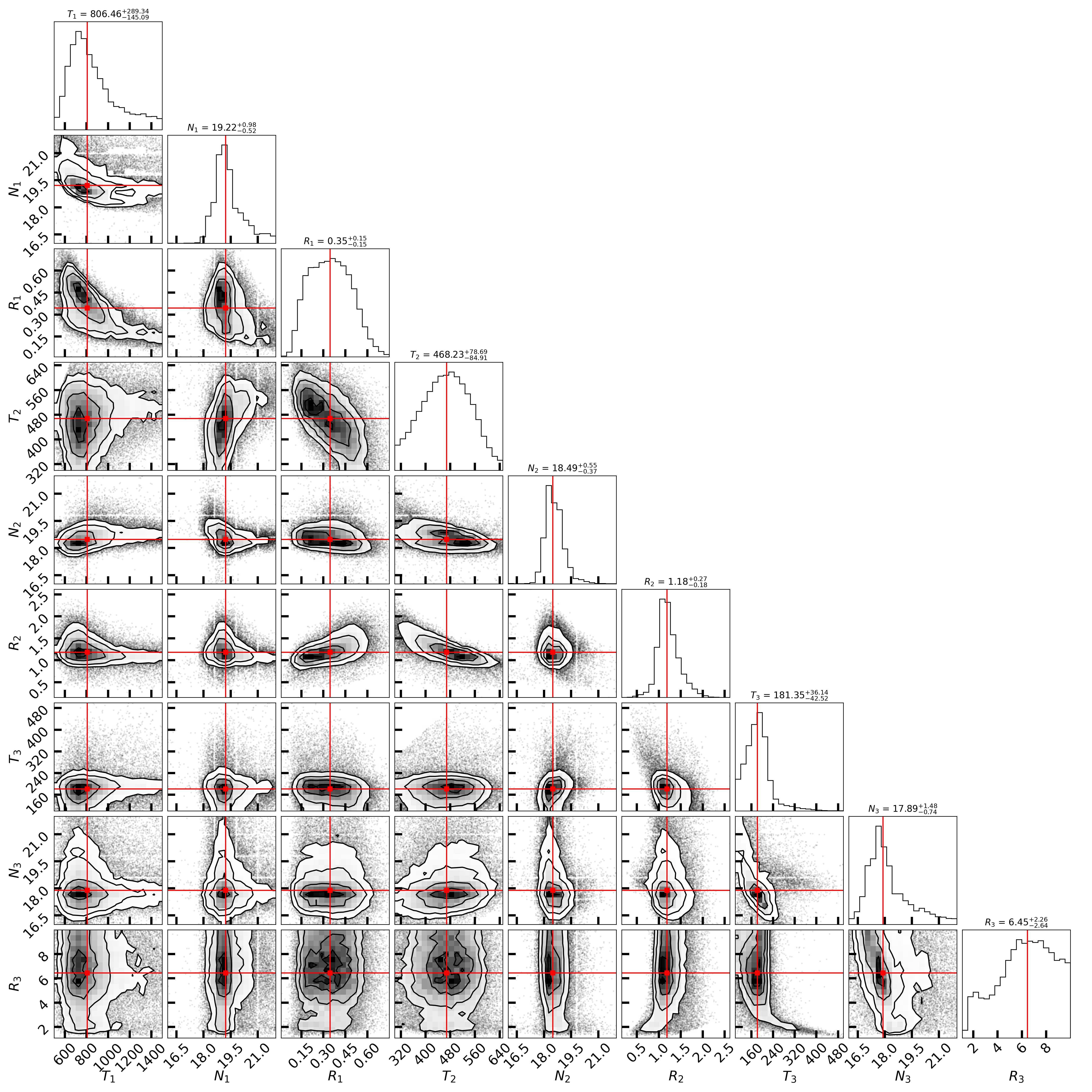}
    \caption{Corner plots for our three component model fits including only a radial temperature gradient (approach I). $T_i$, $N_i$, and $R_i$ denote the excitation temperature (in K), the column density (given in log$_{10}$-space, where $N$ is given in units of cm$^{-2}$), and the emitting radius (in au) of component $i$, respectively.}
    \label{fig:CP1}
\end{figure*}
\begin{figure*}[ht!]
    \centering
    \includegraphics[width=\textwidth]{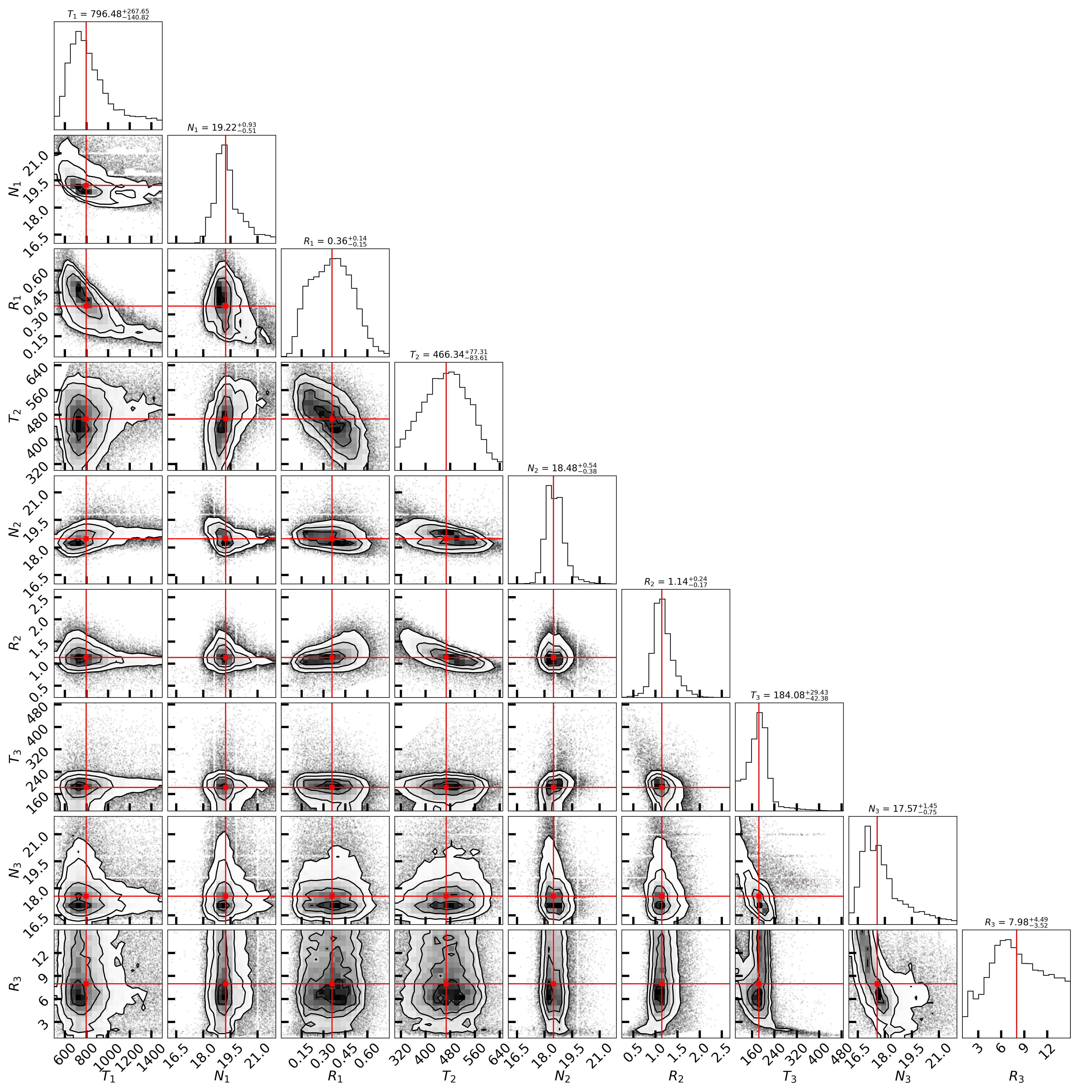}
    \caption{Similar as Figure \ref{fig:CP1}, but for the models with spatial overlap (approach II).}
    \label{fig:CP2}
\end{figure*}
\begin{figure*}[ht!]
    \centering
    \includegraphics[width=\textwidth]{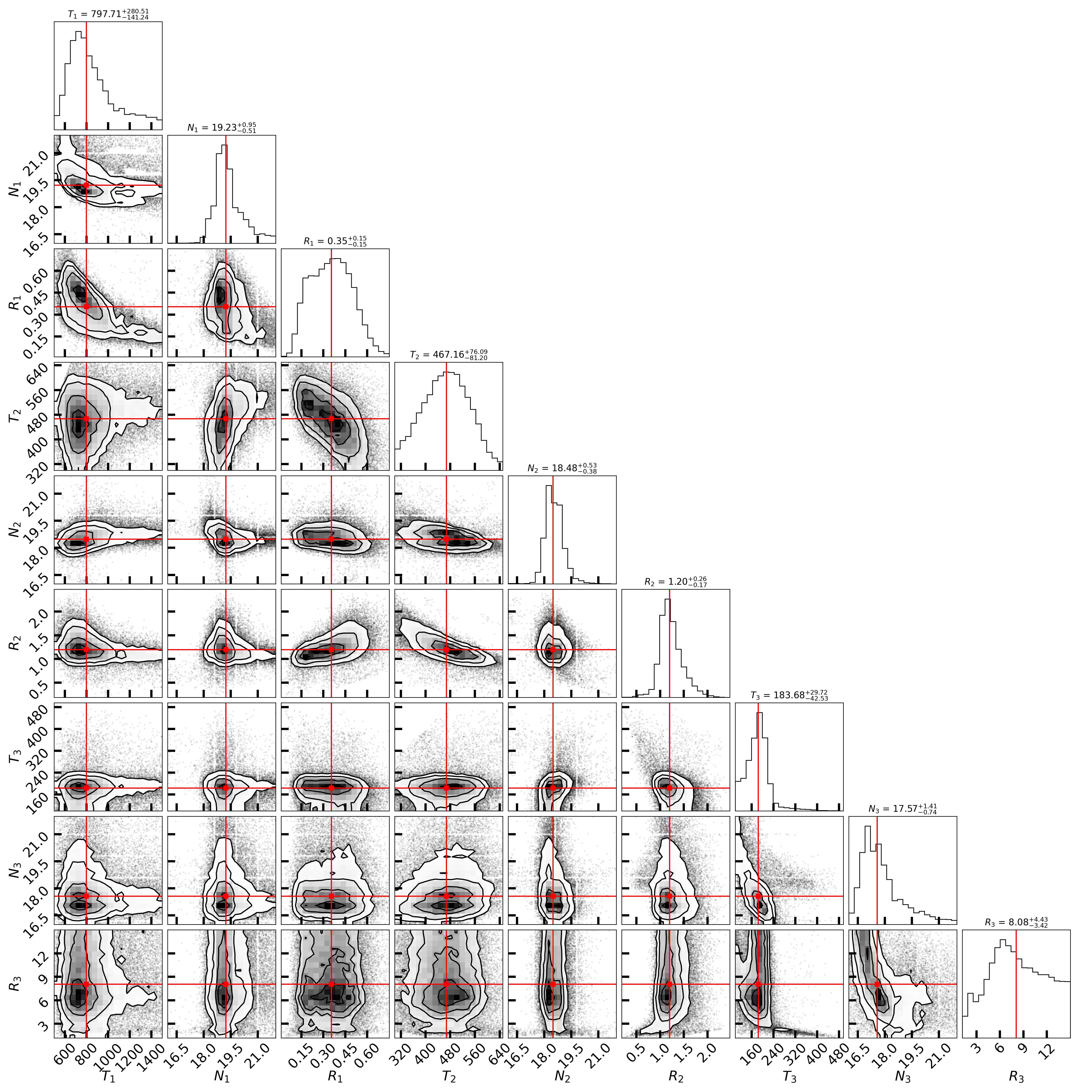}
    \caption{Similar as Figure \ref{fig:CP1}, but for the models with also a vertical temperature gradient and allowing for spatial line shielding (approach III).}
    \label{fig:CP3}
\end{figure*}
\begin{figure*}[ht!]
    \centering
    \includegraphics[width=\textwidth]{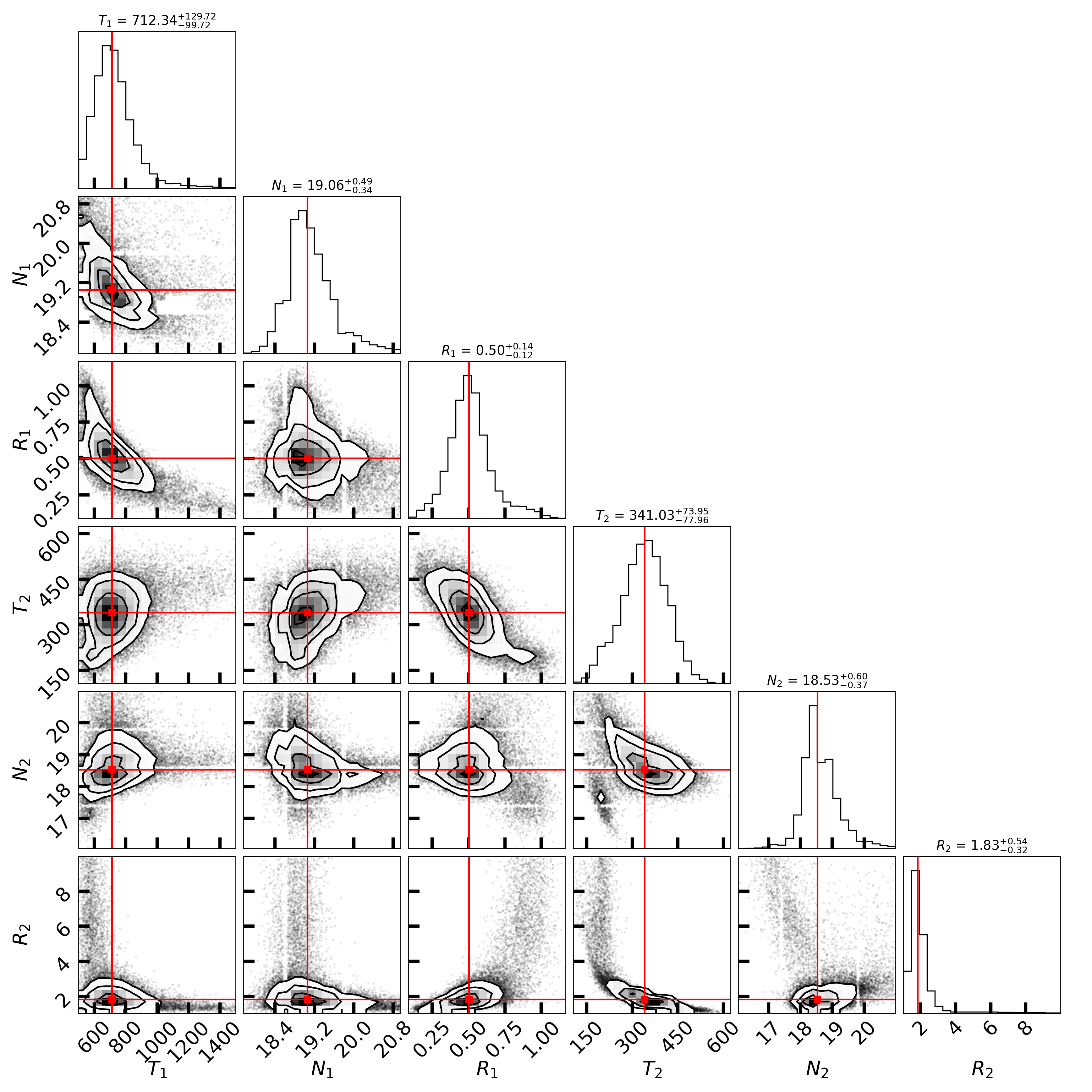}
    \caption{Similar as Figure \ref{fig:CP3}, but for the models with only two components.}
    \label{fig:CP3-2C}
\end{figure*}

\clearpage
\section{\ce{H_2 ^{18}O} over 17.0-27.5 $\mathrm{\mu}$m}
\begin{figure*}[ht!]
    \centering
    \includegraphics[width=\textwidth]{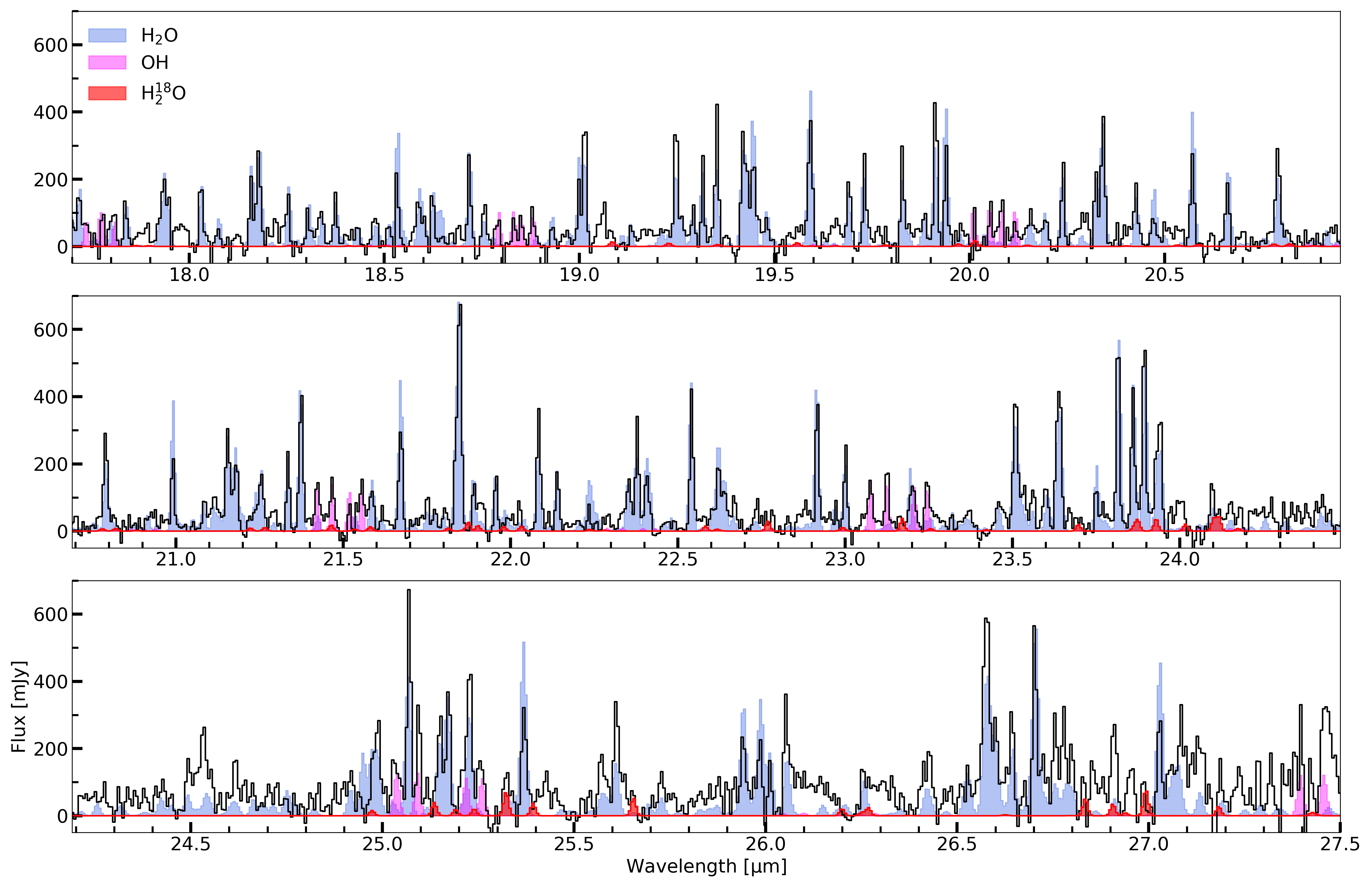}
    \caption{The spectra from subbands 4A to 4C from DR~Tau are shown from top to bottom. A \ce{H_2 ^{18}O} slab model is shown in red, to indicate the locations of the various transitions. The best fitting slab models of \ce{H_2O} and \ce{OH} are shown in, respectively, blue and magenta.}
    \label{fig:H218O-Spectrum}
\end{figure*}

\clearpage
\section{Searching for other molecular species}
\subsection{Atomic and molecular hydrogen}
\begin{figure*}[ht!]
    \centering
    \includegraphics[width=\textwidth]{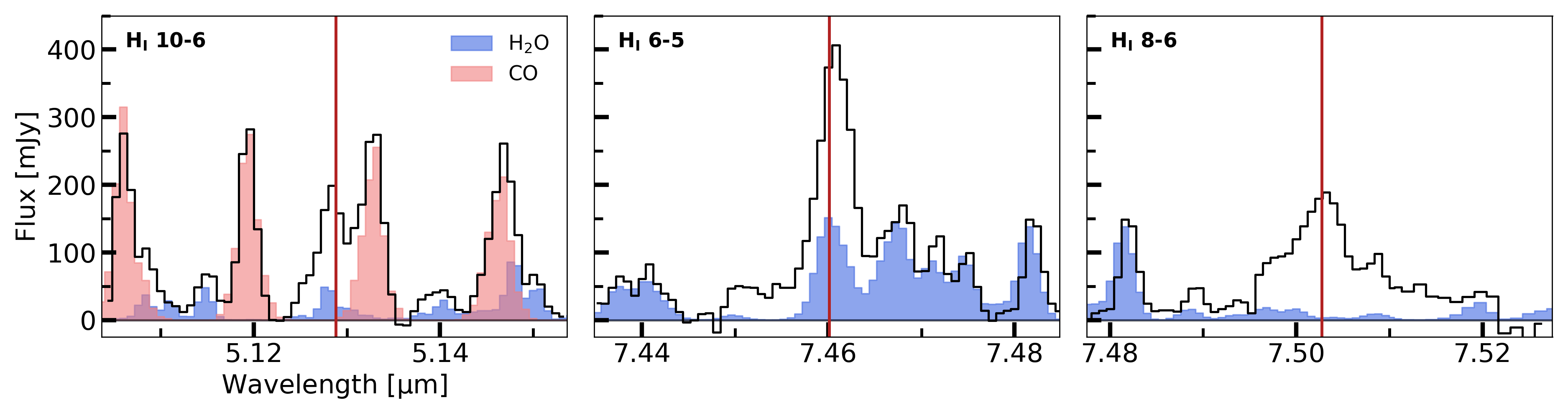}
    \caption{The identified atomic hydrogen transitions in DR~Tau. From left to right we report the detections of the \ce{H_I} 10-6, 6-5, and 8-6 transitions, respectively. The \ce{CO} (see \citealt{TemminkEA24}) and \ce{H_2O} slab models are shown for completeness and to show potential blending effects.}
    \label{fig:AtomicH}
\end{figure*}
\begin{figure*}[ht!]
    \centering
    \includegraphics[width=12cm]{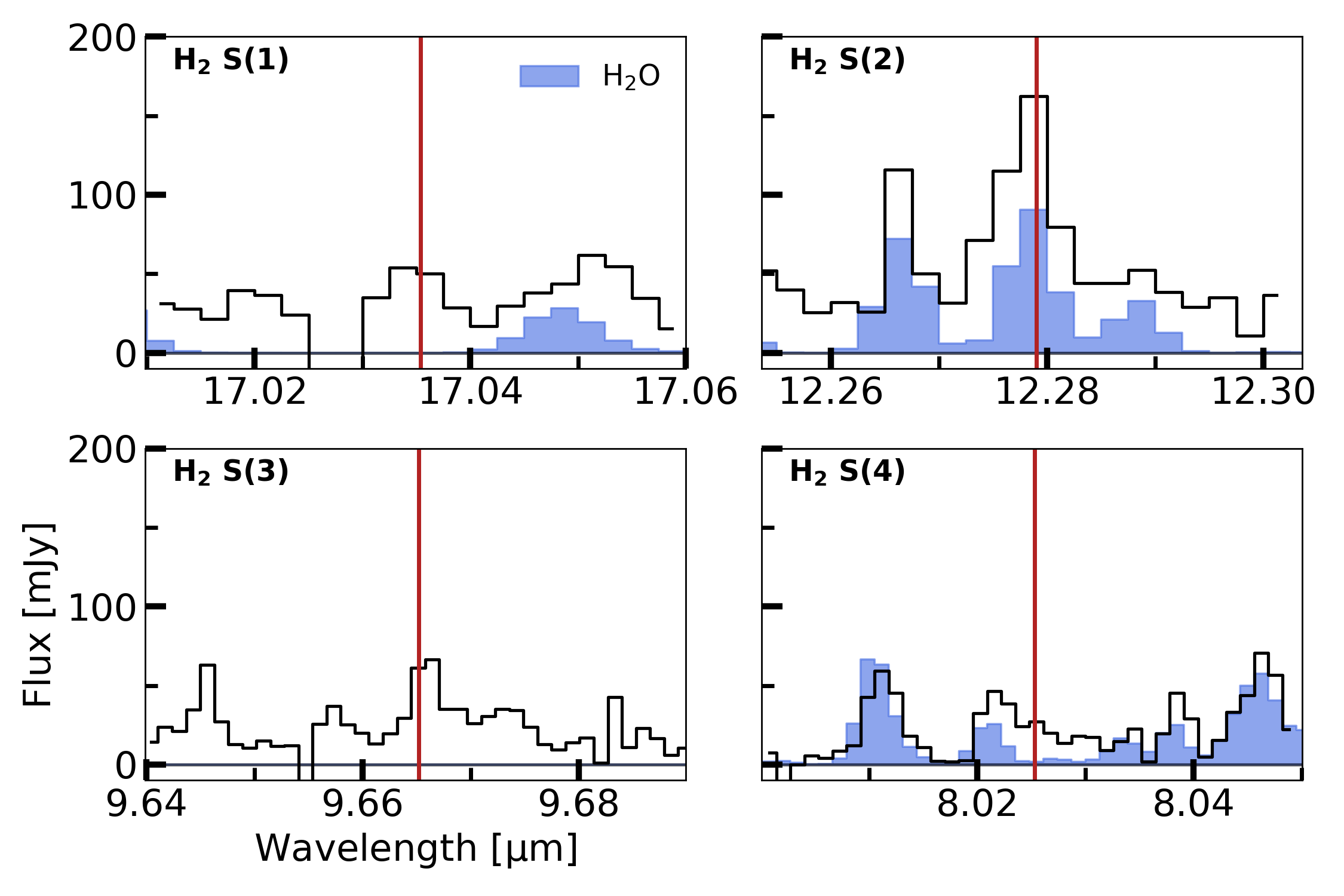}
    \caption{The potential detected \ce{H_2} transitions. The S(1), S(2), S(3), and S(4) transitions are displayed in, respectively, the top left, top right, bottom left, and bottom right panels. The \ce{H_2O} slab models are shown for completeness and to show potential blending effects.}
    \label{fig:MolecularH}
\end{figure*}
\begin{figure*}[ht!]
    \centering
    \includegraphics[width=\textwidth]{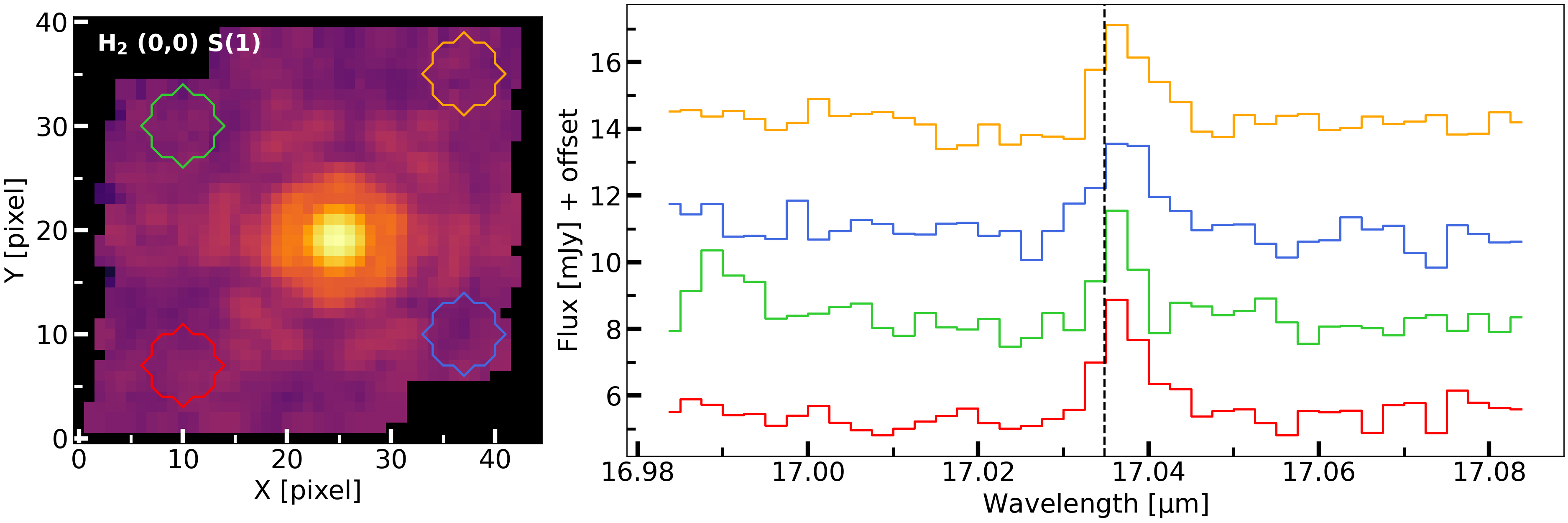}
    \caption{The spectra (right panel) of four off-source locations (as indicated in the left panel), showing the \ce{H_2} S(1) emission of the background cloud. The black dotted, vertical line indicates the unshifted wavelength of the transition, $\lambda_\textnormal{S(1)}$=17.035 $\mathrm{\mu}$m.}
    \label{fig:MolecularH-Spatial}
\end{figure*}

\clearpage
\subsection{Simple molecules}
\begin{figure*}[ht!]
    \centering
    \includegraphics[width=0.9\textwidth]{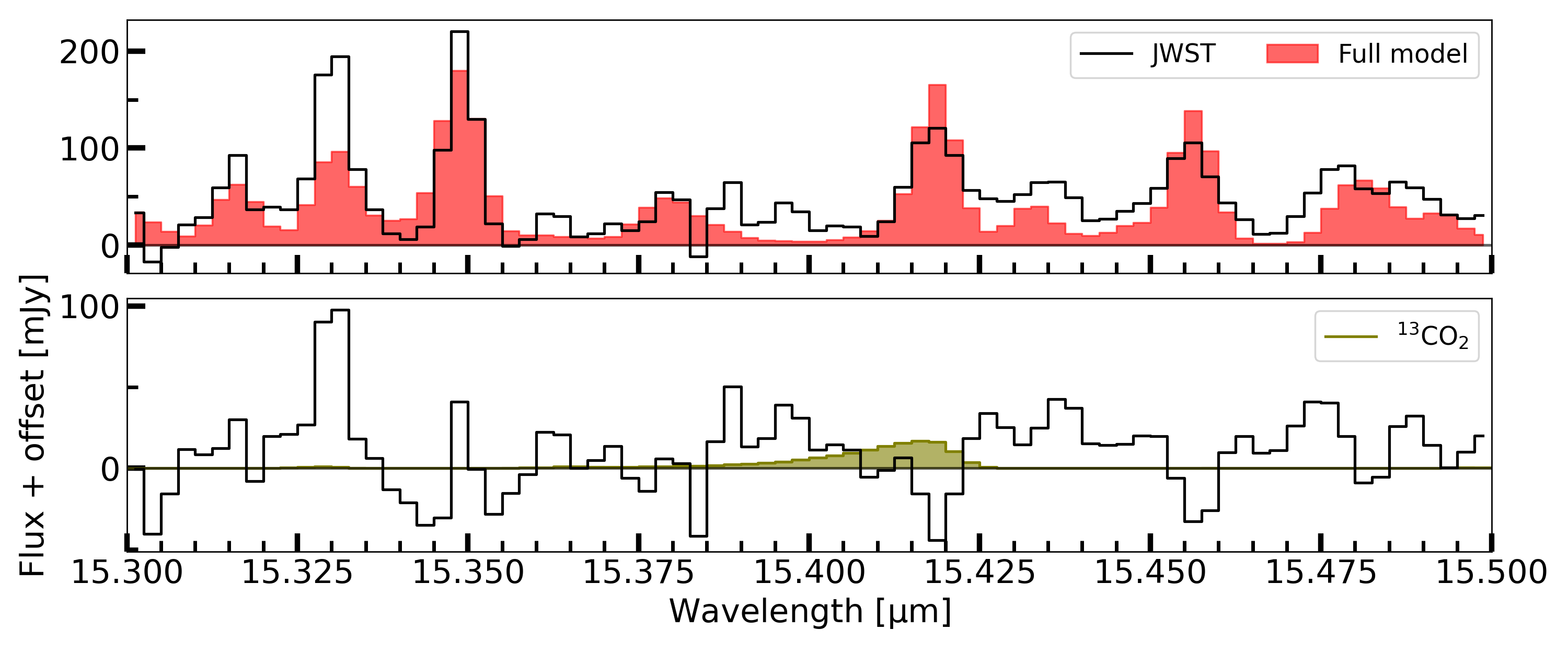}
    \includegraphics[width=0.9\textwidth]{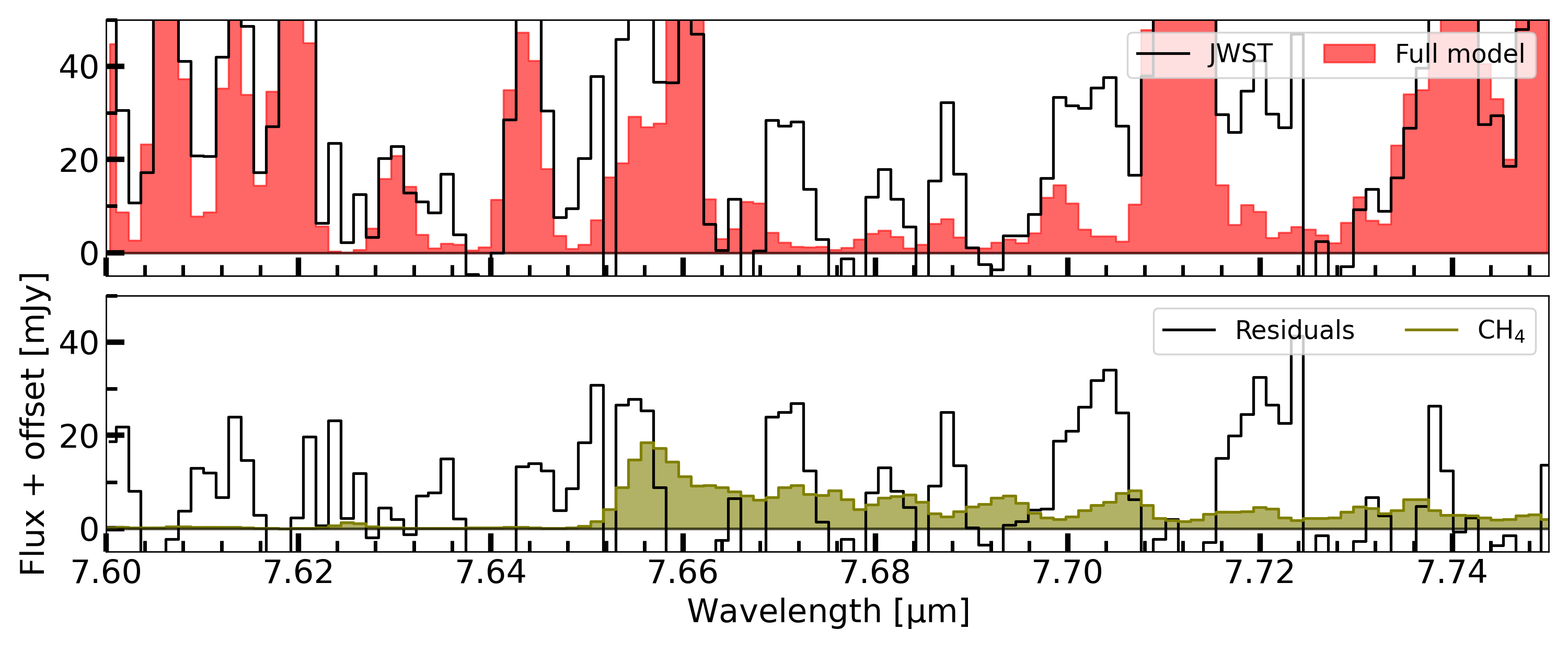}
    \includegraphics[width=0.9\textwidth]{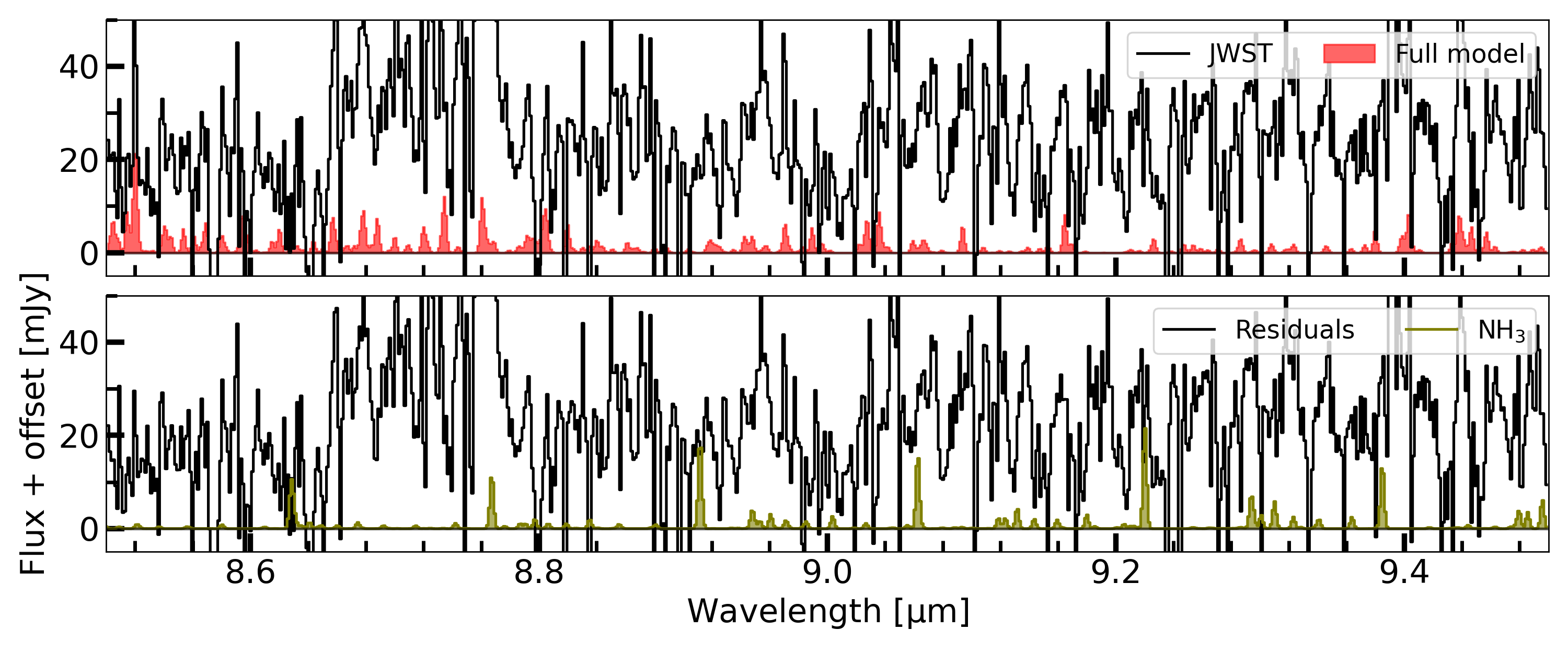}
    \caption{Slab models to investigate the maximum contribution of \ce{^{13}CO_2} (top figure), \ce{CH_4} (middle figure), and \ce{NH_3} (bottom panel) to the spectrum of DR~Tau. We include all confidently detected species to investigate potential blending effects. This include \ce{CO_2}, \ce{HCN}, and \ce{C_2H_2}, which all have been discussed in \citet{TemminkEA24}. The top panel shows the spectrum and the full model consisting of all observed species, while the bottom panel shows the residual spectrum (after subtracting off the full model) and slab models of the targeted species.}
    \label{fig:NDSlabs-1}
\end{figure*}
\begin{figure*}[ht!]
    \ContinuedFloat
    \centering
    \includegraphics[width=0.9\textwidth]{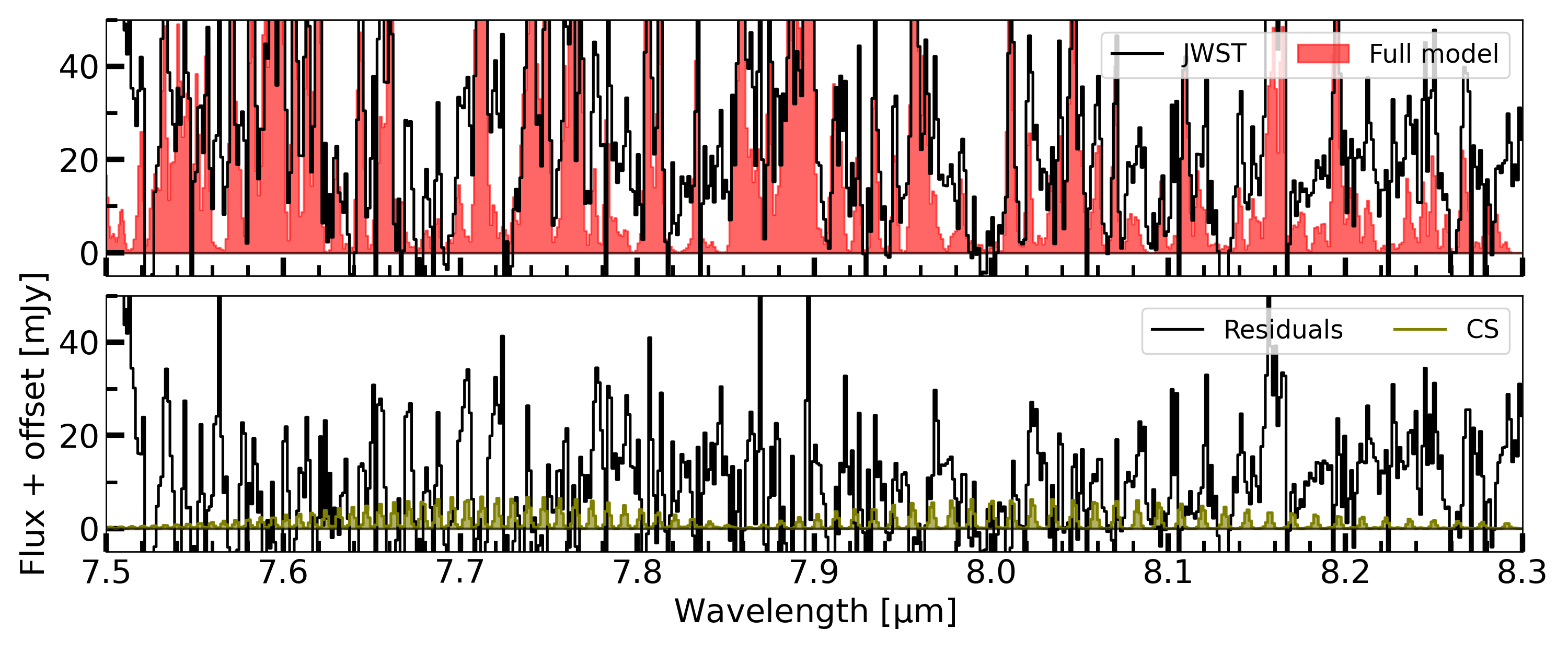}
    \includegraphics[width=0.9\textwidth]{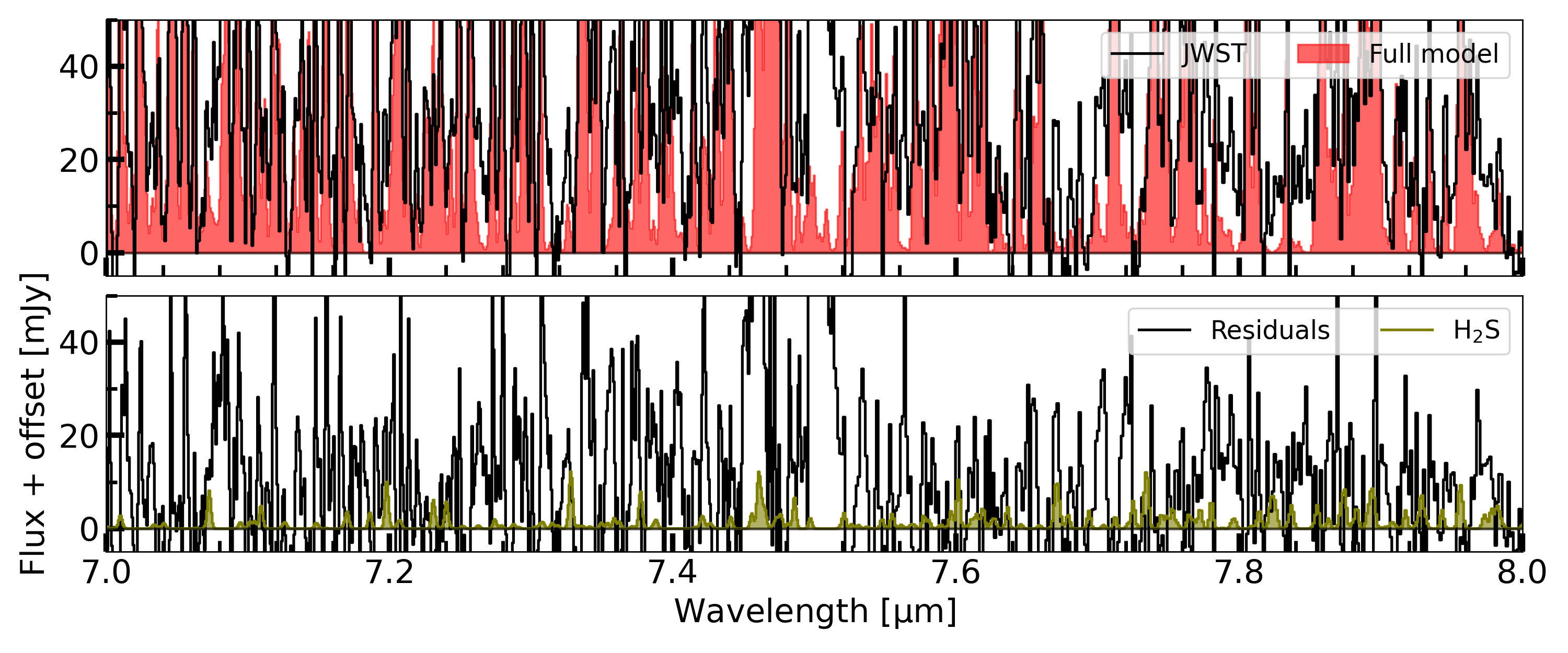}
    \includegraphics[width=0.9\textwidth]{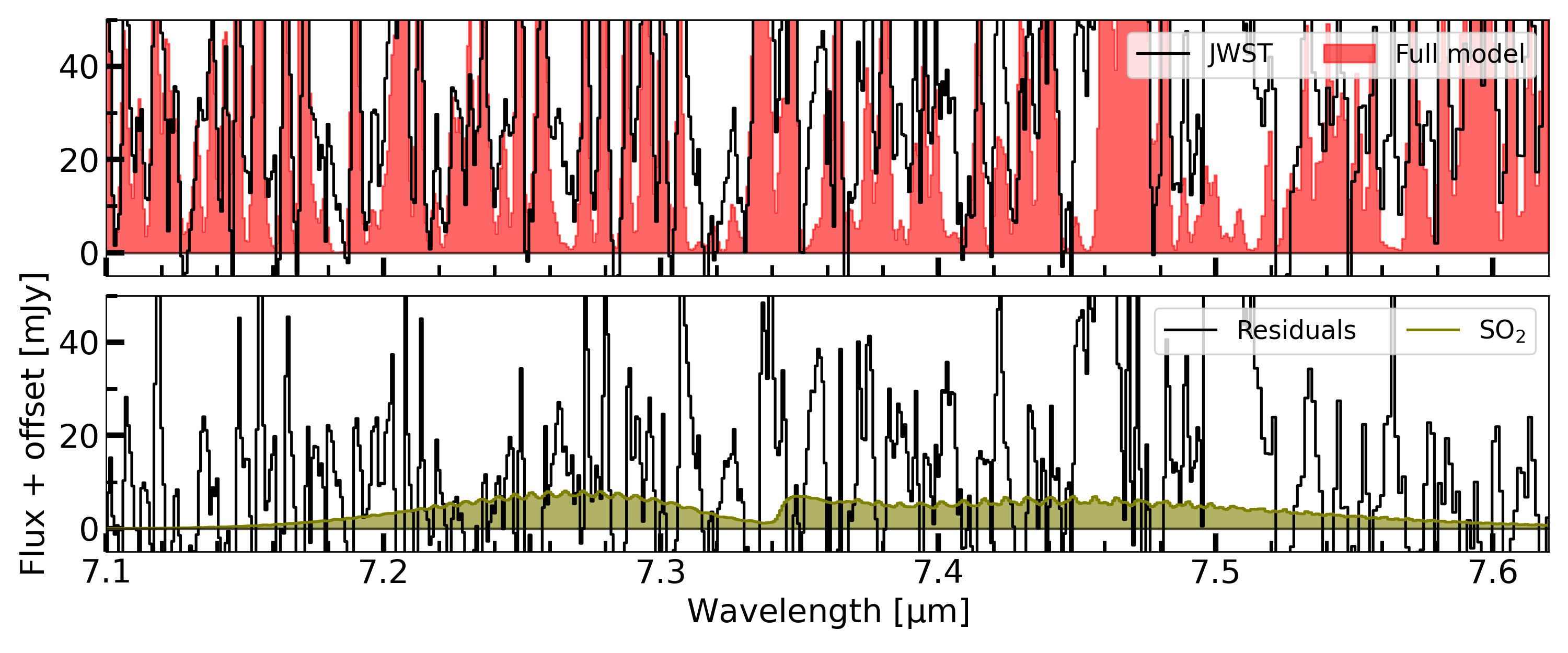}
    \caption{Continuation of Figure \ref{fig:NDSlabs-1}. The molecules shown here in green are \ce{CS} (top panel), \ce{H_2S} (middle panel), and \ce{SO_2} (bottom panel).}
    \label{fig:NDSlabs-2}
\end{figure*}

\clearpage
\subsection{Hydrocarbons}
\begin{figure*}[ht!]
    \centering
    \includegraphics[width=0.9\textwidth]{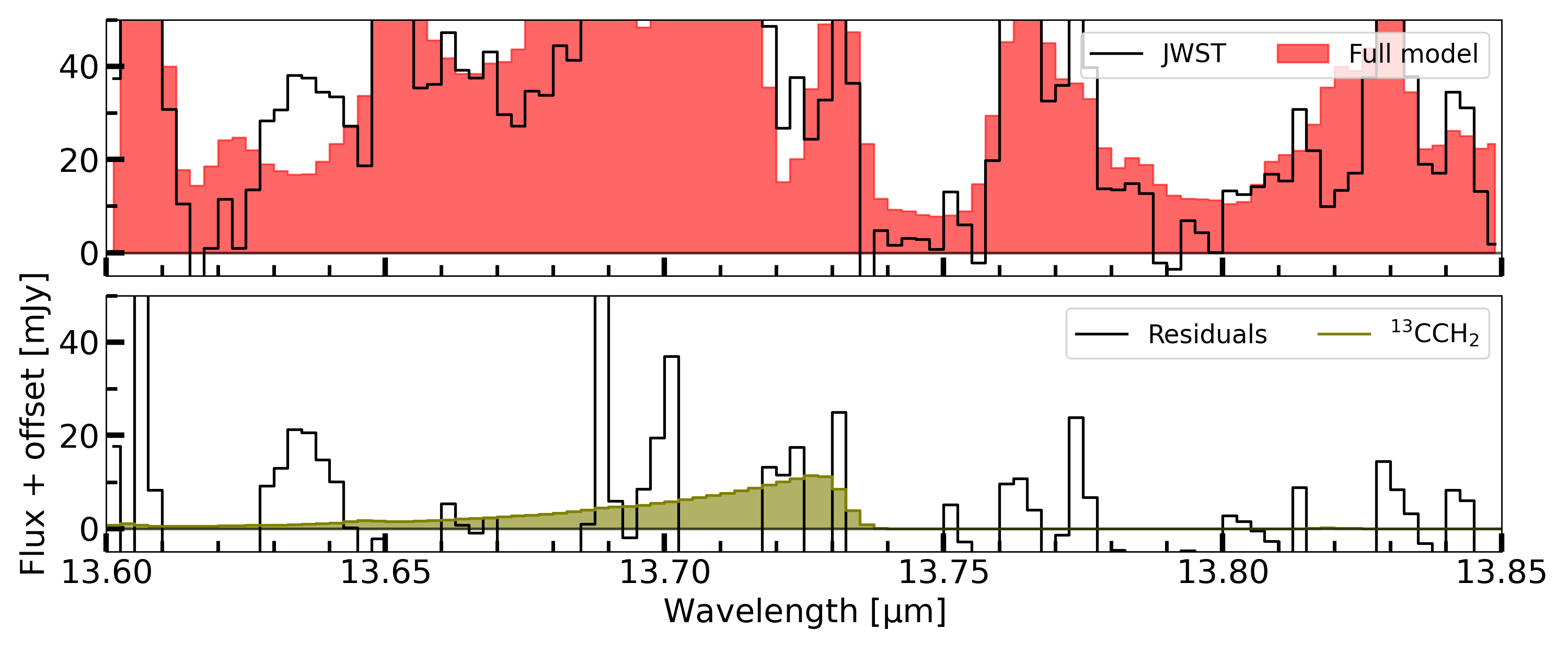}
    \includegraphics[width=0.9\textwidth]{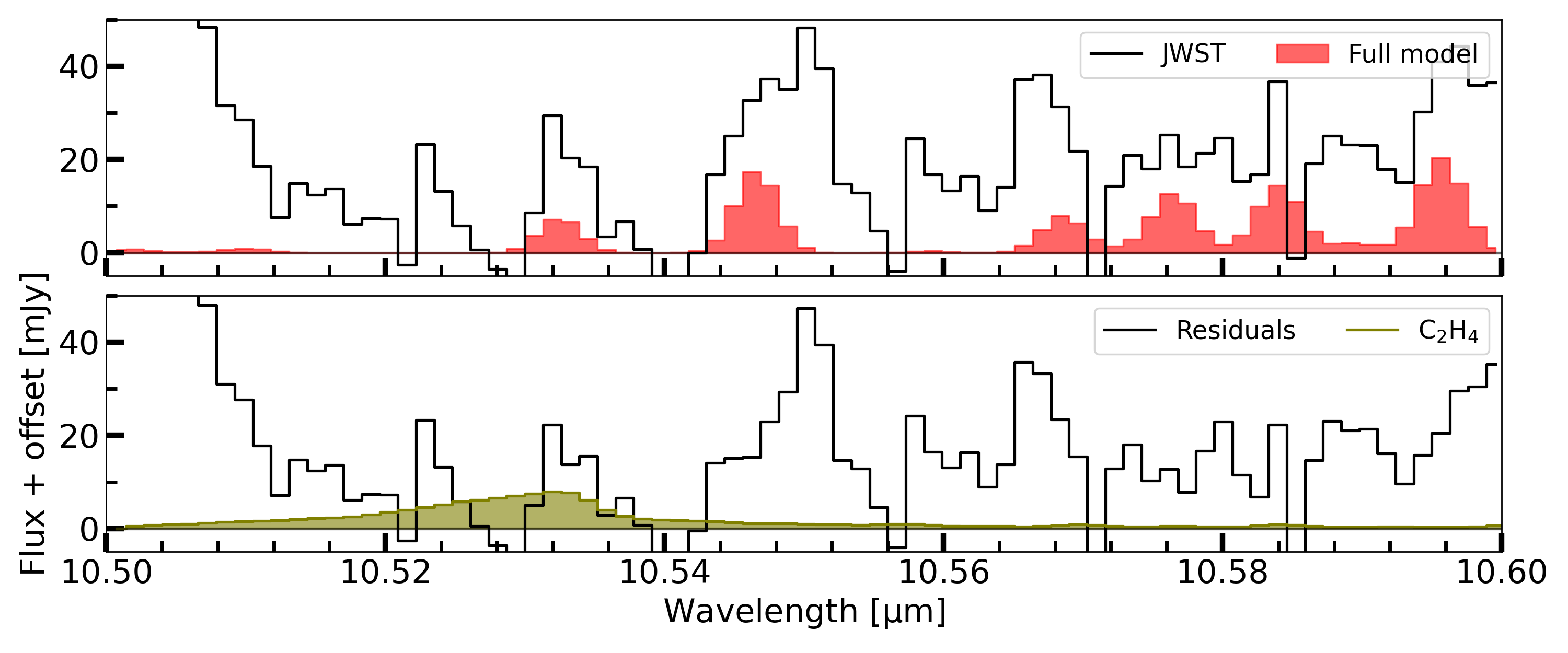}
    \includegraphics[width=0.9\textwidth]{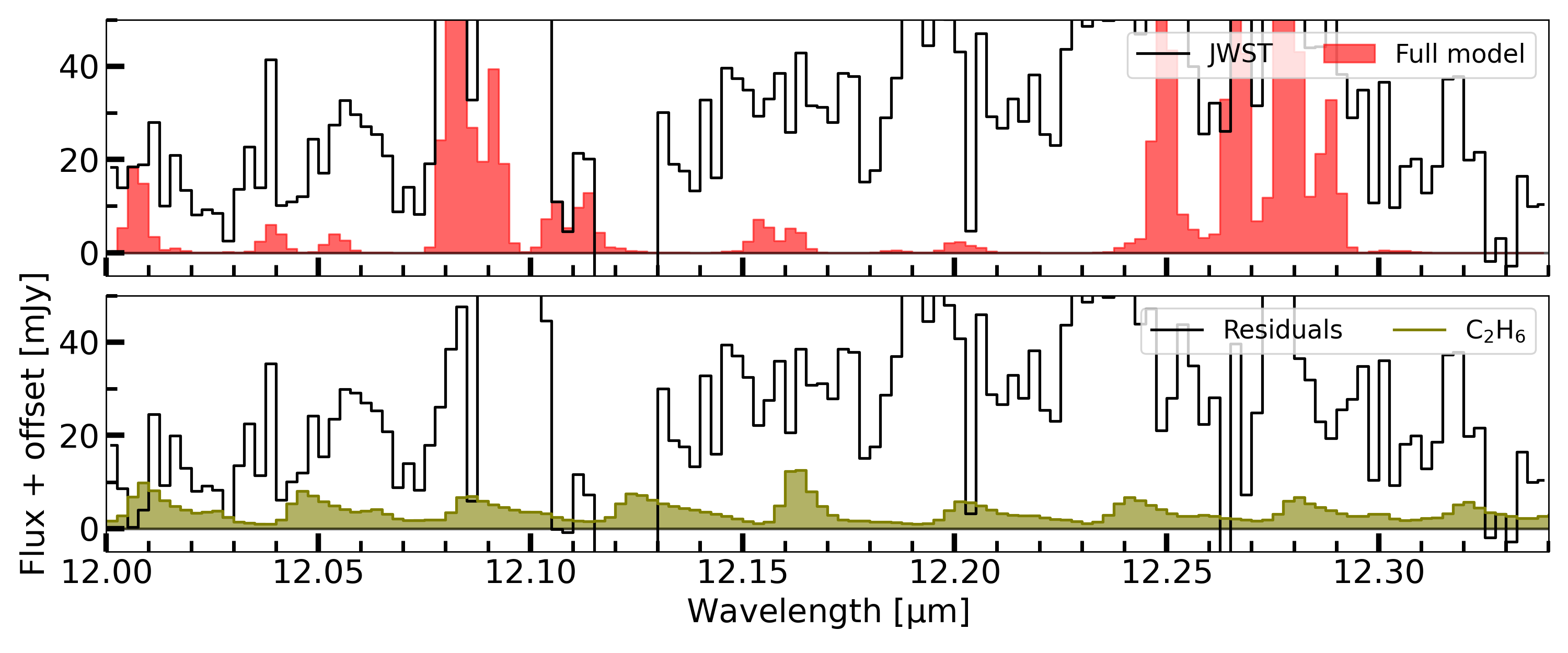}
    \caption{Slab models to investigate the detection of \ce{^{13}CCH_2} (top figure), \ce{C_2H_4} (middle figure), and \ce{C_2H_6} (bottom panel). We include all confidently detected species to investigate potential blending effects. This includes \ce{CO_2}, \ce{HCN}, and \ce{C_2H_2}, which all have been discussed in \citet{TemminkEA24}. The top panel shows the spectrum and the full model consisting of all observed species, while the bottom panel shows the residual spectrum (after subtracting off the full model) and slab models of the targeted species.}
    \label{fig:NDSlabs-3}
\end{figure*}
\begin{figure*}[ht!]
    \ContinuedFloat
    \centering
    \includegraphics[width=0.9\textwidth]{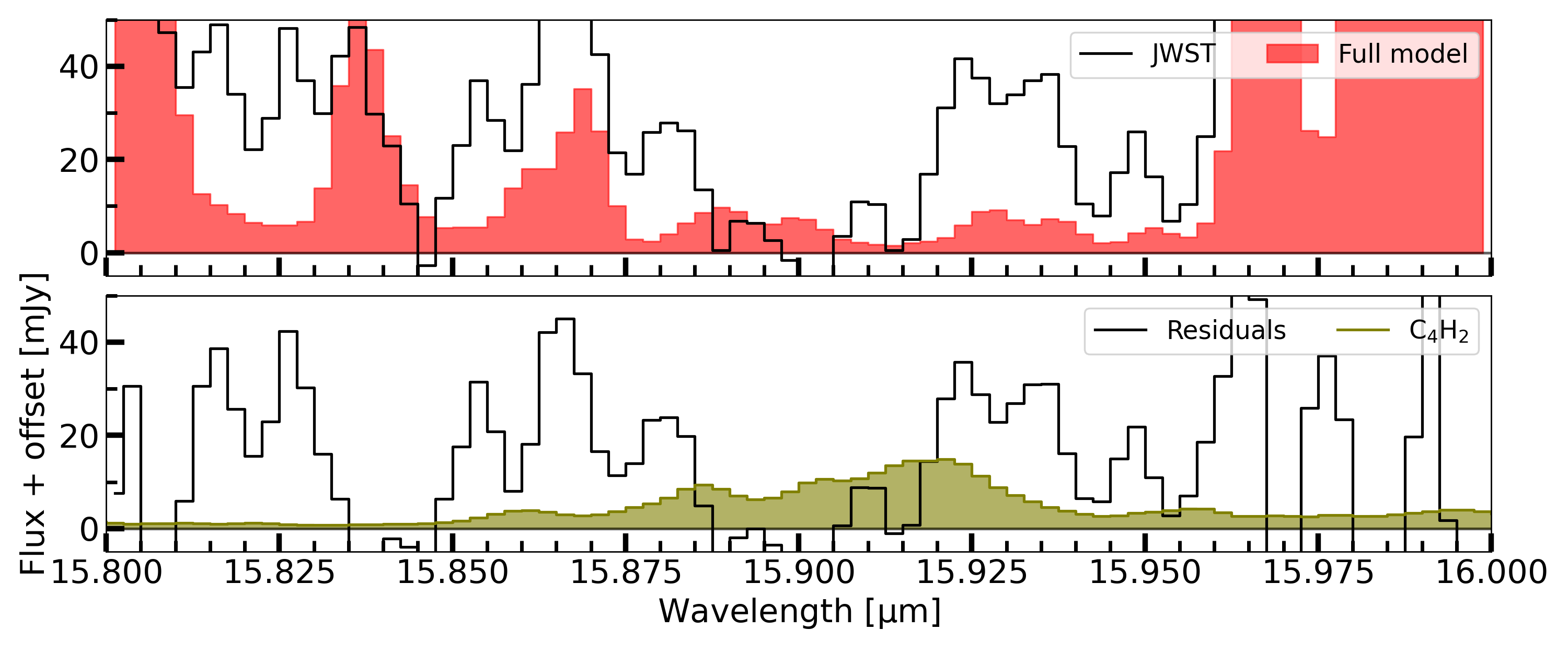}
    \includegraphics[width=0.9\textwidth]{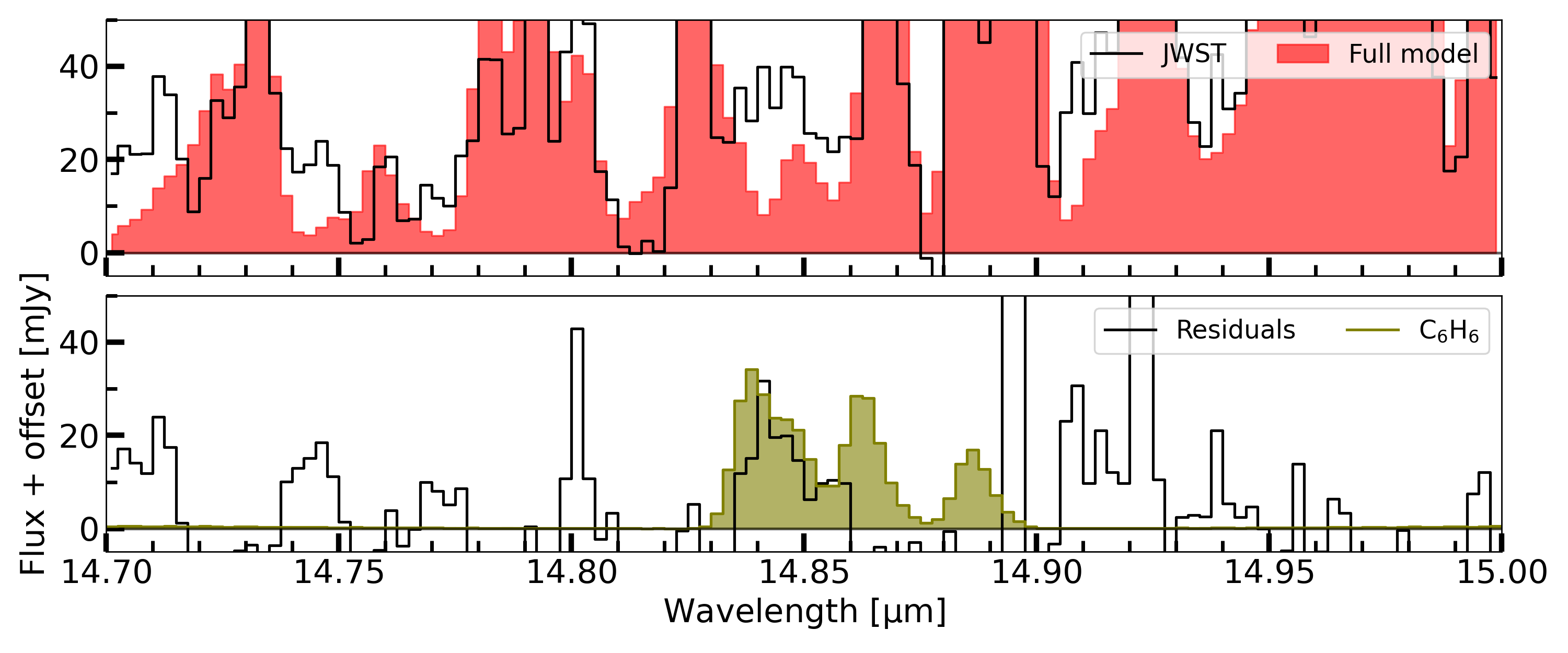}
    \caption{Continuation of Figure \ref{fig:NDSlabs-3}. The molecules shown here are \ce{C_4H_2} (top panel) and \ce{C_6H_6} (bottom panel).}
    \label{fig:NDSlabs-4}
\end{figure*}

\end{appendix}

\end{document}